\documentclass[aps,prd,twocolumn,superscriptaddress,nofootinbib,floatfix]{revtex4-2}

\usepackage[T1]{fontenc}
\usepackage[reqno]{amsmath}
\usepackage{amssymb,bm,mathtools}
\usepackage{graphicx}
\usepackage{xcolor}
\usepackage{tikz}
\usetikzlibrary{arrows.meta,positioning,calc,decorations.pathreplacing}
\usepackage{hyperref}
\hypersetup{colorlinks=true,linkcolor=blue,citecolor=blue,urlcolor=blue}
\allowdisplaybreaks

\def\e{\epsilon}
\def\Bpar{B}
\def\Z{\mathbb{Z}}
\def\fa{f_a}

\begin{document}

\title{\texorpdfstring{Generation as Compositeness:\\
A Subconstituent Interpretation of the\\
$B$-Lattice Flavor Hierarchy}{Generation as Compositeness: A Subconstituent Interpretation of the B-Lattice Flavor Hierarchy}}

\author{Vernon Barger}
\affiliation{Department of Physics, University of Wisconsin--Madison,
Madison, Wisconsin 53706, USA; barger@pheno.wisc.edu}

\date{June 2, 2026}

\begin{abstract}
We interpret the $\Bpar$-lattice flavor framework as a
compositeness hierarchy: all three fermion generations
are elementary chiral fields, but third-generation
Yukawa couplings are undressed ($Q=0$), while lighter
generations acquire their
Yukawa couplings through chains of spin-$0$
subconstituents (``hops'') whose depth is counted by
the $\Z_9$ discrete gauge symmetry.
The $\Z_9$ charge admits a two-index decomposition
$q_9\mapsto(a,b)$ that identifies
two hop species ($\alpha$, $\beta$) and organizes
all fundamental scales from $v_{\rm EW}$ to $M_{\rm Pl}$
on a ``ninths ladder'' $\Lambda\times\e^{n/9}$.
The lattice structure yields the CKM and
PMNS mixing
parameterizations (with all mixing exponents
expressible as charge differences
$\Delta Q$ dressed by a universal Fritzsch--Xing
phase shift of $\pm 1/9$), the integer-exponent
identity $|V_{ub}|\simeq(m_\mu/m_\tau)^2$
connecting quark mixing to charged-lepton masses,
the cross-sector relation
$|V_{ub}|\simeq\sin^3\theta_{13}$ linking the
smallest CKM and PMNS magnitudes,
a seesaw benchmark
$m_3\simeq 51$~meV, the axion mass window
$m_a\sim 7$--$12\;\mu$eV, and the prediction
$\tan\beta\simeq 10$--$16$ (from the chain
internal factor combined with the DFSZ-II
two-Higgs-doublet structure), all from
two parameters ($\Lambda$ and $\e = 14/75$).
Generation-dependent Peccei--Quinn charges restore the
axion--photon coupling ($C_{a\gamma}\simeq 0.6$--$1.0$)
from ``back to invisible'' suppression.
An illustrative UV realization in terms of
hypercolor-confined scalars communicated to the SM
by a gauge-invariant messenger chain is presented
as an existence proof.
\end{abstract}

\maketitle

\section{Introduction}
\label{sec:intro}

\emph{Summary.}---This paper proposes that the three
fermion generations correspond to three levels of
compositeness, with the generation index counting the
number of fundamental subconstituents (``hops'')
in the Yukawa coupling.
The framework yields the following principal results
from two parameters ($\Lambda$ and $\e = 14/75$):
(i)~a ``ninths ladder'' that places every
fundamental scale from $v_{\rm EW}$ to $M_{\rm Pl}$
on a single geometric sequence
$\Lambda\times\e^{n/9}$;
(ii)~an algebraic decomposition of both the
CKM and PMNS mixing matrices in terms of
hop charge differences $\Delta Q$:
the three CKM magnitudes
$|V_{us}|\sim\e^{\Delta Q(Q)_{12}-1/9}$,
$|V_{cb}|\sim\e^{\Delta Q(Q)_{23}-1/9}$,
$|V_{ub}|\sim\e^{\Delta Q(Q)_{13}+Q(d^c_2)}$
yield the Cabibbo master identity
$\e = |V_{us}|^{9/8}$ and the $\Bpar$-lattice
refinement of the Wolfenstein hierarchy
$|V_{cb}| = |V_{us}|^{17/8}$,
$|V_{ub}| = |V_{us}|^{15/4}$;
the three PMNS angles
$\sin\theta_{23}\sim\e^{1/6}$,
$\sin\theta_{12}\sim\e^{1/3}$,
$\sin\theta_{13}\sim\e^{10/9}$ are all expressible
as powers of $m_\mu/m_\tau$ through the identity
$\e = (m_\mu/m_\tau)^{3/5}$;
both sectors share a universal Fritzsch--Xing
phase correction of $\pm 1/9$ (one $\alpha$-hop
quantum);
(iii)~a sharp benchmark for the heaviest
neutrino mass $m_3\simeq 51$~meV
(data: $\sim 50$~meV), from the seesaw once the
Majorana scale is identified with the ninths-lattice
position $M_0 = \Lambda\,\e^{-28/9}$;
(iv)~the axion--photon coupling
$C_{a\gamma}\simeq 0.6$--$1.0$, restored from
``back to invisible'' suppression by the
generation-dependent Peccei--Quinn (PQ) charges;
(v)~the strict prediction of null signals for
TeV-scale compositeness searches, with the axion mass
window $m_a\sim 7$--$12\;\mu$eV as the primary
experimental target;
(vi)~a down-type chain internal tunneling factor
$\e^{7/9}\simeq 0.27$ that suppresses the absolute
bottom Yukawa relative to the top; combined with the
two-Higgs-doublet structure already required by
the DFSZ-II axion model, this predicts
$\tan\beta\simeq 10$--$16$.

A master table of quantitative results is
given in Sec.~\ref{sec:summary}.

\emph{Motivation: a unified attack on multiple SM puzzles.}
Beyond accounting for the flavor hierarchy, the hop
framework offers structural responses to several
long-standing problems of the SM and its extensions.
(a)~\textit{Flavor hierarchy:} all twelve fermion
masses are determined by the electroweak VEV $v$, the
top mass $m_t$, and rational powers of $\Bpar = 75/14$.
(b)~\textit{Strong CP problem:} the flavon $\Phi$ is
simultaneously the Peccei--Quinn field, providing a QCD
axion at $\fa = \langle\Phi\rangle\sim 10^{12}$~GeV with
mass in the $7$--$12\;\mu$eV ADMX-accessible range.
(c)~\textit{Gauge hierarchy problem:}
$\Lambda^2/M_{\rm Pl}\sim 10^{6}$~GeV emerges naturally
as the SUSY-breaking scale, giving mini-split
superpartners that protect the Higgs mass from
quadratic divergences without introducing $\sim 1$~TeV
states excluded by LHC.
(d)~\textit{Dark matter:} the same axion that solves the
strong CP problem provides $\Omega h^2\simeq 0.11$ at the
predicted $\fa$.
(e)~\textit{Neutrino masses:} the parameter-free
prediction $m_3 = \tfrac{1}{2}v\,\e^{17}\simeq 50$~meV
follows from the seesaw with the Majorana scale on the
ninths ladder.
(f)~\textit{Absence of TeV-scale flavor and cLFV
signals:} all messenger-mediated processes are
suppressed by $(M_Z/\Lambda)^4\sim 10^{-40}$, making
the framework consistent with the negative results of
all flavor-precision experiments.
The compositeness interpretation introduced here
provides the underlying physical picture explaining
\emph{why} the FN charges take the values they do, and
identifies a single binding scale $\Lambda$ that ties
together the flavor structure, the axion physics, the
SUSY-breaking scale, and the neutrino sector.

\medskip

The $\Bpar$-lattice program~\cite{PaperI,TwoOverTwo,Companion,LeptonLattice,FlavorInNinths,UFP}
has established that the quark and lepton mass hierarchies,
the Cabibbo--Kobayashi--Maskawa (CKM)~\cite{Cabibbo,KobayashiMaskawa} and
Pontecorvo--Maki--Nakagawa--Sakata (PMNS)
mixing patterns~\cite{MakiNakagawaSakata,Pontecorvo},
and CP violation can be organized by a single expansion
parameter $\e = 1/\Bpar = 14/75\approx 0.187$, with all
Froggatt--Nielsen (FN) exponents quantized in units of
$1/9$~\cite{FN,FN1980}.
The framework is underpinned by a $\Z_{18}$ discrete gauge
symmetry whose $\Z_9$ subgroup enforces the ninths
quantization, while the full $\Z_{18}$ protects the
Peccei--Quinn (PQ) symmetry~\cite{PecceiQuinn1977,PecceiQuinn1977b}
and the quantum
chromodynamics (QCD) axion against
gravitational spoiling up to dimension-18
operators~\cite{FlavorInNinths,BabuGogoladzeWang,BhattiproluMartin}.

The mathematical structure of the framework---additive
charges, a discrete cyclic symmetry, a chain of heavy
vectorlike mediators---admits a natural physical
interpretation that goes beyond the standard FN picture
of flavor as an abstract quantum number.
In this paper we explore the hypothesis that
\emph{the generation index is a compositeness index}:
all three generations are elementary chiral fields,
but their Yukawa couplings differ in depth---the
third generation couples to the Higgs without hop
dressing, while lighter generations acquire Yukawa
couplings through chains of subconstituents whose
presence suppresses their coupling to the Higgs field.

We emphasize at the outset a key distinction:
``compositeness'' in this framework refers to the
structure of the \emph{Yukawa coupling}, not of
the fermion itself.
All three SM generations are elementary chiral fields
in the Lagrangian at all energy scales
(Sec.~\ref{sec:chiral-protection}).
The hops are physical fields---spin-$0$ scalars
carrying hypercolor charge---whose gauge quantum
numbers are specified in
Sec.~\ref{sec:UV-Lagrangian}.
In the illustrative UV realization presented there,
hops propagate above the confinement scale $\Lambda$
as hypercolor-fundamental scalars in a
gauge-invariant messenger chain, and integrating out
the confined sector below $\Lambda$ produces the
Froggatt--Nielsen effective operators.
No spin problem arises because the elementary
spin-$1/2$ core is always present; the spin-$0$ hops
dress the Yukawa coupling without altering the
fermion's Lorentz quantum numbers.

The idea that quarks and leptons might be composites
of more fundamental objects (``preons'') has a long
history~\cite{PatSal,HarariShupe,Shupe}.
The Harari--Shupe rishon model~\cite{HarariShupe,Shupe},
developed into a full dynamical theory by Harari and
Seiberg~\cite{HarariSeiberg1981,HarariSeibergNPB},
postulates two fundamental fermions ($T$ and $V$) whose
combinations reproduce the gauge quantum numbers of the
Standard Model (SM) fermions; the generation problem
was addressed through discrete symmetry
labels~\cite{HarariSeibergGen,HarariSeibergChiral}.
Compositeness models were subsequently developed by many
authors~\cite{Fritzsch1981,Terazawa,AbbottFarhi}, and
phenomenological bounds from contact interactions and
form factors have been
placed~\cite{Eichten1983,PDG2024}.
What distinguishes the present approach from these earlier
proposals is that the compositeness is in the \emph{flavor}
sector, not the gauge sector: all three generations have
identical gauge quantum numbers, and the subconstituent
structure manifests exclusively through mass and mixing
hierarchies. Furthermore, because the hop confinement scale
is rigidly locked to the axion decay constant at
$\Lambda\sim 3\times10^{12}$~GeV, the framework strictly
predicts a ``null signal'' for traditional compositeness
searches. Unlike models that anticipate novel flavor-changing
neutral currents or contact interactions at current colliders,
testability in the hop framework is driven entirely by
cosmological and dark matter observables.
We call the new subconstituents \emph{hops}, since they
are the quanta exchanged at each nearest-neighbor hop of
the vectorlike fermion chain.
The two species are the $\alpha$-hop and the $\beta$-hop,
corresponding to the two independent step directions
in the $(a,b)$ decomposition of the $\Z_9$ charge
(Sec.~\ref{sec:Z3Z3}).
Both are complex scalars carrying hypercolor charge
(Sec.~\ref{sec:UV-Lagrangian}), distinct from the
hypercolor-singlet flavon $\Phi$ that supplies the
Peccei--Quinn breaking.
Since they carry $\Z_9$ charge, the hop
(e.g.\ $\alpha$, $q_9=+1$) and anti-hop
($\bar\alpha$, $q_9=8$) are distinct
particles; similarly
$\beta$ ($q_9=3$) and $\bar\beta$ ($q_9=6$).
This distinction is essential for the
hypercolor bound-state spectrum.

The paper is organized as follows.
Section~\ref{sec:additive} reviews the additive charge
structure of the $\Bpar$-lattice and reinterprets it as a
compositeness counting rule.
Section~\ref{sec:Z3Z3} shows that the $\Z_9$ charge
admits a two-index decomposition identifying two hop
species.
Section~\ref{sec:chain} interprets the vectorlike quark
chain as a confining flux tube between hops.
Section~\ref{sec:hop-masses} derives the hop masses
from the chain propagator structure, obtaining
$m_\alpha\simeq f_a/|V_{us}|$ and
$m_\beta\simeq f_a\,\Bpar^{2/3}$
(here $\Bpar = 75/14$ is the lattice parameter,
not baryon number).
Section~\ref{sec:harari} draws parallels and distinctions
with the Harari--Shupe rishon model.
Section~\ref{sec:BL} shows how baryon and lepton number
are preserved as gauge quantum numbers of the elementary
core, independent of hop content, and explains the
anomaly-freedom of baryon-minus-lepton number ($B\!-\!L$)
as a consistency condition.
Section~\ref{sec:SU5} connects the $SU(5)$ origin of
the lattice parameter $\Bpar=75/14$ to the binding dynamics.
Section~\ref{sec:EWSB} connects the Higgs sector
to the hop framework, identifying the top quark as the
elementary benchmark and showing that the down-type
chain internal factor $\e^{7/9}\simeq 0.27$, combined
with the DFSZ-II two-Higgs-doublet structure, predicts
$\tan\beta\simeq 10$--$16$.
Section~\ref{sec:nuR} shows that right-handed neutrino
hop charges cancel in the seesaw, and that the Majorana
scale $M_0\simeq\Lambda\,\e^{-28/9}$ lies on the ninths
lattice.
Section~\ref{sec:GUT} presents the ``ninths ladder'' of
fundamental scales from $v_{\rm EW}$ to $M_{\rm Pl}$.
Section~\ref{sec:predictions} discusses testable
consequences, emphasizing the predicted absence of
Large Hadron Collider (LHC) and flavor-factory anomalies
in favor of targeted axion searches.
Section~\ref{sec:mixing} derives the CKM and PMNS
mixing angles algebraically from hop charge differences,
establishing a unified charge-arithmetic description
of flavor mixing in both sectors.
Section~\ref{sec:summary} summarizes.

\section{Additive Charges as Compositeness}
\label{sec:additive}

\subsection{The suppression rule revisited}

The central result of the $\Bpar$-lattice
program~\cite{TwoOverTwo,FlavorInNinths} is that
every Yukawa coupling in a given charge sector
(up-type, down-type, or charged-lepton) is given by
\begin{equation}
Y_{ij} = c_{ij}\,\e^{\,p_{ij}},\qquad
p_{ij} = Q(\psi_i) + Q(\psi^c_j) + \Delta_{\rm int},
\label{eq:supp-rule}
\end{equation}
where $Q(\psi_i)$ and $Q(\psi^c_j)$ are the FN charges
of the left- and right-handed fermion fields, $\e = 14/75$,
$c_{ij}$ is an $\mathcal{O}(1)$ complex coefficient,
and $\Delta_{\rm int}$ is the \emph{internal chain factor}:
a universal constant for each charge sector arising from
the tunneling amplitude through the bulk of the
vectorlike fermion chain (Sec.~\ref{sec:chain}).
For the down-type chain, $\Delta_{\rm int}^d = 7/9$;
for the up-type chain, $\Delta_{\rm int}^u = 0$
(the top quark couples directly).
Because $\Delta_{\rm int}$ is common to all entries of
a given Yukawa matrix, it cancels in \emph{mass ratios}
within each sector:
\begin{equation}
\frac{m_i}{m_j}\sim
\e^{[Q(\psi_i)+Q(\psi^c_i)]-[Q(\psi_j)+Q(\psi^c_j)]},
\end{equation}
and the endpoint charges alone control the hierarchy.
However, $\Delta_{\rm int}$ does affect the
\emph{absolute} Yukawa couplings and thereby the ratio
of masses between different sectors---notably
$m_b/m_t$---as shown in Sec.~\ref{sec:EWSB}.

The charges for the three quark doublet generations are
\begin{equation}
Q(Q_1) = 3,\qquad Q(Q_2) = 2,\qquad Q(Q_3) = 0.
\label{eq:QL-charges}
\end{equation}

In the standard FN interpretation, these are abstract
flavor charges assigned under the $\Z_9$ symmetry.
The mass hierarchy $m_1 \ll m_2 \ll m_3$ arises because
larger charges require more flavon insertions, each
contributing a factor of $\e\approx 0.19$.

\subsection{Reinterpretation: charge as structure}

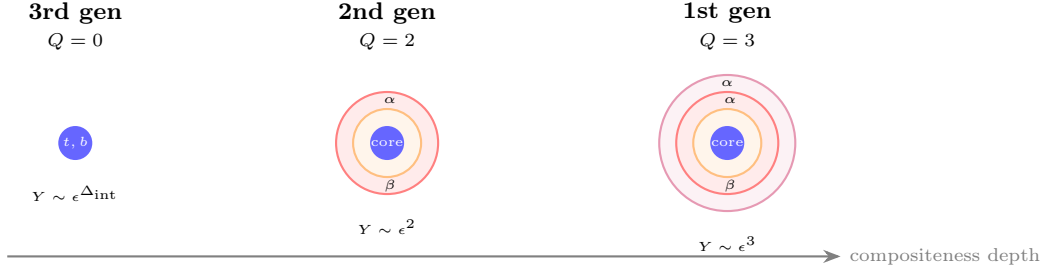
\begin{figure*}[!htbp]
\centering
\begin{tikzpicture}[scale=0.75,>=Stealth]
\begin{scope}[shift={(0,0)}]
  \node[font=\small\bfseries] at (0,2.3) {3rd gen};
  \node[font=\scriptsize] at (0,1.8) {$Q=0$};
  \fill[blue!60] (0,0) circle (3mm);
  \node[font=\tiny,white] at (0,0) {$t,b$};
  \node[font=\tiny,below] at (0,-0.6) {$Y\sim\e^{\Delta_{\rm int}}$};
\end{scope}
\begin{scope}[shift={(5.5,0)}]
  \node[font=\small\bfseries] at (0,2.3) {2nd gen};
  \node[font=\scriptsize] at (0,1.8) {$Q=2$};
  \draw[thick,red!50,fill=red!8] (0,0) circle (9mm);
  \draw[thick,orange!50,fill=orange!8] (0,0) circle (6mm);
  \fill[blue!60] (0,0) circle (3mm);
  \node[font=\tiny,white] at (0,0) {core};
  \node[font=\tiny] at (0.05,0.75) {$\alpha$};
  \node[font=\tiny] at (0.05,-0.75) {$\beta$};
  \node[font=\tiny,below] at (0,-1.2) {$Y\sim\e^2$};
\end{scope}
\begin{scope}[shift={(11.5,0)}]
  \node[font=\small\bfseries] at (0,2.3) {1st gen};
  \node[font=\scriptsize] at (0,1.8) {$Q=3$};
  \draw[thick,purple!40,fill=purple!5] (0,0) circle (12mm);
  \draw[thick,red!50,fill=red!8] (0,0) circle (9mm);
  \draw[thick,orange!50,fill=orange!8] (0,0) circle (6mm);
  \fill[blue!60] (0,0) circle (3mm);
  \node[font=\tiny,white] at (0,0) {core};
  \node[font=\tiny] at (0.05,0.75) {$\alpha$};
  \node[font=\tiny] at (0.05,-0.75) {$\beta$};
  \node[font=\tiny] at (0,1.05) {$\alpha$};
  \node[font=\tiny,below] at (0,-1.5) {$Y\sim\e^3$};
\end{scope}
\draw[->,thick,gray] (-1.2,-2) -- (13.5,-2)
  node[right,font=\scriptsize] {compositeness depth};
\end{tikzpicture}
\caption{Compositeness depth of the three quark-doublet generations.
All three generations are elementary chiral fields;
the blue core represents the undressed third generation
($Q=0$), whose Yukawa coupling is suppressed only by the
sector's internal chain factor (Sec.~\ref{sec:chain}).
Each successive ``shell'' of
hop binding ($\alpha$ and $\beta$ layers) adds one power of
$\e\approx 0.19$ to the wavefunction-overlap suppression.
The left-handed charges $Q(Q_i) = (3,2,0)$ count the total
number of hop layers dressing each generation's Yukawa coupling.}
\label{fig:depth}
\end{figure*}

We propose the following reinterpretation.
The FN charge $Q(\psi_i)$ counts the number of
\emph{units of internal structure} carried by the
fermion $\psi_i$:
\begin{itemize}
\item Third generation ($Q=0$): undressed.
No hop content; the Yukawa coupling to the
Higgs is suppressed only by the sector's internal
chain factor $\e^{\Delta_{\rm int}}$
(unity for up-type, $\e^{7/9}\simeq 0.27$ for
down-type).
\item Second generation ($Q=2$): in the compositeness
picture, an elementary fermion whose Yukawa coupling is
dressed by
two units of subconstituent content.
Relative to the third generation, the Yukawa coupling
is further suppressed by $\e^2\approx 0.035$,
reflecting two additional FN insertions through the
vectorlike quark (VLQ) chain.
\item First generation ($Q=3$): the most deeply
``dressed'' state,
carrying three units.
Relative to the third generation, the Yukawa coupling
is further suppressed by $\e^3\approx 0.0065$.
\end{itemize}
We emphasize that this ``compositeness'' language
describes the structure of the \emph{Yukawa coupling},
not of the fermion: the SM fermions are
elementary chiral fields at all energy scales
(see Sec.~\ref{sec:chiral-protection}).
The ``subconstituent content'' $Q$ counts the
number of hop propagators in the VLQ chain
(Sec.~\ref{sec:UV-Lagrangian}) that generates the
effective Yukawa operator---real physical fields
above $\Lambda$, integrated out below it.
No hop is literally confined \emph{inside} the fermion;
rather, each hop propagator in the chain contributes
one power of $\e$ to the Yukawa coupling, exactly as
in a wavefunction-overlap integral where each layer
of binding introduces a suppression.
This is illustrated schematically in Fig.~\ref{fig:depth}.

\subsection{Mass hierarchy from compositeness depth}

The quark mass ratios follow immediately.
For the down sector, with right-handed charges
$Q(d^c_j) = (10/9,\, 1/3,\, 0)$~\cite{TwoOverTwo}:
\begin{align}
\frac{m_d}{m_b} &\sim \e^{Q(Q_1)+Q(d^c_1)}
= \e^{3+10/9} = \e^{37/9} \approx 0.0010,
\\
\frac{m_s}{m_b} &\sim \e^{Q(Q_2)+Q(d^c_2)}
= \e^{2+1/3} = \e^{7/3} \approx 0.020,
\end{align}
both in excellent agreement with
data~\cite{HuangZhou,PDG2024}.
The compositeness interpretation makes the hierarchy
manifest: the down quark's Yukawa coupling requires
3 left-handed and $10/9$ right-handed units of FN
insertions, for a total
depth of $37/9$; the bottom quark, with zero total depth,
couples to the Higgs at full strength.

The non-integer right-handed charges (like $10/9$ and $1/3$)
indicate that the right-handed subconstituent content is
organized differently from the left-handed content.
This is natural in a chiral theory: the left-handed
doublets and right-handed singlets have different gauge
quantum numbers, and their internal binding dynamics need
not be identical.

\subsection{Binding granularity: resolving fractional hop content}
\label{sec:fractional-hops}

A conceptual subtlety arises regarding the right-handed
flavor charges.
While the left-handed quark doublets carry integer units
of compositeness depth, $Q(Q_i) = (3,2,0)$~\cite{PaperI},
the right-handed down singlets require fractional
assignments such as $Q(d^c_1) = 10/9$ and
$Q(d^c_2) = 1/3$~\cite{TwoOverTwo}.
In a framework where hops are discrete, indivisible
quanta, a fractional subconstituent appears physically
ill-defined.

This tension is resolved by recognizing that the
macroscopic depth parameter $Q=1$ is not the fundamental
quantum of binding.
The true indivisible unit of compositeness is the
$\alpha$-hop, which carries a single unit of $\Z_9$
charge ($q_9=1$) and corresponds to a suppression of
$\e^{1/9}$.
A macroscopic ``layer'' of depth $Q=1$ represents a
saturated, coherent cluster of exactly nine hop quanta
(or a $\Z_9$-equivalent combination of $\alpha$ and
$\beta$ hops).

When expressed in terms of the fundamental hop
count---the ninths numerators $A_i$ and $B_j$
(Sec.~\ref{sec:chain})---the fractional structure
vanishes into pure integers:
\begin{align}
\text{LH hop quanta:}&\quad
A_i = 9\,Q(Q_i) = (27,\,18,\,0),
\\
\text{RH hop quanta:}&\quad
B_j = 9\,Q(d^c_j) = (10,\,3,\,0).
\end{align}
Both left- and right-handed fermions are
characterized by an integer number of discrete
hop quanta that determine their Yukawa suppressions.
A crucial clarification is needed here:
$A_1 = 27$ does \emph{not} mean that the
first-generation doublet is a bound state of
$27$ physical hop particles confined under
the hypercolor (HC) group $SU(N_H)_{\rm HC}$.
As discussed in Sec.~\ref{sec:chiral-protection},
all three SM generations are elementary chiral
fields in the Lagrangian at all scales.
The integer $A_i$ counts the number of
hop propagators in the UV chain
(each contributing a factor
$\langle\Phi\rangle/\Lambda = \e$
to the effective Yukawa coupling).
In terms of the physical FN charge $Q(Q_i)$,
which counts chain links, the first-generation
doublet requires $Q(Q_1) = 3$ links---a modest
number, well within the scope of standard FN models.
The large integer $A_1 = 9\times 3 = 27$ merely
re-expresses this in units of $1/9$
(the smallest hop quantum), not in units of
separate confined constituents.

The distinction between left-handed and
right-handed hop quanta reflects the chiral
structure of the SM gauge group.
The left-handed fields, which carry $SU(2)_L$ weak
isospin, have charges $A_i$ that are exact multiples
of $9$ ($27$, $9$, $0$), corresponding to integer
values of $Q$.
The right-handed fields, being $SU(2)_L$
singlets, can carry charges $B_j$ that are not
multiples of $9$ ($10$, $3$, $0$), corresponding
to fractional values of $Q$ in ninths.

Consequently, $Q(d^c_1) = 10/9$ does not imply a
fraction of a particle; rather, the integer $B_1 = 10$
counts the total number of hop propagator insertions---i.e.,
$10/9$ chain links---needed on the right-handed side
of the Yukawa operator.

\section[The Two-Index Hop Decomposition]{\texorpdfstring{The Two-Index Hop Decomposition}{The Two-Index Hop Decomposition}}
\label{sec:Z3Z3}

\subsection[Algebraic decomposition of Z9]{\texorpdfstring{Algebraic decomposition of $\Z_9$}{Algebraic decomposition of Z9}}

The cyclic group $\Z_9$ is isomorphic to
$\Z_3\times\Z_3$ only if 9 is a product of distinct
primes---which it is not ($9 = 3^2$).
As an abstract group, $\Z_9$ is \emph{not} isomorphic to
$\Z_3\times\Z_3$.
However, the \emph{residues mod 9} can be organized
by a two-index labeling that treats the hop charges
as a two-component basis.

\begin{figure}[!htbp]
\centering
\begin{tikzpicture}[scale=1.1,
  site/.style={circle,draw=blue!70,fill=blue!10,thick,minimum size=7mm,font=\scriptsize},
  hop/.style={rectangle,draw=red!70,fill=red!10,thick,minimum size=6mm,font=\scriptsize},
  empty/.style={circle,draw=gray!40,fill=gray!5,minimum size=6mm,font=\scriptsize\color{gray}}
]
\foreach \a in {0,1,2} {
  \foreach \b in {0,1,2} {
    \pgfmathtruncatemacro{\q}{3*\b+\a}
    \ifnum\q=0 \node[site] (n\a\b) at (\a*1.5, \b*1.5) {$0$};
    \else\ifnum\q=8 \node[site] (n\a\b) at (\a*1.5, \b*1.5) {$8$};
    \else\ifnum\q=6 \node[site] (n\a\b) at (\a*1.5, \b*1.5) {$6$};
    \else\ifnum\q=2 \node[site] (n\a\b) at (\a*1.5, \b*1.5) {$2$};
    \else\ifnum\q=1 \node[hop] (n\a\b) at (\a*1.5, \b*1.5) {$1$};
    \else\ifnum\q=4 \node[hop] (n\a\b) at (\a*1.5, \b*1.5) {$4$};
    \else\ifnum\q=7 \node[hop] (n\a\b) at (\a*1.5, \b*1.5) {$7$};
    \else \node[empty] (n\a\b) at (\a*1.5, \b*1.5) {$\q$};
    \fi\fi\fi\fi\fi\fi\fi
  }
}
\node[below=3mm,font=\small] at (1.5, -0.3) {$\alpha$-hop content $a$};
\node[rotate=90,above=3mm,font=\small] at (-0.5, 1.5) {$\beta$-hop content $b$};
\foreach \a in {0,1,2} \node[below,font=\scriptsize] at (\a*1.5, -0.3) {$\a$};
\foreach \b in {0,1,2} \node[left,font=\scriptsize] at (-0.3, \b*1.5) {$\b$};
\node[site,label=right:{ VLQ site}] at (4.5, 2.7) {};
\node[hop,label=right:{\scriptsize Hop link}] at (4.5, 1.8) {};
\end{tikzpicture}
\caption{The two-index hop grid. Each cell shows the
$\Z_9$ charge $q = 3b + a$, with $a,b\in\{0,1,2\}$. Blue circles: VLQ chain sites
(all carry even $\alpha$ and $\beta$ content---``bosonic'').
Red squares: hop link charges exchanged between adjacent sites
(types 1 and 4; type 2 = two $\alpha$'s is at $q=2$, which
is also a VLQ site charge).
The hop framework populates this grid with the chain topology
of the $\Bpar$-lattice.}
\label{fig:hop-grid}
\end{figure}
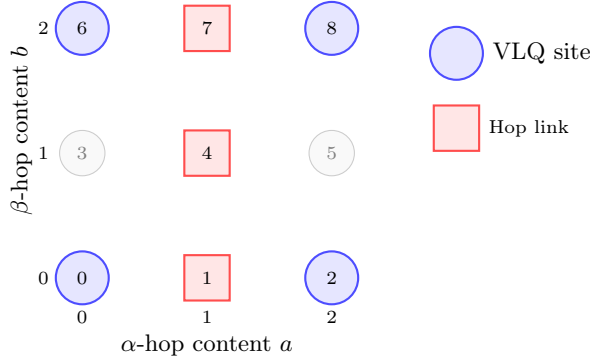

The three hop charges $(1,2,4)$ generate all residues
mod~9 under addition.
An instructive decomposition uses the two basis
elements $g_1 = 1$ and $g_2 = 4$ (note that
$2 = 2\times g_1$ and $4 = g_2$):
\begin{equation}
q \;\mapsto\; (a,b) \equiv (q\!\!\!\mod 3,\,
\lfloor q/3 \rfloor),
\label{eq:ab-decomp}
\end{equation}
so that
\begin{equation}
{\scriptsize
\setlength{\arraycolsep}{2pt}
\renewcommand{\arraystretch}{1.2}
\begin{array}{c|ccccccccc}
q & 0 & 1 & 2 & 3 & 4 & 5 & 6 & 7 & 8 \\
\hline
(a,b) & (0,0) & (1,0) & (2,0) & (0,1) & (1,1) & (2,1)
& (0,2) & (1,2) & (2,2)
\end{array}
}
\label{eq:ab-table}
\end{equation}
This is a relabeling, not an isomorphism of groups
(since $\Z_9$ is cyclic while $\Z_3\times\Z_3$ is not).
But it suggests a physical picture in which each unit
of $\Z_9$ charge is built from two species of
subconstituent---call them $\alpha$ (carrying
$q_9 = 1$, i.e.\ one unit along the $a$-axis)
and $\beta$ (carrying $q_9 = 3$, i.e.\ one unit
along the $b$-axis)---with the total $\Z_9$ charge
determined by the $\alpha$ and $\beta$ content.
Since the anti-particle carries the $\Z_9$
conjugate charge $q_9(\bar X) = 9 - q_9(X)$,
the complete set of hop charges is:
\begin{equation}
\boxed{\;
q_9(\alpha) = 1,\quad
q_9(\bar\alpha) = 8,\quad
q_9(\beta) = 3,\quad
q_9(\bar\beta) = 6.
\;}
\label{eq:hop-charges}
\end{equation}
The hop--anti-hop distinction underlies the
scalar potential structure of
Sec.~\ref{sec:UV-Lagrangian} and the
hypercolor bound-state spectrum of
Sec.~\ref{sec:chiral-protection}.
Figure~\ref{fig:hop-grid} displays this decomposition
as a two-dimensional grid, with the VLQ sites and hop
links identified.

\subsection{Hop content of the hop types}

Under this decomposition, the three hop types become:
\begin{align}
\text{Type-1 hop}\;(q=1)&:\quad (a,b)=(1,0)
\notag\\
&\quad\to\;\text{exchange of one }\alpha,
\\[3pt]
\text{Type-2 hop}\;(q=2)&:\quad (a,b)=(2,0)
\notag\\
&\quad\to\;\text{exchange of two }\alpha\text{'s},
\\[3pt]
\text{Type-4 hop}\;(q=4)&:\quad (a,b)=(1,1)
\notag\\
&\quad\to\;\text{exchange of one }\alpha
\text{ plus one }\beta.
\end{align}
This is reminiscent of meson exchange in nuclear physics:
the chain links mediate transitions between
adjacent vectorlike quark sites by exchanging quanta
of the hop fields $\alpha$ and $\beta$.

\subsection{Site charges and bosonic structure}

The four VLQ site charges
$q_9(D_a) = (0,8,6,2)$~\cite{FlavorInNinths,UFP}
decompose as
\begin{equation}
\renewcommand{\arraystretch}{1.2}
\begin{array}{c|cccc}
D_a & D_1 & D_2 & D_3 & D_4 \\
\hline
q_9 & 0 & 8 & 6 & 2 \\
(a,b) & (0,0) & (2,2) & (0,2) & (2,0)
\end{array}
\end{equation}
Every site carries an \emph{even} number of each hop
type.
In a hop language, this means the VLQ sites are
``bosonic''---they carry an even number of
fundamental hop quanta, analogous to mesons
composed of an even number of constituents in QCD.
The nearest-neighbor hops then correspond to the exchange
of ``fermionic'' quanta (odd hop number) between
adjacent bosonic sites, a structure that is natural in
confining gauge theories.

\emph{Uniqueness of the VLQ chain.}
The four-pair chain with charges $(0,8,6,2)$ is not the
unique VLQ configuration consistent with the framework,
but it is the \emph{minimal} choice that simultaneously
satisfies four requirements:
(i)~reproduces the observed CKM hierarchy with rational
ninths exponents and the empirically determined
$\Bpar = 75/14$;
(ii)~satisfies all $\Z_{18}$ discrete gauge anomaly
cancellation conditions;
(iii)~preserves Peccei--Quinn quality at dimension~18;
(iv)~places the messenger sites at integer
$q_9$ values (necessary for a discrete-gauge-invariant
chain Lagrangian).
Configurations with five or six VLQ pairs at alternative
intermediate charges can be constructed, but they
require additional fine-tuning of $\mathcal{O}(1)$
Wilson coefficients to reproduce the CKM elements
or violate one of the discrete-anomaly conditions.
The chosen four-pair chain is therefore the unique
minimal solution; the framework's flavor predictions
are robust against this choice in the sense that any
extension preserving the four conditions above gives
the same leading-order mass-ratio predictions.

\section{The Chain as a Confining Flux Tube}
\label{sec:chain}

\subsection{Chain topology and tunneling}

\begin{figure*}[!tbp]
\centering
\begin{tikzpicture}[scale=1.1,
  sm/.style={rectangle,draw=black,fill=yellow!20,thick,
    minimum width=14mm,minimum height=9mm,font=\small},
  vlq/.style={circle,draw=blue!70,fill=blue!10,thick,
    minimum size=8mm,font=\small},
  >=Stealth
]
\node[sm] (qL) at (0,0) {$q_{L,i}$};
\node[sm] (dR) at (12,0) {$d_{R,j}$};
\node[vlq] (D4) at (2.8,0) {$D_4$};
\node[vlq] (D3) at (5.2,0) {$D_3$};
\node[vlq] (D2) at (7.6,0) {$D_2$};
\node[vlq] (D1) at (10,0) {$D_1$};
\draw[->,thick,dashed,red!70!black] (qL) -- node[above,font=\scriptsize] {$\e^{A_i/9}$} (D4);
\draw[->,thick,blue!70] (D4) -- node[above,font=\scriptsize] {$\e^{4/9}$} (D3);
\draw[->,thick,blue!70] (D3) -- node[above,font=\scriptsize] {$\e^{2/9}$} (D2);
\draw[->,thick,blue!70] (D2) -- node[above,font=\scriptsize] {$\e^{1/9}$} (D1);
\draw[->,thick,dashed,red!70!black] (D1) -- node[above,font=\scriptsize] {$\e^{B_j/9}$} (dR);
\draw[decorate,decoration={brace,amplitude=5pt,mirror},thick,red!70!black]
  ([yshift=-7mm]qL.south) -- ([yshift=-7mm]D4.south)
  node[midway,below=6pt,font=\scriptsize,red!70!black] {Entrance};
\draw[decorate,decoration={brace,amplitude=5pt,mirror},thick,blue!70]
  ([yshift=-7mm]D4.south) -- ([yshift=-7mm]D1.south)
  node[midway,below=6pt,font=\scriptsize,blue!70] {Internal: $\e^{7/9}\simeq 0.27$};
\draw[decorate,decoration={brace,amplitude=5pt,mirror},thick,red!70!black]
  ([yshift=-7mm]D1.south) -- ([yshift=-7mm]dR.south)
  node[midway,below=6pt,font=\scriptsize,red!70!black] {Exit};
\foreach \x/\q in {2.8/2, 5.2/6, 7.6/8, 10/0}
  \node[below=2mm,font=\scriptsize,gray] at (\x,0) {$q_9\!=\!\q$};
\end{tikzpicture}
\caption{The hop flux tube. SM quarks (yellow boxes) couple to the
endpoints of the VLQ chain (blue circles) with hop-dressed
suppressions $\e^{A_i/9}$ (entrance) and $\e^{B_j/9}$ (exit).
The internal chain provides a common factor $\e^{7/9}$. The total
Yukawa suppression is the product of all three segments. Mass
ratios depend only on the endpoint dressings, since the internal
factor cancels.}
\label{fig:flux-tube}
\end{figure*}
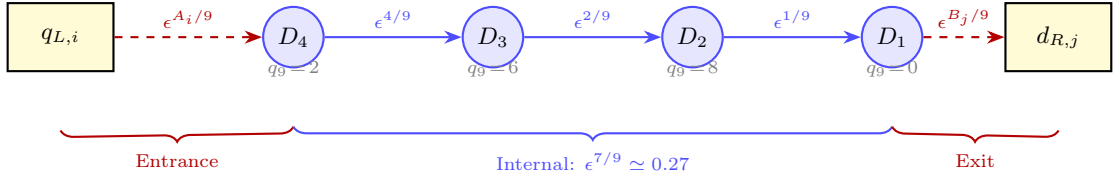

In the $\Bpar$-lattice
framework~\cite{FlavorInNinths,UFP}, the effective
Yukawa coupling between SM quarks $q_{L,i}$ and $d_{R,j}$
is generated by integrating out the heavy VLQ chain.
The full Yukawa suppression factorizes into three
segments (Table~\ref{tab:factorization}):
entrance $\times$ internal $\times$ exit $= \e^{(A_i+7+B_j)/9}$,

\begin{table*}[!tbp]
\caption{Three-factor decomposition of the Yukawa
suppression $Y_{ij}\sim\e^{(A_i+7+B_j)/9}$.
The entrance and exit factors depend on the
generation indices $i,j$; the internal factor is
universal.
$A_i = 9\,Q(Q_i)$ and $B_j = 9\,Q(d^c_j)$ are the
ninths numerators of the endpoint
dressings~\cite{FlavorInNinths,UFP}.}
\label{tab:factorization}
\begin{ruledtabular}
\begin{tabular}{lcccc}
Segment & Suppression & Physical origin & Dependence & Value \\
\hline
Entrance ($q_{L,i}\!\to\! D_4$) & $\e^{A_i/9}$ & LH hop cloud & $A_i = (27,18,0)$ & $\e^3,\;\e^2,\;1$ \\
Internal ($D_4\!\to\!\cdots\!\to\! D_1$) & $\e^{7/9}$ & Flux-tube tunneling & Universal & $0.27$ \\
Exit ($D_1\!\to\! d_{R,j}$) & $\e^{B_j/9}$ & RH hop cloud & $B_j^d = (10,3,0)$ & $\e^{10/9},\;\e^{1/3},\;1$ \\
\hline
Total ($q_{L,i}\!\to\! d_{R,j}$) & $\e^{(A_i+7+B_j)/9}$ & Full chain & $p_{ij}\!=\!Q_i\!+\!Q^c_j\!+\!7/9$ & \\
\end{tabular}
\end{ruledtabular}
\end{table*}

\noindent
where $A_i = 9\,Q(Q_i)$ and $B_j = 9\,Q(d^c_j)$ are the
entrance and exit endpoint dressings, and the internal
factor $\e^{7/9}\simeq 0.27$ is the product of the three
hop suppressions
$\e^{1/9}\times\e^{2/9}\times\e^{4/9}$~\cite{FlavorInNinths}.
Figure~\ref{fig:flux-tube} illustrates this three-factor
structure.

In the compositeness picture, this factorization has a
transparent physical meaning.
The SM quark $q_{L,i}$ couples to the chain endpoint
with a suppression $\e^{A_i/9}$ determined by its FN
charge; the deeper the hop content (larger $A_i$),
the smaller the effective coupling.
The chain itself is a \emph{confining flux tube}---a
one-dimensional sequence of bound states (the VLQ sites)
connected by hop exchange---through which the left-handed
and right-handed quarks communicate.
The internal suppression $\e^{7/9}$ is the tunneling
amplitude through the bulk of the tube.
The right-handed quark $d_{R,j}$ has its own wavefunction
overlap $\e^{B_j/9}$ at the exit endpoint.

\subsection{Why mass ratios are endpoint-dominated}

Because the internal factor $\e^{7/9}$ is common to all
entries of the Yukawa matrix, it cancels in mass ratios:
\begin{equation}
\frac{m_i}{m_j} \sim
\e^{(A_i+B_i-A_j-B_j)/9}.
\end{equation}
In the flux-tube language, this means that mass ratios
are determined entirely by the \emph{boundary conditions}
(the compositeness depth of the external quarks), not by
the bulk of the confining tube.
This is a well-known property of tunneling amplitudes
in quantum mechanics: the ratio of tunneling rates through
barriers of identical shape but different boundary
conditions depends only on the boundary terms.

\subsection{Nearest-neighbor locality as confinement}

The $\Z_2^{(\mathrm{NN})}$ symmetry that forbids
non-nearest-neighbor couplings between VLQ
sites~\cite{UFP} has a natural interpretation as
\emph{confinement locality}: hop exchange is short-ranged
in theory space, just as gluon exchange is short-ranged
in position space in a confining gauge theory.
This locality simultaneously explains the ninths
quantization of exponents (each hop contributes a discrete
quantum of charge) and the suppression of flavor-changing
neutral currents (FCNCs): dangerous FCNC amplitudes require
long-range correlations between distant chain sites, which
are exponentially suppressed by the same confinement
mechanism that generates the mass hierarchy.

\subsection{Vectorlike quark mass spectrum}
\label{sec:VLQ-masses}

The four VLQ pairs $D_a+\bar{D}_a$ ($a=1,2,3,4$)
acquire bare masses $M_a\,D_a\bar D_a$ that must be
$\Z_{18}$-invariant.
In the minimal realization~\cite{UFP,FlavorInNinths},
the bare masses are all of order $\Lambda$:
\begin{equation}
M_a \;\sim\; \Lambda \;\simeq\;
3.2\times10^{12}\;\text{GeV}
\qquad (a = 1,\ldots,4).
\end{equation}
This is a sharp and distinctive prediction.
The VLQ masses are fixed by the \emph{same} scale that
sets the axion decay constant
($\Lambda = \fa/\e$, Sec.~\ref{sec:binding}),
so any attempt to lower the VLQ masses to collider-accessible
energies would simultaneously destroy the axion mass
prediction and the fermion mass hierarchy.

In detail, the four VLQ bare masses need not be
exactly degenerate.
Their $\Z_9$ site charges $(0,8,6,2)$
allow mass splittings of order $\e^{|q_i-q_j|/9}$,
giving a spread
$M_{\rm max}/M_{\rm min}\sim\e^{-8/9}\sim 7$.
However, all four masses remain firmly in the
$10^{12}$--$10^{13}$~GeV range---nine orders of
magnitude above the current LHC bound
$M_D\gtrsim 1.5$~TeV~\cite{PDG2024} and six
orders above any conceivable future
collider~\cite{BernreutherDobrescu2025}.

After electroweak symmetry breaking, the VLQ
chain mixes with the SM down-type quarks through
the endpoint Yukawa
couplings~\cite{UFP,BargerBergerPhillips}.
The resulting shifts to the light-quark masses
and CKM elements are of order
$(v/M_a)^2\sim(246/3.2\times10^{12})^2
\sim 10^{-20}$---entirely negligible.
The VLQ sector is therefore
\emph{completely decoupled} from collider and
precision flavor physics, and its only low-energy
footprint is the Yukawa texture it generates.

This prediction stands in stark contrast to the
conventional VLQ literature, which typically places
vectorlike quarks in the $1$--$3$~TeV
range~\cite{BargerBergerPhillips,PDG2024}
as solutions to the little hierarchy problem or
as low-scale Froggatt--Nielsen
mediators~\cite{ArkaniHamedHall}.
At such masses, VLQs produce spectacular
collider signatures: pair production via QCD
with decays $D\to bZ,\;bH,\;tW$ yielding
multi-top and multi-$b$ final
states~\cite{BernreutherDobrescu2025}, or
single production through electroweak mixing
with the third-generation quarks.
In the hop framework, \emph{none of these
signatures exist}.
The VLQ masses are not a free parameter to be
scanned over; they are locked to
$\Lambda\sim 10^{12}$~GeV by the same discrete
$\Z_{18}$ symmetry that generates the flavor
hierarchy and protects the axion.
The absence of TeV-scale VLQ signals at the
LHC is therefore not a constraint on the model
but a \emph{confirmed prediction}.

\section{Hop Mass Estimates from Virtual Propagators}
\label{sec:hop-masses}

\subsection{Propagator model}

The hop suppressions at each chain link arise from the
exchange of hop quanta between adjacent VLQ sites.
If each exchange is mediated by a virtual propagator with
suppression factor $m_{\rm hop}/\Lambda$, the three hop
types (Sec.~\ref{sec:Z3Z3}) yield three equations:
\begin{align}
\text{Type-1}\;(q=1,\;\text{one }\alpha)&:\quad
\e^{1/9} = \frac{m_\alpha}{\Lambda},
\label{eq:type1}
\\[4pt]
\text{Type-2}\;(q=2,\;\text{two }\alpha\text{'s})&:\quad
\e^{2/9} = \left(\frac{m_\alpha}{\Lambda}\right)^{\!2},
\label{eq:type2}
\\[4pt]
\text{Type-4}\;(q=4,\;\text{one }\alpha+\text{one }\beta)
&:\quad
\e^{4/9} = \frac{m_\alpha}{\Lambda}\,
\frac{m_\beta}{\Lambda}.
\label{eq:type4}
\end{align}
Equation~\eqref{eq:type2} is automatically the square of
Eq.~\eqref{eq:type1}, providing a non-trivial
self-consistency check of the two-hop picture:
the Type-2 hop, which exchanges two $\alpha$-hops,
produces exactly the square of the Type-1 single-$\alpha$
suppression.

\subsection{Solving for the hop masses}

From Eq.~\eqref{eq:type1}:
\begin{equation}
m_\alpha = \Lambda\,\e^{1/9}.
\label{eq:m-alpha}
\end{equation}
Substituting into Eq.~\eqref{eq:type4}:
\begin{equation}
m_\beta = \Lambda\,\e^{4/9}\Big/\e^{1/9}
= \Lambda\,\e^{1/3}.
\label{eq:m-beta}
\end{equation}
The mass ratio is
\begin{equation}
\frac{m_\beta}{m_\alpha}
= \frac{\e^{1/3}}{\e^{1/9}}
= \e^{2/9} \simeq 0.69,
\label{eq:mass-ratio}
\end{equation}
so the $\beta$-hop is approximately $30\%$ lighter
than the $\alpha$-hop.
Expressed as fractions of the confinement scale:
\begin{equation}
\frac{m_\alpha}{\Lambda} = \e^{1/9}\simeq 0.83,
\qquad
\frac{m_\beta}{\Lambda} = \e^{1/3}\simeq 0.57.
\label{eq:hop-fractions}
\end{equation}
Both masses lie just below $\Lambda$, analogous to
constituent quark masses in QCD
($m_q\sim\Lambda_{\rm QCD}$), consistent with the
picture of tightly bound composites.

\subsection{Numerical estimates}

Using the axion-window relation
$\Lambda = \fa/\e$~\cite{FlavorInNinths} with
$\fa\sim(5$--$8)\times10^{11}$~GeV:
\begin{equation}
\renewcommand{\arraystretch}{1.3}
\begin{array}{lccc}
\hline\hline
& \fa = 5\times10^{11} & 6\times10^{11}
& 8\times10^{11}\;\text{GeV} \\
\hline
\Lambda & 2.7 & 3.2 & 4.3 \\
m_\alpha & 2.2 & 2.7 & 3.6 \\
m_\beta & 1.5 & 1.8 & 2.4 \\
\hline\hline
\end{array}
\label{eq:hop-numerics}
\end{equation}
(all entries in units of $10^{12}$~GeV).
The predicted mass window is
\begin{equation}
m_\alpha \simeq (2\text{--}4)\times10^{12}\;\text{GeV},
\qquad
m_\beta \simeq (1.5\text{--}2.5)\times10^{12}\;\text{GeV}.
\end{equation}

\subsection{Connection to the Cabibbo angle}

The $\alpha$-hop mass can be re-expressed in terms of
the axion decay constant and the Cabibbo angle.
Since $m_\alpha = \Lambda\,\e^{1/9}
= (\fa/\e)\,\e^{1/9} = \fa\,\e^{-8/9}$ and
$|V_{us}|\sim\e^{8/9}$~\cite{FlavorInNinths,Companion},
\begin{equation}
\boxed{\;m_\alpha \,\simeq\,
\frac{\fa}{|V_{us}|}\;\simeq\;
2.7\times10^{12}\;\text{GeV}\;}
\label{eq:m-alpha-Vus}
\end{equation}
for $\fa\simeq 6\times10^{11}$~GeV.
This remarkable relation links the lightest hop mass
directly to two of the most precisely measured quantities
in the flavor sector: the axion decay constant and the
Cabibbo angle.
In the compositeness picture, the Cabibbo angle
\emph{is} the ratio $\fa/m_\alpha$---the overlap
between the axion scale and the $\alpha$-hop mass.

Similarly, the $\beta$-hop mass satisfies
\begin{equation}
m_\beta = \fa\,\Bpar^{2/3}
= \fa\times\left(\frac{75}{14}\right)^{\!2/3}
\simeq 3.1\,\fa,
\label{eq:m-beta-B}
\end{equation}
connecting it to the fundamental $\Bpar$-lattice parameter.

\subsection{Interpretation as constituent masses}

In QCD, the constituent quark mass $m_q\sim 300$~MeV is
close to $\Lambda_{\rm QCD}\sim 200$~MeV, and represents
the dynamical mass acquired through chiral symmetry
breaking in the confining vacuum.
The analogous picture here is that the $\alpha$- and
$\beta$-hops acquire dynamical masses of order $\Lambda$
through the confining dynamics of the hypercolor
interaction.
The fact that $m_\alpha/\Lambda\simeq 0.83$ and
$m_\beta/\Lambda\simeq 0.57$ (rather than, say, $10^{-2}$)
indicates that the hops are \emph{not} weakly coupled
constituents but are deeply confined, with most of their
mass arising from the binding energy---again paralleling
the QCD situation where the proton mass ($938$~MeV) far
exceeds the sum of current quark masses
($\sim 10$~MeV).

The mass splitting $m_\alpha - m_\beta\simeq 0.9\times
10^{12}$~GeV, while large in absolute terms, is only
$\sim 30\%$ of the confinement scale.
This is comparable to the splitting between strange and
non-strange constituent quarks in QCD
($m_s^{\rm const}/m_u^{\rm const}\sim 1.5$), suggesting
that the two hop species play analogous roles to
the light and strange quarks in the hadron spectrum.

\subsection{The spin of the hops}

The spin of the hops is determined by the field content
of the UV Lagrangian (Sec.~\ref{sec:UV-Lagrangian}).
The hop fields $\alpha^a$ and $\beta^a$ are complex
scalars---\emph{spin-$0$} particles---carrying
hypercolor charge.
At each chain link, the messenger coupling involves
a hop field and a flavon insertion $\Phi$, both of
which are spin-$0$:
\begin{equation}
\mathcal{L}_{\rm chain}\supset
\kappa_r\,\Phi\,\bar S_{r-1}\,h^\dagger_{I_r\,a}\,F_r^a
+ \text{h.c.},
\end{equation}
where $S_r$ and $F_r^a$ are spin-$1/2$ vectorlike
messengers and $h_{I_r}\in\{\alpha,\beta\}$ is the
hop field at the $r$th step.
The hops therefore have \emph{spin $0$}.

The complete quantum numbers of the hops are remarkably
minimal: each hop carries one unit of $\Z_9$ flavor charge
(``hopness'') and \emph{nothing else}.
Hops are singlets under $SU(3)_C$, $SU(2)_L$, and
$U(1)_Y$---they have zero electric charge, zero color,
zero weak isospin, zero baryon number, and zero lepton
number.
Their only quantum number beyond spin is hopness.
This is why the compositeness is purely in the flavor
sector: the hop cloud modifies only the Yukawa couplings
of the SM fermion it dresses, without altering any gauge
quantum number.
It is also why hops are so difficult to detect---they
interact with ordinary matter only through the
$\Z_9$-mediated exchange, which is suppressed by
$1/\Lambda^4$.

This identification has several important consequences:

\emph{No Pauli exclusion.}
Multiple hops can occupy the same quantum state,
consistent with $Q(Q_1) = 3$ (three flavon insertions
dressing a
single first-generation quark doublet).
By contrast, the rishon model's spin-$1/2$ preons
require antisymmetrization to avoid a spin-$3/2$ ground
state~\cite{HarariSeibergNPB}.

\emph{Automatic spin-$1/2$ composites.}
In the hop picture, the SM fermion's Yukawa coupling
is that of an elementary
spin-$1/2$ field dressed by spin-$0$ flavon insertions:
schematically,
$|\psi_i\rangle = |\text{core}\rangle_{1/2}
\otimes|\text{hop dressing}\rangle_0$,
yielding spin $1/2$ without any angular-momentum
gymnastics.
(This is a description of the effective operator
structure, not of a literal multi-particle bound
state; see Sec.~\ref{sec:chiral-protection}.)
The absence of unwanted spin-$3/2$ composites is a
structural advantage over preon models.

\emph{Bosonic site charges explained.}
The VLQ site charges decompose as
$(a,b)=(0,0),\,(2,2),\,(0,2),\,(2,0)$---all with even
hop numbers (Sec.~\ref{sec:Z3Z3}).
Since each hop is a boson, an even number of bosons
naturally forms a bosonic composite, consistent with
the integer-spin VLQ sites of the chain.

\section{Comparison with the Harari--Seiberg Rishon Model}
\label{sec:harari}

\subsection{Structural parallels}

The Harari--Shupe
model~\cite{HarariShupe,Shupe}, subsequently developed
into a full dynamical theory by Harari and
Seiberg~\cite{HarariSeiberg1981,HarariSeibergNPB},
postulates two fundamental spin-$1/2$ fermions:
\begin{itemize}
\item $T$ (``third''), carrying electric charge $+1/3$,
\item $V$ (``vafer''), electrically neutral,
\end{itemize}
with SM fermions built as three-preon composites:
$e^+ = TTT$, $u = TTV$, $\bar{d} = TVV$,
$\nu = VVV$, and their antiparticles
$e^- = \bar{T}\bar{T}\bar{T}$,
$\bar{u} = \bar{T}\bar{T}\bar{V}$,
$d = \bar{T}\bar{V}\bar{V}$,
$\bar\nu = \bar{V}\bar{V}\bar{V}$.
The rishons are confined by an $SU(3)_H$ hypercolor
interaction, and only hypercolor-singlet composites
appear below the confinement scale
$\Lambda_H$~\cite{HarariSeiberg1981}.
The model reproduces the correct electric charges and
$SU(3)_C\times SU(2)_L\times U(1)_Y$ quantum numbers,
and an approximate
$SU(2)_L\times SU(2)_R\times U(1)_{B-L}$ symmetry
emerges at the composite level~\cite{HarariSeibergNPB}.

The present framework shares several features:
\begin{itemize}
\item Two species of subconstituent ($\alpha$- and
$\beta$-hops in our notation, $T$ and $V$ rishons in
Harari--Shupe--Seiberg).
\item Additive quantum numbers: the FN charge is the sum
of subconstituent charges, just as the rishon electric
charge is the sum of $T$ and $V$ charges.
\item A discrete symmetry ($\Z_9$ with a two-index
hop structure, versus the $\Z_3$ baryon number
of the rishon model) organizing the composite spectrum.
\item Confinement by a non-abelian gauge interaction
($SU(3)_H$ hypercolor for rishons; $SU(N_H)_{\rm HC}$
at scale $\Lambda$ in our framework).
\end{itemize}

\subsection{The generation problem: labels versus dynamics}

A central difficulty of the rishon model, recognized
by Harari and Seiberg~\cite{HarariSeibergGen}, is the
\emph{generation problem}: since all up-type quarks have
rishon content $TTV$ regardless of generation, the model
provides no intrinsic distinction between $u$, $c$, and $t$.
The generation index must be imposed externally---for
example, as a radial excitation quantum number or a
``generation label'' carried by one of the preons.
Harari and Seiberg introduced discrete symmetries as
generation labels~\cite{HarariSeibergGen} and showed
that, under certain assumptions, exactly three
generations can emerge~\cite{HarariSeibergChiral}.
However, they could not derive the \emph{mass hierarchy}
between generations from the model's dynamics.

The hop framework resolves this problem.
The generation structure is \emph{built into the charge
assignments}: the first generation carries the most hop
content ($Q=3$), the second carries an intermediate amount
($Q=2$), and the third is undressed ($Q=0$).
The mass hierarchy $m_1\ll m_2\ll m_3$ is a direct,
quantitative consequence of compositeness depth---each
unit of hop content costs a factor $\e\approx 0.19$
in the Yukawa coupling---not an additional assumption.
The discrete $\Z_9$ symmetry that Harari and Seiberg
sought as a generation label~\cite{HarariSeibergGen}
is realized in the hop framework as a gauged flavor
symmetry with a precise dynamical role.

\subsection{Key distinctions}

Despite these parallels, the two frameworks differ in
fundamental ways:

\emph{Gauge versus flavor compositeness.}
In the rishon model, the substructure determines the
\emph{gauge} quantum numbers of the SM fermions---electric
charge, color, and weak isospin all arise from rishon
content.
In the $\Bpar$-lattice, all three generations have
identical gauge quantum numbers; the substructure
determines only the \emph{Yukawa couplings} (masses and
mixing angles).
This is a weaker form of compositeness that evades the
strongest experimental constraints.

\emph{Baryon number.}
In the rishon model, $B$ and $L$ are not separately
conserved---only $B-L$ is an exact symmetry---and
proton decay is generically
predicted~\cite{HarariMohapatraSeiberg}.
In the hop framework, $B$ and $L$ are preserved as gauge
quantum numbers of the elementary core
(Sec.~\ref{sec:BL}), and the hop sector introduces no
new proton-decay operators.

\emph{Chiral asymmetry.}
The rishon model assigns the same rishon content to
left- and right-handed components of a given quark
(up to CP conjugation).
The $\Bpar$-lattice has different charges for
left-handed doublets and right-handed singlets:
$Q(Q_i) = (3,2,0)$ but $Q(d^c_j) = (10/9, 1/3, 0)$.
This chiral asymmetry is natural in a theory where the
binding dynamics respects the gauge structure of the SM,
and it is responsible for the non-trivial pattern of
right-handed mixing angles that differs from the
left-handed (CKM) pattern~\cite{UFP}.

\emph{Quantitative predictions.}
The rishon model provides a qualitative framework but no
quantitative mass predictions.
The hop framework, by contrast, predicts all quark and
lepton masses and mixing angles in terms of two
parameters ($\Lambda$ and $\e$), and extends to predict
the hop masses ($m_\alpha\simeq\fa/|V_{us}|$,
$m_\beta\simeq\fa\,\Bpar^{2/3}$) and the axion mass
window.

\emph{Subconstituent spin.}
The rishon model's spin-$1/2$ preons require
antisymmetrization of three identical fermions to avoid
spin-$3/2$ composites~\cite{HarariSeibergNPB}---a
non-trivial constraint that restricts the allowed
representations.
In the hop framework, the hops are spin-$0$
hypercolor-fundamental scalars
(Sec.~\ref{sec:UV-Lagrangian}), so
the SM fermion is simply an elementary spin-$1/2$ core
dressed by a scalar cloud, yielding spin $1/2$
automatically with no spin-$3/2$ problem.

Table~\ref{tab:comparison} summarizes the key features of the
three compositeness frameworks.

\begin{table*}[!tbp]
\caption{Comparison of three compositeness frameworks.
The hop framework addresses the generation and mass-hierarchy
problems that remain open in the earlier models, while
introducing no new proton-decay operators.}
\label{tab:comparison}
\begin{ruledtabular}
\begin{tabular}{lccc}
Feature & Harari--Seiberg~\cite{HarariSeibergNPB}
& Dobrescu $SU(15)$~\cite{Dobrescu2022}
& Hop framework \\
\hline
Subconstituents & 2 rishons ($T$, $V$) & 19 preons & 2 hops ($\alpha$, $\beta$) \\
Confining group & $SU(3)_H$ & $SU(15)$ & $SU(N_H)_{\rm HC}$ \\
Compositeness type & Gauge \& flavor & Gauge \& flavor & Flavor only \\
Generation mechanism & Radial excitations & Chiral prebaryons & $\Z_9$ charge (hop depth) \\
Mass hierarchy & Not explained & Not explained & $m_i/m_3\sim\e^{Q_i}$ \\
Higgs origin & Not addressed & Composite (di-prebaryon) & Elementary \\
Proton decay & Predicted & $\Lambda_{\rm pre}\gtrsim 10^3$~TeV & No hop-induced ($B$ preserved) \\
Confinement scale & $\Lambda_H\gg$~TeV & $10^3$--$10^4$~TeV & $3\times10^{12}$~GeV \\
$\alpha_s$ UV behavior & Not addressed & Loses asympt.\ freedom & $\Delta b_3\!=\!+8/3$, nearly freezes \\
DM candidate & None & VL lepton (possible) & Axion ($7$--$12\;\mu$eV) \\
Free parameters & $\Lambda_H$ & $\Lambda_{\rm pre}$, couplings & $\Lambda$ and $\e=14/75$ \\
\end{tabular}
\end{ruledtabular}
\end{table*}

\section{Baryon Number, Lepton Number, and Hop Content}
\label{sec:BL}

\subsection{Gauge versus flavor quantum numbers}

A central feature of the hop framework is the sharp
separation between \emph{gauge} quantum numbers (baryon
number $B$, lepton number $L$, electric charge $Q_{\rm em}$,
color) and \emph{flavor} quantum numbers ($\Z_{18}$ charge,
hop content).
Baryon number $B=1/3$ for every quark regardless of
generation; lepton number $L=1$ for every lepton regardless
of generation.
Hop content, by contrast, depends on the generation index
but not on $B$ or $L$.

This is the fundamental distinction from the Harari--Shupe
rishon model~\cite{HarariShupe,Shupe,HarariSeibergNPB},
where $B$ and $L$
are \emph{determined by} the preon content ($B = 1/3$ arises
from three rishons carrying $B=1/9$ each).
In the hop framework, $B$ and $L$ are properties of the
\emph{elementary chiral field}---present in all three
generations---and are unaffected by the hop dressing.

The QCD analogy is precise: electric charge is a property
of quarks, and binding quarks into hadrons does not change
$Q_{\rm em}$ but does change the mass.
Similarly, $B$, $L$, $Q_{\rm em}$, and color are properties
of the elementary fermion (all generations), and hop binding
changes only the Yukawa couplings (hence the mass) without
modifying the gauge structure.
This is why the compositeness is purely in the flavor
sector: hops are SM-singlets that add flavor structure
without touching gauge quantum numbers.

\subsection{Joint anomaly constraints}

Although $B$, $L$, and $\Z_{18}$ are logically independent
quantum numbers, they are not physically independent:
all three must be consistent with the same set of anomaly
cancellation conditions.
The $\Z_{18}$ discrete gauge anomalies require
\begin{equation}
\sum_i q_{18}(\text{quarks}_i) +
\sum_i q_{18}(\text{leptons}_i) \equiv 0
\pmod{18},
\end{equation}
which is a \emph{joint} constraint linking the quark
sector ($B\neq 0$) to the lepton sector ($L\neq 0$).
Meanwhile, $B-L$ is the unique anomaly-free global
symmetry of the SM fermion spectrum.

In the hop picture, the anomaly-freedom of $B-L$ acquires
a structural explanation.
Since hops are SM-singlets, they carry neither $B$ nor $L$.
The $\Z_{18}$ symmetry (which \emph{is} anomaly-free by
construction, as verified in
Ref.~\cite{FlavorInNinths}) must coexist with $B-L$ in
a consistent quantum theory.
This coexistence \emph{requires} $B-L$ to be anomaly-free:
if $B-L$ had a mixed anomaly with any SM gauge factor,
the $\Z_{18}$ anomaly conditions (which involve both quarks
and leptons) could not be simultaneously satisfied.

In other words, $B-L$ anomaly freedom is not an
accident of the SM spectrum but a \emph{consistency
condition} for flavor compositeness to coexist with
gauge structure.

\subsection{Proton stability}

The stability of the proton in the hop framework follows
from a double argument.
First, $B$ is an accidental symmetry of the SM Lagrangian
(broken only by dimension-6 operators suppressed by
$M_{\rm GUT}^2$), and hops do not introduce any new
$B$-violating interactions since they are SM-singlets.
Second, the lightest baryon (the proton) carries
$B=1$ but zero hop content ($Q_3$ has $q_9 = 0$).
Even if hop-mediated processes could in principle transfer
$\Z_9$ charge between quarks, they cannot change $B$
because the $\Z_9$ charge is a flavor index, not a baryon
number.
The proton is therefore at least as stable in the hop
framework as in the SM.
This statement is specific to hop-mediated processes.
Standard GUT-mediated proton decay---through
dimension-6 operators from heavy gauge boson exchange,
suppressed by $M_{\rm GUT}^2\simeq
(2\times10^{16}\;\text{GeV})^2$---is entirely
compatible with the hop framework.
The GUT-breaking sector operates at
$M_{\rm GUT}\sim 10^{16}$~GeV, four orders of
magnitude above the hop confinement scale $\Lambda$,
and the $B$-violating operators propagate through
the same quark fields regardless of their hop dressing.

\section[The SU(5) Origin of B=75/14 and Binding Dynamics]{\texorpdfstring{The $SU(5)$ Origin of $\Bpar=75/14$ and Binding Dynamics}{The SU(5) Origin of B=75/14 and Binding Dynamics}}
\label{sec:SU5}

\subsection[Group-theoretic origin of B]{\texorpdfstring{Group-theoretic origin of $\Bpar$}{Group-theoretic origin of B}}
\label{sec:SU5-charges}

The expansion parameter $\Bpar = 75/14$ was originally
determined phenomenologically from the CKM
elements~\cite{PaperI}:
\begin{equation}
|V_{us}|^{-9/8} = |V_{cb}|^{-9/17} = 5.357 \approx
\frac{75}{14}.
\label{eq:B-determination}
\end{equation}
The numerator 75 is the dimension of the adjoint-antisymmetric
representation of $SU(5)$~\cite{PaperI}.
This suggests that the fundamental scale of the flavor
hierarchy is set by an $SU(5)$ grand unified theory (GUT)
in which the flavon $\Phi$ transforms as a component of
the $\mathbf{75}$.

In the compositeness interpretation, the $SU(5)$ connection
offers a suggestive---but not dynamically
derived---heuristic.
If the subconstituents ($\alpha$- and $\beta$-hops) are
bound by an $SU(5)$ hypercolor interaction, then the
ratio $\e = 14/75$ can be read as a ratio of
representation dimensions:
$75 = \dim(\mathbf{75}_{SU(5)})$ and
$14 = \dim(\mathbf{14}_{G_2})$.
We stress that this identification is a mnemonic, not
a derivation: the physical determination of
$\Bpar = 75/14$ comes entirely from the CKM
elements~\eqref{eq:B-determination}, and the
group-theoretic label does not by itself determine
the confining gauge group, the beta function, or
the confinement scale.
What \emph{is} established phenomenologically is that
$\Bpar = 75/14$ simultaneously fits all nine
quark masses, all four CKM parameters, the
charged-lepton mass ratios, and the neutrino
mixing angles~\cite{PaperI,TwoOverTwo,FlavorInNinths,UFP}.

\emph{FN--Yukawa master formula.}
Every Dirac Yukawa coupling in the model obeys the
Froggatt--Nielsen rule~\cite{FN}
\begin{equation}
Y_{ij} = \lambda_{ij}\,
\Bigl(\frac{\langle\Phi\rangle}{M}\Bigr)^{\!|Q_i+Q_j+Q_H|}
= \lambda_{ij}\,\Bpar^{-|Q_i+Q_j+Q_H|}
\label{eq:FN-master}
\end{equation}
where $Q_i,\,Q_j$ are the flavor charges of the left-
and right-handed fermions, $Q_H$ is the Higgs charge,
and $\lambda_{ij}\sim\mathcal{O}(1)$.
With a single flavon $\Phi(-1\bmod 9)$ and
Higgs fields that carry \emph{no} net FN charge
($Q_{H_u} = Q_{H_d} = 0$; the chain internal
factor $\Delta_{\rm int}$ provides the sector-dependent
suppression previously attributed to $Q_H$),
the $SU(5)$ $\mathbf{10}$-plet charge is simply
the quark-doublet charge:
\begin{equation}
Q_{10} = Q(Q_i) = Q(u^c_i) = (3,\,2,\,0)
\label{eq:SU5-charges}
\end{equation}
for generations $(1,\,2,\,3)$, consistent
with the top quark having zero total FN depth
($Q_{10_3} = 0$, $Y_t\simeq 1$).
This identification is direct: both the quark
doublet $Q_i$ and the up-type singlet $u^c_i$ reside
in the $\mathbf{10}$ of $SU(5)$, and the hop framework
assigns them equal FN charges.

The up-type mass formula is then
\begin{equation}
Y^u_{ij}\;\sim\;\e^{\;Q_{10_i}+Q_{10_j}},
\label{eq:SU5-up}
\end{equation}
which gives $Y_t \sim \e^0 = 1$,
$Y_c \sim \e^4$, $Y_u\sim\e^6$ as required.
The down-type and charged-lepton Yukawas include
the chain internal factor $\Delta_{\rm int}^{d,\ell} = 7/9$:
\begin{align}
Y^d_{ij} &\sim
\e^{\;Q_{10_i}+Q(d^c_j)+\Delta_{\rm int}^d},
\nonumber\\
Y^\ell_{ij} &\sim
\e^{\;Q(L_i)+Q(e^c_j)+\Delta_{\rm int}^\ell},
\label{eq:SU5-mass-formulas}
\end{align}
with the SM-level right-handed and lepton-doublet
charges
\begin{equation}
\renewcommand{\arraystretch}{1.3}
\begin{array}{lcccc}
 & Q_1 & Q_2 & Q_3 & \text{multiplet} \\
\hline
Q(Q_i) = Q(u^c_i) & 3 & 2 & 0 & \mathbf{10} \\
Q(d^c_i) & 10/9 & 1/3 & 0 & \overline{\mathbf{5}} \\
Q(L_i) & 1 & 1/2 & 0 & \overline{\mathbf{5}} \\
Q(e^c_i) & 23/6 & 7/6 & 0 & \mathbf{10}
\end{array}
\label{eq:all-charges}
\end{equation}

\emph{Georgi--Jarlskog structure.}
In a minimal $SU(5)$ GUT, all fields in the same
multiplet would carry identical FN charges.
The fractional charges in
Eq.~\eqref{eq:all-charges} violate this:
$Q(d^c_i)\neq Q(L_i)$ in the $\overline{\mathbf{5}}$
and $Q(e^c_i)\neq Q(Q_i)$ in the $\mathbf{10}$.
These splittings have a well-known
origin~\cite{GeorgiJarlskog}: the down-type
and lepton Yukawas couple to different Higgs
representations ($\mathbf{5}_H$ vs.\
$\mathbf{45}_H$) whose $SU(5)$ Clebsch--Gordan
coefficients distinguish quarks from leptons
within the same multiplet.
In the hop picture, this corresponds to different
wavefunction overlaps for colored versus colorless
composites bound by the same hypercolor interaction.
The key empirical consequence is $b$-$\tau$ Yukawa
unification: $Q(d^c_3) = Q(L_3) = 0$ gives
$Y_b = Y_\tau$ at leading order (both equal to
$\e^{7/9}$), with the residual splitting at low
energies given by the golden ratio
$m_b/m_\tau = \varphi$~\cite{UFP}.

These formulas generate every mass-exponent entry
in Table~\ref{tab:fermion-ratios}, every CKM and
PMNS magnitude in Table~\ref{tab:4mag}, and the
$\alpha$-hop census of
Eq.~\eqref{eq:alpha-table}~\cite{FlavorInNinths,UFP}.
In the compositeness picture, the charges count the
hop content of each fermion, and the Yukawa suppression
$\e^p$ is the wavefunction-overlap penalty for $p$
units of subconstituent structure.

\emph{Quark--lepton unification in the hop framework.}
Because the $SU(5)$ multiplets $\mathbf{10}$ and
$\overline{\mathbf{5}}$ contain quarks and leptons
together, the hop-binding interpretation must be
consistent in both sectors simultaneously.
Three structural consequences follow:
(i)~the down-quark and charged-lepton sectors share
the same $\overline{\mathbf{5}}$ embedding, which is
precisely why $b$-$\tau$ Yukawa unification holds at
the GUT scale and produces the golden-ratio relation
$m_b/m_\tau = \varphi$ at $M_Z$~\cite{UFP}.
(ii)~The $\mathbf{10}$ embedding shared by up-quarks
and the charged-lepton singlets is split by the
Georgi--Jarlskog Clebsches: the $e^c$ charges
$(23/6,\,7/6,\,0)$ differ from the quark charges
$(3,\,2,\,0)$ by $\pm 5/6$ in generations~1 and~2,
a fractional shift that requires the full $\Z_{18}$
embedding (not just $\Z_9$).
(iii)~The hop-binding interaction $SU(5)_{\rm HC}$
acts on both quarks and leptons through their shared
$SU(5)$ structure: the $\alpha$- and $\beta$-hops are
$SU(5)$-singlets that bind colored and colorless
multiplets uniformly, with the colored ones receiving
additional QCD interactions as well.

\subsection{Binding scale and the axion connection}
\label{sec:binding}

The flavon vacuum expectation value (VEV) is
$\langle\Phi\rangle = \e\,\Lambda$, where
$\Lambda$ is the messenger mass scale.
In the $\Bpar$-lattice, the flavon is simultaneously the
PQ field, so $\langle\Phi\rangle = \fa$, the axion decay
constant~\cite{FlavorInNinths}.
This gives
\begin{equation}
\Lambda = \frac{\fa}{\e}
\sim \frac{(5\text{--}8)\times10^{11}\;\text{GeV}}{0.187}
\sim (3\text{--}4)\times 10^{12}\;\text{GeV}.
\label{eq:Lambda}
\end{equation}
In the compositeness picture, $\Lambda$ is the
\emph{confinement scale} of the hop-binding interaction.
The fact that $\Lambda\sim 10^{12}$~GeV---intermediate
between the electroweak and Planck scales---is consistent
with the scale at which a new confining gauge theory could
become strong, analogous to the QCD confinement scale
$\Lambda_{\rm QCD}\sim 200$~MeV but at much higher
energies.

The axion mass $m_a\sim 7$--$12\;\mu$eV predicted by the
framework~\cite{FlavorInNinths} is then directly related
to the hop confinement scale, since
$m_a\propto 1/\fa = \e/\Lambda$.
This provides an intriguing link: the mass of the lightest
pseudo-Goldstone boson in the theory (the axion) is
controlled by the same confinement dynamics that generates
the fermion mass hierarchy.

\subsection{The flavon VEV and the axion connection}

In the UV Lagrangian (Sec.~\ref{sec:UV-Lagrangian}),
the flavon $\Phi$ is an independent hypercolor-singlet
scalar whose VEV $\langle\Phi\rangle = \fa$
spontaneously breaks the Peccei--Quinn symmetry.
The expansion parameter
$\e = \langle\Phi\rangle/\Lambda = \fa/\Lambda$ is the
ratio of the PQ-breaking scale to the hop confinement
scale.
The Yukawa coupling $Y_{ij}\sim\e^{p_{ij}}$ arises
from a chain of $p_{ij}$ messenger links, each
contributing one power of $\e$ after integrating
out the heavy messengers and the hypercolor sector.

The Goldstone boson of the PQ breaking is the
QCD axion; the radial mode is the saxion.
The hop confinement scale and the PQ-breaking
scale are tied together by
$\Lambda = \fa/\e\simeq 5.4\,\fa$, so
the axion decay constant is a direct measure
of the confinement scale.

An alternative possibility---not assumed in
this paper but worth noting---is that $\Phi$
is itself a composite of hop fields
(e.g.\ $\Phi\sim\alpha^\dagger_a\alpha^a/\Lambda$,
an HC-singlet bilinear).
In that case, PQ breaking would be a
consequence of hop dynamics rather than an
independent sector, and the flavon VEV would
literally be a hop condensate.
This identification has implications for the
axion quality problem and is left as an open
question for future work.

The cosmological history involves
a phase transition at $T\sim\Lambda$:
above $\Lambda$, the hypercolor sector is
deconfined and the messenger chains are active;
below $\Lambda$, hop confinement integrates
out the hypercolor sector and the effective
Yukawa hierarchy turns on.
PQ symmetry breaking ($\langle\Phi\rangle\neq 0$)
may occur at a similar scale, since
$\fa = \e\,\Lambda\sim 0.19\,\Lambda$.
Cosmological constraints favor
this transition occurring \emph{before} or \emph{during}
inflation, so that the PQ symmetry is never restored
in the post-inflationary universe.
In this pre-inflationary scenario, the axion relic
density arises from vacuum misalignment
alone~\cite{PreskillWiseWilczek,AbbottSikivie,DineFischler}, with the initial
misalignment angle $\theta_i$ as the single
cosmological free parameter.
For $\theta_i\sim\mathcal{O}(1)$ and
$\fa\sim(5$--$8)\times10^{11}$~GeV, the predicted
axion density is
$\Omega_a h^2\simeq 0.12$~\cite{FlavorInNinths,BaerBargerSengupta},
and the entire dark matter abundance is a relic of
pre-inflationary PQ breaking and axion vacuum
misalignment.

\subsection{The nature of ``compositeness'' and anomaly consistency}
\label{sec:chiral-protection}

A central conceptual question must be addressed:
what does ``compositeness'' mean in the hop framework,
and how does it evade the constraints of
't~Hooft anomaly matching~\cite{tHooftAnomaly}?

\emph{Compositeness of the Yukawa coupling, not the
fermion.}
In the rishon model~\cite{HarariSeiberg1981}, quarks
and leptons are \emph{literally} bound states of
fermionic preons, and the UV theory has different chiral
degrees of freedom from the infrared (IR) theory.
In the hop framework, by contrast, all three generations
of Standard Model chiral fermions exist as
\emph{elementary fields in the Lagrangian at all scales}.
Their gauge quantum numbers
($SU(3)_C\times SU(2)_L\times U(1)_Y$) and their
$\Z_{18}$ charges are fixed by construction and do not
change across any threshold.
The ``compositeness'' resides in the
\emph{Yukawa coupling generation mechanism}:
the effective Yukawa matrix element $Y_{ij}$ arises from
$p_{ij}$ hop propagators in the VLQ chain
(Sec.~\ref{sec:UV-Lagrangian}), and
$p_{ij}$---the ``hop content''---counts the number of
these propagators, not the number of constituents
confined inside the fermion.

In this picture, the hypercolor $SU(N_H)_{\rm HC}$
confines the \emph{hops} (the scalar fields
$\alpha^a$, $\beta^a$)
into hypercolor-singlet bound states
(hyper-mesons, hyperglueballs),
but the SM fermions
are hypercolor singlets and are
\emph{spectators to the confinement}.
The chiral fermion spectrum---three generations
of quarks and leptons, each with fixed gauge and
$\Z_{18}$ charges---is identical above and below
$\Lambda$.
The vectorlike quark chain
(masses $\sim\Lambda$) drops out of the low-energy
theory, but its anomaly contributions cancel identically
(vectorlike pairs, $X_{D_a}+X_{\bar D_a}=0$).

\emph{Anomaly cancellation, not anomaly matching.}
Because $\Z_{18}$ is a \emph{gauged} discrete
symmetry~\cite{KraussWilczek,BanksDine,IbanezRoss},
the relevant consistency condition is anomaly
\emph{cancellation}---the discrete anomaly must
vanish identically---not 't~Hooft anomaly
\emph{matching} between UV and IR, which applies
only to global symmetries.
Since the chiral fermion content is unchanged across
the confinement transition and the hops are scalars
(which do not contribute to fermionic triangle diagrams
at any scale), the $\Z_{18}$--$[SU(3)]^2$,
$\Z_{18}$--$[SU(2)]^2$, $\Z_{18}$--$[\text{grav}]^2$,
and $\Z_{18}^3$ anomaly cancellation conditions that
are satisfied in the UV remain satisfied in the IR
automatically.
Should any residual discrete anomaly survive
(for example from an incomplete vectorlike spectrum),
a Green--Schwarz mechanism~\cite{GreenSchwarz}
involving the axion field $a(x)$ could cancel it
through the coupling $a\,F\widetilde F$, precisely
as in string-theoretic constructions of discrete
gauge symmetries; in the present framework, however,
the anomaly cancels identically without invoking
this mechanism.

\emph{Chiral protection.}
Why don't the light fermions acquire dynamical masses
of order $\Lambda$?
Because their left- and right-handed components carry
different $SU(2)_L\times U(1)_Y$ quantum numbers
(inherited entirely from the elementary core, since
the hops are SM singlets), a direct mass term
$m\,\bar\psi_L\psi_R\sim\Lambda$ would break the
unbroken electroweak gauge symmetry.
Furthermore, the asymmetric hop content
($Q(\psi_L)\neq Q(\psi_R)$) means such a mass term
would also violate the exact $\Z_{18}$ gauge symmetry
prior to flavon condensation.
The chiral fermions are therefore protected by exact
gauge symmetries and remain strictly massless until
the electroweak scale, where the Higgs VEV and the
flavon condensate $\langle\Phi\rangle = \fa$
simultaneously bridge the $SU(2)_L$ and $\Z_9$
quantum number gaps.

This distinction from the rishon model is fundamental:
in the rishon model, chiral anomaly matching is the
dynamical mechanism that \emph{requires} massless
composites to appear in the IR, and a mismatch would
signal an inconsistency.
In the hop framework, the SM fermions are already
present as elementary fields---they do not need to
be ``produced'' by confinement---and anomaly
consistency is maintained trivially because the
chiral spectrum never changes.

\subsection{UV Lagrangian and EFT matching}
\label{sec:UV-Lagrangian}

A central issue is whether the ``hop'' variables can
be interpreted as more than bookkeeping devices in the
low-energy Froggatt--Nielsen (FN) operators.
A conservative but more explicit answer is to regard
them as fields of an ultraviolet hypercolor sector
whose effects are communicated to the Standard Model
by heavy vectorlike messenger chains.
The low-energy FN operators then arise only after
both the messenger sector and the hypercolor sector
have been integrated out.
In this way the same underlying sector admits two
complementary descriptions: a propagating UV
description above the confinement scale and an
effective Yukawa description below it.

\emph{Field content and symmetries.}
Let the hypercolor gauge group be $SU(N_H)_{\rm HC}$.
We introduce two complex scalar hop fields
$\alpha^a$, $\beta^a$ ($a=1,\ldots,N_H$), each
transforming in the fundamental $\mathbf{N_H}$,
together with a hypercolor-singlet flavon $\Phi$.
All three are singlets under the SM gauge group:
\begin{equation}
\renewcommand{\arraystretch}{1.2}
\begin{array}{lccc}
\hline\hline
\text{Field} & SU(N_H)_{\rm HC} & \Z_9
& \text{SM} \\
\hline
\alpha^a & \mathbf{N_H} & +1 & \text{singlet} \\
\beta^a  & \mathbf{N_H} & +3 & \text{singlet} \\
\Phi     & \mathbf{1}   & -1 & \text{singlet} \\
\hline\hline
\end{array}
\label{eq:UV-fields}
\end{equation}
That is, the hops and flavon carry zero color,
zero weak isospin, and zero hypercharge---they
interact with ordinary matter only through
$\Z_9$ and hypercolor.
with $\langle\Phi\rangle\equiv\fa$ and
$\e\equiv\fa/\Lambda$, where $\Lambda$ denotes the
characteristic heavy scale of the
messenger/hypercolor sector.
For illustrative purposes one may later consider
$N_H = 5$, but the discussion here does not
require fixing the group.
The messenger multiplicities are assumed small enough
that $SU(N_H)_{\rm HC}$ remains asymptotically free
and confining at the scale $\Lambda$; for $N_H = 5$
with two scalar fundamentals, the one-loop
coefficient remains in the asymptotically free
regime.

\emph{Scalar potential and species stability.}
The $\Z_9$ symmetry constrains the scalar potential
$V(\Phi,\alpha,\beta)$.
Since $q_9(\alpha)=1$ and $q_9(\beta)=3$,
the quadratic mixing term
$\alpha^\dagger_a\beta^a$ carries $\Z_9$ charge
$+2$ and is \emph{forbidden}.
The species-diagonal mass terms
$m_\alpha^2\,\alpha^\dagger_a\alpha^a$ and
$m_\beta^2\,\beta^\dagger_a\beta^a$ are allowed
($q_9=0$), as are all quartic couplings built
from $|\alpha|^2$, $|\beta|^2$, and $|\Phi|^2$
(portal couplings such as
$|\alpha|^2|\beta|^2$, $|\Phi|^2|\alpha|^2$,
$|\Phi|^2|\beta|^2$).
These portals do not mix the two species.
The lowest-dimension $\alpha$--$\beta$ mixing
operator allowed by $\Z_9$ is the quartic
$\lambda_{\alpha\beta}\,\Phi^2\,\alpha^\dagger_a\beta^a$
($q_9 = -2-1+3 = 0$),
which after PQ breaking induces an off-diagonal
mass-squared term
$\delta m^2_{\alpha\beta}
\sim\lambda_{\alpha\beta}\,\fa^2$.
Relative to diagonal masses of order $\Lambda^2$,
the resulting mixing is naturally of order
$\lambda_{\alpha\beta}\,\e^2\simeq 0.035\,
\lambda_{\alpha\beta}$---a
small perturbation that does not destabilize
the species distinction at leading order.
The two hop species are therefore protected
as independent degrees of freedom by the
$\Z_9$ charge algebra.

To communicate the hop sector to the SM in a
manifestly gauge-invariant way, we introduce
two classes of heavy vectorlike messenger fermions
for a chain of $m$ steps:
\begin{itemize}
\item $m+1$ hypercolor-singlet messengers
$S_0,\,S_1,\,\ldots,\,S_m$
(with vectorlike partners $\bar S_r$);
\item $m$ hypercolor-fundamental messengers
$F_1^a,\,\ldots,\,F_m^a$
(with partners
$\bar F_{1\,a},\,\ldots,\,\bar F_{m\,a}$
in the $\overline{\mathbf{N_H}}$).
\end{itemize}
The messengers are chosen vectorlike under the SM
gauge group with whatever electroweak/color quantum
numbers are needed to couple to a given SM bilinear.
The detailed SM representations depend on whether
one is generating up-quark, down-quark,
charged-lepton, or neutrino operators; this is
standard FN model-building~\cite{FN,FN1980} and is
not the focus here.
In particular, for up-type operators the Higgs
field $H$ in the endpoint coupling is replaced by
$\widetilde H = i\sigma_2 H^*$, and the SM quantum
numbers of the HC-singlet endpoint messengers $S_0$,
$S_m$ are chosen to match the corresponding
right-handed fermion representation so that each
vertex is fully SM-gauge-invariant.
What matters for present purposes is the
\emph{hypercolor structure}.

\emph{Manifestly gauge-invariant chain Lagrangian.}
For definiteness, consider a chain generating an
effective Yukawa operator connecting a left-handed
SM fermion $\psi_{Li}$ and a right-handed
SM fermion $\psi_{Rj}$.
A gauge-invariant UV Lagrangian is
\begin{equation}
\mathcal{L}_{\rm UV}
= \mathcal{L}_{\rm SM}
+ \mathcal{L}_{\rm HC}
+ \mathcal{L}_{\Phi,h}
+ \mathcal{L}_{\rm mess}
+ \mathcal{L}_{\rm chain},
\end{equation}
with
\begin{align}
\mathcal{L}_{\Phi,h}
&= |D_\mu \alpha|^2 + |D_\mu \beta|^2
+ |\partial_\mu \Phi|^2
- V(\Phi,\alpha,\beta),
\\[3pt]
\mathcal{L}_{\rm mess}
&= -\sum_{r=0}^{m}
M_{S_r}\,\bar S_r S_r
-\sum_{r=1}^{m}
M_{F_r}\,\bar F_{r\,a}\,F_r^a,
\end{align}
and
\begin{align}
\mathcal{L}_{\rm chain}
&= -y^{(L)}_{i}\,\bar\psi_{Li}\, H\, S_0
  - y^{(R)}_{j}\,\bar S_{m}\,\psi_{Rj}
\nonumber\\[3pt]
&\quad
-\sum_{r=1}^{m}
\Bigl[
\kappa_r\,\Phi\,\bar S_{r-1}\,
h^\dagger_{I_r\,a}\,F_r^a
+ \widetilde\kappa_r\,
\bar F_{r\,a}\,h_{I_r}^a\,S_{r}
\Bigr]
\nonumber\\
&\quad + \text{h.c.},
\label{eq:chain-vertex}
\end{align}
where $h_{I_r}\in\{\alpha,\beta\}$ denotes whichever
hop species is required at the $r$th step.
Every term in Eq.~\eqref{eq:chain-vertex} is
manifestly invariant under hypercolor: in each
step coupling, the fundamental index $a$ is
contracted between $h^\dagger_{I_r\,a}$ and $F_r^a$,
or between $\bar F_{r\,a}$ and $h_{I_r}^a$.
The endpoint couplings
$\bar\psi_{Li}\,H\,S_0$ and $\bar S_m\,\psi_{Rj}$
involve only hypercolor-singlet messengers,
so they are also hypercolor invariant.
Discrete invariance of each link under $\Z_9$ is
ensured by assigning charges recursively along the
chain: once the external fermion charges $q_9(\psi_i)$,
$q_9(\psi_j)$ and the hop species sequence $\{I_r\}$
are specified, the messenger charges $q_9(S_r)$ and
$q_9(F_r)$ are fixed uniquely (up to an overall shift)
by requiring each vertex to be $\Z_9$-neutral.

\begin{figure*}[!tbp]
\centering
\begin{tikzpicture}[
  x=1.65cm, y=1.0cm,
  sing/.style={circle,draw=blue!70,fill=blue!10,thick,
    minimum size=8mm,font=\scriptsize},
  fund/.style={circle,draw=red!70,fill=red!10,thick,
    minimum size=8mm,font=\scriptsize},
  sm/.style={rectangle,draw=black!70,fill=green!8,thick,
    rounded corners=2pt,minimum width=9mm,
    minimum height=6mm,font=\scriptsize},
  hop/.style={->,thick,orange!80!black,>=stealth},
  link/.style={-,very thick,black!60},
  lbl/.style={font=\scriptsize,inner sep=1pt},
  >=stealth
]

\node[sm] (psiL) at (-0.8,0) {$\bar\psi_{Li}$};
\node[sm] (H) at (0.0,0.85) {$H$};
\node[sm] (psiR) at (8.8,0) {$\psi_{Rj}$};

\node[sing] (S0) at (0.7,0) {$S_0$};
\node[fund] (F1) at (2.0,0) {$F_1^a$};
\node[sing] (S1) at (3.3,0) {$S_1$};
\node[fund] (F2) at (4.6,0) {$F_2^a$};
\node[sing] (S2) at (5.9,0) {$S_2$};
\node[sing] (Sm) at (7.8,0) {$S_m$};

\node at (6.85,0) {$\cdots$};

\draw[link] (psiL) -- (S0);
\draw[link] (S0) -- (F1);
\draw[link] (F1) -- (S1);
\draw[link] (S1) -- (F2);
\draw[link] (F2) -- (S2);
\draw[link] (Sm) -- (psiR);

\draw[link,dashed] (H) -- (S0);

\draw[hop] (1.35,0.15) -- ++(0,1.0)
  node[above,lbl,orange!80!black]
  {$h^\dagger_{I_1,a}\;(\Phi)$};
\node[lbl,blue!60] at (1.35,-0.6)
  {$\kappa_1\,\Phi\,\bar S_0\,
   h^\dagger_{I_1,a}\,F_1^a$};

\draw[hop] (2.65,1.15) -- ++(0,-1.0);
\node[above,lbl,orange!80!black] at (2.65,1.15)
  {$h_{I_1}^a$};
\node[lbl,red!60] at (2.65,-0.6)
  {$\widetilde\kappa_1\,\bar F_{1,a}\,
   h_{I_1}^a\,S_1$};

\draw[hop] (3.95,0.15) -- ++(0,1.0)
  node[above,lbl,orange!80!black]
  {$h^\dagger_{I_2,a}\;(\Phi)$};

\draw[hop] (5.25,1.15) -- ++(0,-1.0);
\node[above,lbl,orange!80!black] at (5.25,1.15)
  {$h_{I_2}^a$};

\node[lbl,blue!70] at (0.7,-1.1) {\textbf{1}};
\node[lbl,red!70] at (2.0,-1.1) {$\mathbf{N_H}$};
\node[lbl,blue!70] at (3.3,-1.1) {\textbf{1}};
\node[lbl,red!70] at (4.6,-1.1) {$\mathbf{N_H}$};
\node[lbl,blue!70] at (5.9,-1.1) {\textbf{1}};
\node[lbl,blue!70] at (7.8,-1.1) {\textbf{1}};

\node[lbl,gray] at (4.0,-1.5)
  {$SU(N_H)_{\rm HC}$ representations};

\node[lbl] at (-0.4,0.4) {$y^{(L)}_i$};
\node[lbl] at (8.3,0.35) {$y^{(R)}_j$};

\draw[blue!70,thick] (0.2,-2.1) -- ++(0.4,0)
  node[right,lbl,black] {HC singlet ($S_r$)};
\draw[red!70,thick] (3.2,-2.1) -- ++(0.4,0)
  node[right,lbl,black] {HC fundamental ($F_r^a$)};
\draw[hop] (6.2,-2.1) -- ++(0.4,0)
  node[right,lbl,black] {hop ($h_{I_r}$)};

\end{tikzpicture}
\caption{The alternating messenger chain.
Blue circles: hypercolor-singlet messengers
$S_0,\ldots,S_m$.
Red circles: hypercolor-fundamental messengers
$F_1^a,\ldots,F_m^a$.
Orange arrows: hop fields emitted ($h^\dagger_{I_r,a}$)
or absorbed ($h_{I_r}^a$) at each link, with the
HC index $a$ contracted between the hop and the
adjacent $F_r^a$ or $\bar F_{r,a}$.
The SM fields (green) couple only to the
HC-singlet endpoints $S_0$ and $S_m$.
Each step also involves one flavon insertion $\Phi$.
The species $I_r\in\{\alpha,\beta\}$ at each step
is fixed by $\Z_9$ charge conservation.}
\label{fig:chain-UV}
\end{figure*}
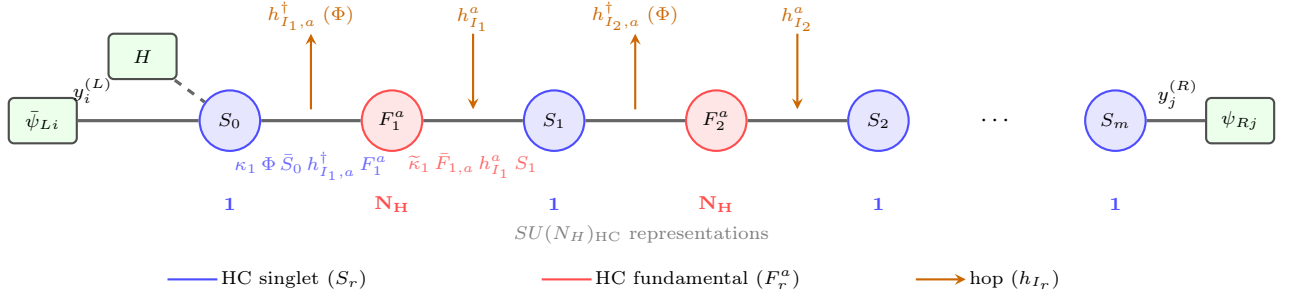

The chain thus alternates
$S_0$--$F_1$--$S_1$--$F_2$--$S_2$--$\cdots$--$F_m$--$S_m$,
with each $S_{r-1}\to F_r$ transition emitting a
hop $h^\dagger_{I_r}$ and each
$F_r\to S_r$ transition absorbing one $h_{I_r}$.
The SM fields couple only to the HC-singlet
endpoints $S_0$ and $S_m$, avoiding the problem
of coupling an HC-charged state directly to a
purely SM operator.
Figure~\ref{fig:chain-UV} displays this structure.

\emph{Tree-level matching above confinement.}
At scales $\mu$ above the hypercolor confinement
scale, the fields $\alpha$, $\beta$ may be treated
as propagating degrees of freedom.
Integrating out the heavy messengers at tree level
produces a nonlocal operator connecting the SM
bilinear to a string of hop insertions.
Parametrically one finds
\begin{align}
\mathcal{L}_{\rm eff}^{(\mu>\Lambda)}
&\sim
y^{(L)}_i\,y^{(R)}_j\;
\bar\psi_{Li}\, H\, \psi_{Rj}\;
\prod_{r=1}^{m}
\biggl(
\frac{\kappa_r\widetilde\kappa_r}
{M_{S_{r-1}}M_{F_r}M_{S_{r}}}
\biggr)
\nonumber\\
&\quad\times
\Phi^m\,
\prod_{r=1}^{m}
\bigl(
h^\dagger_{I_r\,a}\,h_{I_r}^a
\bigr)
+ \text{h.c.},
\label{eq:chain-Yukawa}
\end{align}
up to index-ordering conventions and order-one
combinatorial factors.
The inverse messenger masses in the denominator
are the standard result of integrating out heavy
fields at tree level: each messenger propagator
contributes a factor $1/M$ to the effective
operator.
When all messenger masses are of order $\Lambda$,
each step contributes a net suppression
$\sim\langle\Phi\rangle/\Lambda = \e$, recovering
the FN counting $y_{ij}\sim\e^m$ for a chain
of $m$ steps.
The precise Wilson coefficient is model-dependent,
but the structure is fixed: each chain step
contributes one power of $\Phi$ and one
hypercolor-singlet contraction built from hop fields.

\emph{Matching through confinement.}
Below $\Lambda$, the hypercolor sector confines and
the bilinears
$\mathcal{O}_I\equiv h^\dagger_{I\,a}\,h_I^a$
($I=\alpha,\beta$) must be matched onto
hypercolor-singlet low-energy operators.
By naive dimensional analysis~(NDA), the
strong-dynamics matching coefficient for each
bilinear scales as
\begin{equation}
\mathcal{O}_I \;\to\; \xi_I\,\Lambda^2
\qquad\text{(low-energy matching)},
\end{equation}
with $\xi_I$ dimensionless hadronic coefficients
expected to be of order unity,
modulo group-theory and wavefunction factors.
This is a matching relation in the effective field
theory (EFT) sense
(analogous to $\langle\bar qq\rangle\sim\Lambda_{\rm QCD}^3$
in chiral perturbation theory;
the power $\Lambda^2$ rather than $\Lambda^3$
reflects the mass dimension of a scalar bilinear
versus a fermion bilinear),
not a claim about
the vacuum structure of the hypercolor theory.
Substituting these matching coefficients into
Eq.~\eqref{eq:chain-Yukawa} yields effective
Yukawa operators of Froggatt--Nielsen form,
\begin{equation}
\mathcal{L}_{\rm eff}^{(\mu<\Lambda)}
= c_{ij}\,\e^{\,p_{ij}}
\,\bar\psi_i\,H\,\psi_j
+ \text{h.c.},
\label{eq:EFT-Yukawa}
\end{equation}
where $p_{ij}$ counts the total number of chain steps and
$c_{ij}$ collects the products of endpoint couplings
$y^{(L)}_i$, $y^{(R)}_j$,
chain couplings $\kappa_r$, $\widetilde\kappa_r$,
mass ratios $M_{S_r}/\Lambda$, $M_{F_r}/\Lambda$,
and nonperturbative factors $\xi_{I_r}$.
At the parametric level, the low-energy hierarchy is
controlled by the total number of chain steps $p_{ij}$;
the nonperturbative hypercolor dynamics enter only
through the $\mathcal{O}(1)$ coefficients $c_{ij}$.

The predictions of the $\Bpar$-lattice
(mass ratios, mixing angles, $m_3$, $m_a$) depend
only on Eq.~\eqref{eq:EFT-Yukawa} and are
insensitive to the details of the UV construction.

\emph{Species dependence and its limitations.}
In the UV chain, the species label $I_r\in\{\alpha,\beta\}$
at each step is fixed by the $\Z_9$ charge flow:
the messengers at the $r$th link must carry
$\Z_9$ charges compatible with absorbing an $\alpha$
(advancing $q_9$ by $+1$) or a $\beta$ (advancing
by $+3$).
Thus the \emph{number} of $\alpha$- and $\beta$-type
steps in a given chain is determined by the
$\Z_9$ charges of the external fermions, and
different Yukawa entries involve chains with
distinct species sequences.
However, after matching through confinement,
the hypercolor bilinears
$\mathcal{O}_\alpha = \alpha^\dagger_a\alpha^a$
and $\mathcal{O}_\beta = \beta^\dagger_a\beta^a$
are both $\Z_9$-neutral:
the species information survives in the low-energy
EFT only through the $\mathcal{O}(1)$
nonperturbative coefficients $\xi_\alpha$ and
$\xi_\beta$, not through a symmetry-enforced
distinction.
In other words, the UV chain \emph{selects}
which species appears at each step, but the
confinement matching does not \emph{derive}
the species-dependent suppression hierarchy
from first principles---it parametrizes it.
A complete derivation would show how the additive
$(\alpha,\beta)$ lattice structure maps
step-by-step onto a fully specified messenger
$\Z_9$ charge table; this remains an open problem.

\emph{Interpretation and scope.}
In light of this limitation,
this UV construction should be read as an
\emph{existence proof}---a demonstration that
a gauge-invariant renormalizable messenger-chain
embedding of the hop fields is possible:
it shows that the hop fields can in principle be
genuine hypercolor-charged degrees of freedom
communicated to the SM by a renormalizable
messenger chain, with the low-energy FN
operators emerging after messenger decoupling
and hypercolor confinement.
What has been specified is:
(i)~a manifestly hypercolor-gauge-invariant UV chain,
(ii)~a consistent alternation between HC-singlet and
HC-fundamental messengers, and
(iii)~the parametric EFT matching onto low-energy
Yukawa operators.
What has \emph{not} been derived from first
principles is the species-dependent matching
(i.e.\ the distinction between $\xi_\alpha$ and
$\xi_\beta$), the detailed
spectrum of confined states, or their decay widths
and relic abundances.
Those belong to the illustrative hypercolor
realization and are left for future work.

\section{The Higgs Sector}
\label{sec:EWSB}

\subsection{The top quark as the elementary benchmark}

In the $\Bpar$-lattice, the top quark is the unique
fermion with zero hop content on both sides of the Yukawa
bilinear: $Q(Q_3)=Q(u^c_3)=0$, giving $p^u_{33}=0$
and $Y_t = \mathcal{O}(1)$.
The top mass is therefore
\begin{equation}
m_t^{\overline{\rm MS}}(M_Z)
= c_t\,Y_t\,\frac{v}{\sqrt{2}}\simeq 169\;\text{GeV},
\end{equation}
with $c_t\simeq 0.97$ absorbing the small departure
of $Y_t$ from unity (the pole mass
$m_t^{\rm pole}\simeq 173$~GeV corresponds to
$c_t Y_t\simeq 0.99$).
Either way, this fixes the Higgs vacuum expectation
value (VEV) $v = 246$~GeV as the
\emph{unsuppressed Yukawa scale}---the
energy at which an elementary fermion couples to the Higgs
at full strength.

In the hop-compositeness picture, this acquires a
structural meaning: $v$ is the scale at which the
elementary top quark interacts with the
electroweak sector without the penalty of wavefunction
overlap through a hop cloud.
The top is unique in this respect: among all SM
fermions, only the top has $Q(Q_3) = Q(u^c_3) = 0$,
i.e.\ vanishing FN charges on \emph{both} sides of
the Yukawa bilinear and a trivial up-type chain
internal factor $\Delta_{\rm int}^u = 0$
(with $\sin\beta\simeq 1$ at $\tan\beta\simeq 16$
contributing no additional suppression).
This is the converse of top-condensation
models~\cite{Hill2025}, where the Higgs boson is a
$\bar tt$ bound state and the top Yukawa is dynamically
generated; in the hop framework, both the Higgs and
the top are elementary, and the lighter-generation
Yukawas (including the bottom quark and the entire
charged-lepton sector) are suppressed by hop
compositeness rather than the top Yukawa being
enhanced by condensation.

\emph{Origin of the top--tau hierarchy.}
A natural question is how the framework, in which
all fundamental Yukawas start at $\mathcal{O}(1)$,
generates the large hierarchy
$m_t/m_\tau\sim 100$ between the two third-generation
charged fermions.
The hierarchy has \emph{two independent structural
sources}:
(i)~\emph{Chain internal factor (compositeness).}
The up-type Yukawa receives the trivial internal
factor $\Delta_{\rm int}^u = 0$, while the down-type
and lepton Yukawas receive
$\Delta_{\rm int}^{d,\ell} = 7/9$, giving a relative
suppression $\e^{7/9}\simeq 0.27$ of $Y_\tau$
relative to $Y_t$.
This is a property of the confining VLQ chain and
is independent of the Higgs sector.
(ii)~\emph{Two-Higgs-doublet-model (2HDM) VEV ratio.}
In DFSZ-II, $m_t\propto v\sin\beta$ while
$m_b,\,m_\tau\propto v\cos\beta$.
At $\tan\beta\simeq 16$ (Sec.~\ref{sec:tanb}),
$\sin\beta\simeq 1$ and
$\cos\beta/\sin\beta = 1/\tan\beta\simeq 0.06$
provides an additional Higgs-sector suppression.
This is a property of the two-Higgs-doublet
structure and is independent of the chain.
Combining the two independent effects,
\begin{equation}
\frac{m_\tau}{m_t}\;\sim\;
\frac{c_\tau}{c_t}\,
\frac{\e^{7/9}}{\tan\beta}
\;\simeq\;\frac{1}{\sqrt{3}}(0.27)(0.06)
\;\simeq\; 0.01,
\end{equation}
matching the observed ratio.
Here the $\mathcal{O}(1)$ factor
$c_\tau/c_t\simeq 1/\sqrt{3}$ is the
Georgi--Jarlskog Clebsch--Gordan ratio responsible
for the $b$--$\tau$ unification relation at the
GUT scale (Sec.~\ref{sec:alpha-census}).
Thus the top--tau hierarchy is generated by the
combination of hop compositeness ($\e^{7/9}$) and
2HDM VEV alignment ($1/\tan\beta$); the two
suppressions are conceptually distinct and
multiply.
No explicit hierarchy among $\mathcal{O}(1)$
Yukawa coefficients is required.

The mass of every lighter fermion is obtained by
multiplying the appropriate VEV prefactor by the
hop suppression $\e^{p_{ii}}$:
\begin{equation}
m_f = c_f\,\e^{p_f}\times\frac{v_f}{\sqrt{2}},
\qquad
v_f = \begin{cases}
v\sin\beta & f\in\{u,c,t\}\\
v\cos\beta & f\in\{d,s,b,e,\mu,\tau\}\\
v\sin\beta & f\in\{\nu_1,\nu_2,\nu_3\}
\end{cases}
\end{equation}
so the entire SM mass spectrum is the Higgs VEV (with
appropriate DFSZ-II projection) filtered through hop
compositeness.

\emph{Higgs charges and the lepton chain.}
As established in Sec.~\ref{sec:SU5-charges}, the Higgs fields carry
no net FN charge ($Q_{H_u} = Q_{H_d} = 0$); the
sector-dependent suppression enters through the
chain internal factor
$\Delta_{\rm int}^{d,\ell} = 7/9$ rather than through
$Q_H$.
The lepton charge assignments are
$Q(L_i) = (1,\,1/2,\,0)$ and
$Q(e^c_i) = (23/6,\,7/6,\,0)$~\cite{LeptonLattice},
generating the charged-lepton mass formula
$m_\ell\propto\e^{Q(L_i)+Q(e^c_j)+\Delta_{\rm int}^\ell}$
where $\Delta_{\rm int}^\ell = 7/9$ is the lepton-chain
internal factor (identical to the down-type quark value).
Unlike the colored VLQ chain that mediates quark Yukawas,
the lepton chain uses hypercolor-singlet messengers
(Sec.~III); the topology and counting are identical, but
the messengers carry no $SU(3)_c$ charge.

\subsection{The bottom Yukawa and the chain internal factor}
\label{sec:tanb}

The top Yukawa $Y_t\sim\mathcal{O}(1)$ reflects the
absence of hop dressing on both sides of the up-type
bilinear \emph{and} the trivial internal factor
$\Delta_{\rm int}^u=0$ of the up-type chain.
The bottom quark, however, propagates through the
down-type VLQ chain, whose three hop links
$\e^{4/9}\!\times\!\e^{2/9}\!\times\!\e^{1/9}
= \e^{7/9}$
(Sec.~\ref{sec:chain}) contribute a universal internal
tunneling factor.
Although $Q(Q_3)=Q(d^c_3)=0$, so the endpoint
dressings vanish, the \emph{absolute} bottom Yukawa is
\begin{equation}
Y_b = c_{33}^d\;\e^{\Delta_{\rm int}^d}
= c_{33}^d\;\e^{7/9}
\simeq 0.27,
\label{eq:Yb-chain}
\end{equation}
with $c_{33}^d = \mathcal{O}(1)$.
In the SM with a single Higgs doublet,
$m_b/m_t = Y_b/Y_t \simeq \e^{7/9}\simeq 0.27$;
the measured ratio
$\overline{m}_b(M_Z)/m_t\simeq 0.017$~\cite{PDG2024,HuangZhou}
is smaller by a factor $\sim 16$, indicating
that additional structure is required.

Precisely this structure is already present in the
framework: the DFSZ-II axion model adopted
throughout this paper
(Appendix~\ref{app:EN}) requires two Higgs
doublets $H_u$ and $H_d$ with independent
PQ charges.
In such a Type-II two-Higgs-doublet model,
$m_t = Y_t\,v\sin\beta/\sqrt{2}$ and
$m_b = Y_b\,v\cos\beta/\sqrt{2}$, so
\begin{equation}
\frac{m_b}{m_t}
= \frac{Y_b}{Y_t}\,\frac{1}{\tan\beta}
= \frac{\e^{7/9}}{\tan\beta},
\end{equation}
which gives
\begin{equation}
\tan\beta \;\simeq\;
\frac{\e^{7/9}}{m_b/m_t}
\;\simeq\; \frac{0.27}{0.017}
\;\simeq\; 16,
\label{eq:tanb}
\end{equation}
consistent with $\tan\beta\simeq 10$--$16$
depending on the renormalization-group running
between $M_Z$ and the heavy Higgs
scale~\cite{PDG2024}.
This is a \emph{self-consistency} of the
framework: the two-Higgs-doublet structure
that the DFSZ-II axion model requires
(and that is used to compute $C_{a\gamma}$
and $C_{ae}$ in the appendix) automatically
provides the $\tan\beta$ needed to reconcile
the chain internal factor with the observed
$b/t$ mass ratio.
The key lattice result is that the chain
internal factor $\e^{7/9}$, which cancels
in all within-sector mass ratios, survives
in the \emph{inter-sector} ratio $Y_b/Y_t$
and, combined with the DFSZ-II Higgs sector,
predicts $\tan\beta$.

\section{Right-Handed Neutrinos and the Seesaw}
\label{sec:nuR}

\subsection{Hop dressing of the seesaw}

Neutrino masses in the $\Bpar$-lattice arise from the
dimension-five Weinberg operator with exponents
$p^\nu_{ij}=Q(L_i)+Q(L_j)$, giving a normal-ordered
spectrum $m_1:m_2:m_3\sim\e^2:\e:1$~\cite{FlavorInNinths,UFP}.
In the hop picture, each factor of $\e$ reflects one unit
of left-handed lepton compositeness.

If the Weinberg operator is UV-completed by the Type-I
seesaw
mechanism~\cite{Minkowski,Yanagida,GellMannRamondSlansky,MohapatraSenjanovic}
with three right-handed neutrinos $N_i$,
both the Dirac Yukawa coupling and the Majorana mass
acquire hop dressing:
\begin{align}
(Y_\nu)_{ij} &= c_{ij}\,\e^{Q(L_i)+Q(N_j)},
\\
(M_R)_{ij} &= M_0\,c'_{ij}\,\e^{Q(N_i)+Q(N_j)},
\label{eq:MR-ij}
\end{align}
where $Q(N_j)$ are the FN charges of the right-handed
neutrinos and $M_0$ is the bare Majorana scale.

\subsection{Cancellation of right-handed hop content}

The seesaw formula
$m_\nu = Y_\nu\,M_R^{-1}\,Y_\nu^T\,v^2/2$
has a remarkable property.
For the diagonal entries (which set the mass eigenvalue
hierarchy to leading order), the contribution from the
$i$-th generation is
\begin{equation}
(m_\nu)_{ii} \sim
\frac{(Y_\nu)_{ii}^2\,v^2}{2\,(M_R)_{ii}}
\sim \frac{\e^{2[Q(L_i)+Q(N_i)]}\,v^2}
{2\,M_0\,\e^{2Q(N_i)}}
= \frac{\e^{2Q(L_i)}\,v^2}{2\,M_0}.
\end{equation}
The right-handed hop charges $Q(N_i)$ cancel exactly:
each power of $\e^{Q(N_i)}$ from the Dirac Yukawa is
compensated by a power of $\e^{-Q(N_i)}$ from the
Majorana mass inverse.
The neutrino mass hierarchy therefore depends
\emph{only on the left-handed hop content}:
\begin{equation}
\frac{m_i}{m_3} = \e^{2Q(L_i)},
\end{equation}
giving $m_2/m_3 = \e^{2\times1/2}=\e$ and
$m_1/m_3 = \e^{2\times1}=\e^2$, consistent with the
Weinberg-operator result.

This cancellation has a beautiful hop interpretation:
in the seesaw, each light neutrino mass receives two
Dirac Yukawa insertions (one for each lepton doublet)
and one Majorana mass insertion (for the intermediate
right-handed neutrino).
The hop content of the right-handed neutrino enters
symmetrically in the Yukawa and the Majorana mass, so it
drops out of the ratio.
The seesaw ``undoes'' the right-handed compositeness,
leaving only the left-handed hop content to control
the hierarchy---a natural consequence of the fact that
we observe left-handed neutrinos, not right-handed ones.

\subsection[The Majorana scale and a prediction of m3]{\texorpdfstring{The Majorana scale and a prediction of $m_3$}{The Majorana scale and a prediction of m3}}

If the Majorana scale $M_0$ sits on the ninths lattice,
its position is determined by $\Lambda$ and $\e$---the
same two parameters that control the entire quark and
charged-lepton spectrum.
The ninths ladder
(Sec.~\ref{sec:GUT}) assigns
\begin{equation}
\boxed{\;M_0 = \Lambda\times\e^{-28/9}
\simeq 6\times10^{14}\;\text{GeV}\;}
\label{eq:M0-ninths}
\end{equation}
where $28/9$ lies on the ninths lattice and
$\e^{-28/9}\simeq 190$.
That is, the Majorana scale is the hop confinement scale
$\Lambda$ enhanced by $28$ units (in ninths) of
``inverse hop content,'' placing it between $\Lambda$
and $M_{\rm Pl}$ on the geometric staircase.

\emph{Bare versus physical Majorana mass.}
$M_0$ is the \emph{bare} (Lagrangian-level) Majorana
mass parameter that sits on the ninths ladder; it
multiplies the $\e^{Q(N_i)+Q(N_j)}$ hop dressing
in Eq.~\eqref{eq:MR-ij} to give the
\emph{physical} (dressed) Majorana mass
$(M_R)_{33} = M_0\,\e^{2Q(N_3)} = M_0\,\e^{28/9}
\simeq \Lambda \simeq 3\times 10^{12}$~GeV.
The hop dressing thus pulls the physical right-handed
neutrino mass back down to the hop confinement scale,
even though $M_0$ itself lives two orders of magnitude
above $\Lambda$.
The factor $\e^{Q(N_i)}$ that suppresses the Dirac
Yukawa coupling is exactly compensated by the
factor $\e^{-Q(N_i)}$ from
$(M_R)^{-1}$ in the seesaw, so only $M_0$ (not
the physical $M_R$) appears in the final formula
for $m_3$:
\begin{equation}
m_3 = \frac{v^2}{2\,M_0}
= \frac{v^2}{2\,\Lambda\,\e^{-28/9}}
\simeq \frac{(246\;\text{GeV})^2}
{2\times 5.9\times10^{14}\;\text{GeV}}
\simeq 51\;\text{meV}.
\label{eq:m3-prediction}
\end{equation}
The experimental value
$\sqrt{\Delta m^2_{\rm atm}}\simeq 50$~meV
agrees to $\sim 2\%$.
Once the Majorana scale is placed on the ninths
lattice, no additional parameter is adjusted:
$\Lambda = \fa/\e$ is fixed by the axion decay
constant, $\e = 14/75$ is fixed by the quark mass
ratios, and $v = 246$~GeV is the Higgs VEV.
The placement of $M_0$ at $n = -28/9$ is itself
motivated by the $\Z_9$ charge algebra
(the hop charges of the right-handed neutrinos
sum to $28/9$; see above), but is not uniquely
derived from first principles---it remains the
one structural input specific to the neutrino sector.

The full predicted spectrum is
\begin{align}
m_3 &\simeq 51\;\text{meV}, \nonumber\\
m_2 &= m_3\,\e \simeq 9.5\;\text{meV}
\quad(\text{data:}\;\sim 8.6\;\text{meV}),
\nonumber\\
m_1 &= m_3\,\e^2 \simeq 1.8\;\text{meV},
\label{eq:nu-spectrum}
\end{align}
with $\sum m_i \simeq 62$~meV.
This prediction is sharply testable:
the conservative cosmological bound
$\sum m_i < 120$~meV~\cite{PDG2024} is comfortably
satisfied, but recent Dark Energy Spectroscopic
Instrument (DESI) baryon-acoustic-oscillation (BAO)
data combined with
cosmic microwave background (CMB) measurements yield
$\sum m_\nu < 64$~meV (95\% CL) under
$\Lambda$CDM~\cite{DESI2025}---placing the hop-framework prediction
directly at the current sensitivity frontier.
The bound relaxes substantially
($\sum m_\nu < 160$~meV) in extended dark
energy models ($w_0 w_a$CDM), so the prediction
remains viable, but further tightening of the
$\Lambda$CDM bound would provide a decisive test.
The effective Majorana mass for neutrinoless
double-beta decay
$m_{\beta\beta}\sim m_1\simeq 2$~meV,
below current sensitivity but within reach of
next-generation ton-scale experiments.

Alternatively,
$M_0\simeq M_{\rm Pl}\times\e^{53/9}$, placing the
Majorana scale between $\Lambda$ and $M_{\rm Pl}$ on
the ninths lattice:
\begin{equation}
\Lambda\;\ll\; M_0 = \Lambda\,\e^{-28/9}\;\ll\; M_{\rm Pl}.
\end{equation}

The prediction acquires a strikingly compact form
when expressed on the ninths ladder.
Since $v = \Lambda\,\e^{125/9}$
(Eq.~\eqref{eq:scale-ladder}),
the confinement scale cancels:
\begin{equation}
\boxed{\;m_3 = \tfrac{1}{2}\,v\,\e^{17}\;}
\label{eq:m3-compact}
\end{equation}
(the factor $1/2 = (1/\sqrt{2})^2$ is the standard
electroweak normalization
$\langle H^0\rangle = v/\sqrt{2}$, which enters
the seesaw squared through the Dirac Yukawa
$m_D = Y_\nu\,v/\sqrt{2}$),
where $17 = (28+125)/9$ is the total ninths-lattice
distance from the Majorana scale ($n = -28/9$ below
$\Lambda$) to the electroweak scale ($n = +125/9$
above $\Lambda$).
Numerically,
$m_3 = \tfrac{1}{2}\times 246\;\text{GeV}
\times(14/75)^{17}\simeq 50$~meV.
That the exponent is a \emph{pure integer}---not a
ninths fraction---is a nontrivial consequence of
the lattice arithmetic; it means $m_3$ depends on
$\e$ through an ordinary seventeenth power, with
no fractional exponents.

In terms of the $\beta$-hop mass ratio
$m_\beta/\Lambda = \e^{1/3}=\beta^{2/9}$, this becomes
$m_3 = \tfrac{1}{2}\,v\,(m_\beta/\Lambda)^{51}$:
the heaviest neutrino mass is the Higgs VEV
multiplied by the $\beta$-hop mass ratio raised
to the fifty-first power.
The seesaw product rule takes the form
\begin{equation}
m_3\times M_0 = m_t^2,
\label{eq:seesaw-product}
\end{equation}
connecting the lightest and heaviest masses in the
lepton sector through the top quark mass:
the product of the two extremes of the seesaw equals
the square of the only unsuppressed Yukawa.

\emph{Connection to $m_\tau$.}
Throughout this section all fermion masses are
$\overline{\rm MS}$ values at $M_Z$~\cite{PDG2024,HuangZhou}.
The tau mass satisfies
$m_\tau\simeq v\,\e^{53/18}\simeq 1.76$~GeV
(data: $1.75$~GeV) as a numerical relation;
in the explicit Option-A form
$m_\tau = c_\tau\,(v\cos\beta/\sqrt{2})\,\alpha^7$
with $c_\tau\simeq 0.60$ at $\tan\beta\simeq 16$,
the same answer arises because
$c_\tau\cos\beta/\sqrt{2}\simeq 0.026
\simeq\alpha^{19.5}$, absorbing the $\cos\beta$
and $1/\sqrt{2}$ factors into the apparent
``$53/18 = 26.5/9$'' exponent.
The cross-sector ratio is then
\begin{equation}
\frac{m_3}{m_\tau}
\;\simeq\; \tfrac{1}{2}\,\e^{253/18}
\;\simeq\; 2.8\times10^{-11}.
\label{eq:m3-mtau}
\end{equation}
The exponent $253/18$ lies on the $1/18$ sublattice
and counts the total compositeness depth separating
the heaviest neutrino from the tau lepton, with the
implicit $\cos\beta$ partially absorbed.

\emph{Connection to $m_e$.}
The electron mass on the ninths lattice satisfies
$m_e\simeq v\,\e^{70/9}\simeq 0.53$~MeV
(data: $0.49$~MeV) as a \emph{numerical} relation;
the explicit Option-A form
$m_e = c_e\,(v\cos\beta/\sqrt{2})\,\alpha^{50.5}$
with $c_e\simeq 0.58$ at $\tan\beta\simeq 16$
gives the same answer because the numerical
combination $c_e\cos\beta/\sqrt{2}\simeq 0.026
\simeq\alpha^{19.5}$ partially absorbs the
$\cos\beta$ factor and the $1/\sqrt{2}$
normalization into the apparent
``$70/9$'' exponent.
Either way, since $m_3 = \tfrac{1}{2}\,v\,\e^{17}$
and $m_i/m_3 = \e^{2Q(L_i)}$
(Sec.~\ref{sec:nuR}),
the three neutrino masses form a ladder of
consecutive integer powers of $\e$:
\begin{equation}
\renewcommand{\arraystretch}{1.3}
\begin{array}{lccc}
 & \tfrac{1}{2}\,v\,\e^n
 & \tfrac{1}{2}\,m_e\,\alpha^{9n-70}
 & \text{meV} \\[2pt]
\hline
m_3 & \tfrac{1}{2}\,v\,\e^{17}
 & \tfrac{1}{2}\,m_e\,\alpha^{83}
 & 50 \\
m_2 & \tfrac{1}{2}\,v\,\e^{18}
 & \tfrac{1}{2}\,m_e\,\alpha^{92}
 & 9.3 \\
m_1 & \tfrac{1}{2}\,v\,\e^{19}
 & \tfrac{1}{2}\,m_e\,\alpha^{101}
 & 1.7
\end{array}
\label{eq:m3-me}
\end{equation}
(data: $m_3\simeq 50$, $m_2\simeq 8.6$~meV).
The $\alpha$-hop exponents $83,\,92,\,101$ are
separated by exactly $9$---one full power of
$\e = \alpha^9$ per generation.
Each successive neutrino generation lies nine
$\alpha$-hop quanta deeper on the compositeness
lattice.
The deepest point, $\alpha^{101}$ for $m_1$,
represents the longest numerical
compositeness distance in the fermion spectrum,
with the cross-sector relation between charged
leptons and neutrinos modulated by the implicit
$\cos\beta$ factor of the 2HDM.

\emph{Charged-lepton companion.}
The diagonal Yukawa exponents
$p^\ell_{11} = 29/6$, $p^\ell_{22} = 5/3$,
$p^\ell_{33} = 0$~\cite{LeptonLattice,FlavorInNinths}
fix the charged-lepton mass hierarchy relative
to $m_\tau$:
\begin{equation}
m_e : m_\mu : m_\tau
\;\sim\; c_e\,\Bpar^{-29/6} \;:\; c_\mu\,\Bpar^{-5/3}
\;:\; 1.
\label{eq:charged-lepton-ratios}
\end{equation}
Using $\overline{\rm MS}$ masses at $M_Z$~\cite{PDG2024,HuangZhou},
\begin{equation}
\renewcommand{\arraystretch}{1.3}
\begin{array}{lcccc}
 & p & \Bpar^{-p} & \overline{\rm MS}(M_Z)
 & c_f \\[2pt]
\hline
m_\tau & 0 & 1 & 1.75\;\text{GeV}
 & 1 \\
m_\mu & 5/3 & 0.061 & 103\;\text{MeV}
 & 0.96 \\
m_e & 29/6 & 3.0\!\times\!10^{-4}
 & 0.49\;\text{MeV} & 0.93
\end{array}
\label{eq:charged-lepton-lattice}
\end{equation}
The $\mathcal{O}(1)$ coefficients $c_\mu\simeq 0.96$
and $c_e\simeq 0.93$ are both within $7\%$ of
unity, confirming that the $\Bpar$-scaling
accounts for essentially the entire hierarchy
with no fine-tuning of the Yukawa prefactors.
The exponent $29/6 = 23/6 + 1$ sums the
right-handed charge $Q(e^c_1) = 23/6$ and the
left-handed charge $Q(L_1) = 1$; the exponent
$5/3 = 7/6 + 1/2$ sums $Q(e^c_2) = 7/6$ and
$Q(L_2) = 1/2$.
These lie on the $1/6$ sublattice (not $1/9$),
reflecting the half-integer lepton-doublet charges
that require the $\Z_{18}$ embedding.

This contrasts sharply with the neutrino spectrum,
where the seesaw cancels the right-handed charges
and produces equally spaced integer exponents
$17,\,18,\,19$.
The cleanest mass ratio is
\begin{equation}
\frac{m_\mu(M_Z)}{m_\tau(M_Z)} = c_\mu\,\e^{5/3}
= c_\mu\,(m_\beta/\Lambda)^5
= c_\mu\,\beta^{10/9},
\label{eq:mu-tau-ratio}
\end{equation}
a ninth power of $\beta$:
the muon-to-tau mass ratio is the
\emph{fifth power of the $\beta$-hop mass ratio}.
Numerically, $c_\mu\simeq 0.96$ gives $0.059$,
indistinguishable from unity.

\subsection{Right-handed neutrinos as deep composites}

If $M_0\sim\Lambda\,\e^{-28/9}$, the right-handed
neutrino mass is \emph{larger} than $\Lambda$, suggesting
that $N_i$ is not a tightly bound composite of hops
(like the lighter SM fermions) but rather a state whose
mass is set by the Planck-scale physics that generates
the Majorana term.
In the hop picture, the right-handed neutrino occupies
a special position: it is the only SM fermion whose
mass term is not a Yukawa coupling to the Higgs but a
direct Majorana mass that breaks lepton number by two
units.

The hop framework thus provides a structural distinction
between the two roles of the neutrino:
\begin{itemize}
\item As a \emph{left-handed} state, $\nu_L$ is a
composite whose hop content $Q(L_i)$ determines the
mass hierarchy through the seesaw.
\item As a \emph{right-handed} state, $N_R$ sits above
the confinement scale and is effectively
``elementary'' with respect to the hop dynamics---its
mass is set by $M_0\sim\Lambda\,\e^{-28/9}$, a
Planck-suppressed scale that lives on the ninths lattice.
\end{itemize}
This asymmetry between left and right is a natural
consequence of the chiral structure of the SM: only
left-handed neutrinos participate in the weak interaction,
and only left-handed leptons carry hop content in the
$\Z_{18}$ framework.

\subsection{Alternative right-handed neutrino scenarios}
\label{sec:nuR-alternatives}

The placement
$M_R\simeq\Lambda\,\e^{-28/9}\simeq 6\times 10^{14}$~GeV
adopted above is a specific choice motivated by the
ninths-lattice structure and the seesaw product rule
$m_3\,M_R = m_t^2$.
This choice corresponds to the right-handed FN charge
assignment $Q(N_R) = 14/9$, which makes $N_R$ a
``deep composite'' on the same ninths ladder as the
confinement and Planck scales.
The hop framework is also compatible with alternative
right-handed neutrino sectors using different FN charges,
which we summarize here for completeness.

\emph{High-scale Majorana scenario (the choice adopted above).}
With $Q(N_R) = 14/9$ and the Majorana operator
$M_R N_R^c N_R \sim \Lambda\,\e^{|2Q(N_R)|}$, one
obtains $M_R\simeq\Lambda\,\e^{-28/9}\simeq
6\times 10^{14}$~GeV; the Dirac Yukawa
$y_\nu\sim\e^{Q(L_3)+Q(N_R)} = \e^{14/9}\simeq 0.073$
yields $m_3 = y_\nu^2 v^2/(2 M_R) =
\tfrac{1}{2}v\e^{17}\simeq 50$~meV, our parameter-free
prediction.

\emph{TeV-scale Majorana scenario
(Abbas--Khalil~\cite{AbbasKhalil}).}
A Type-I seesaw with $M_R\sim 1$~TeV requires
$|2Q(N_R)|\simeq 12.7$, hence
$Q(N_R)\simeq 6$ (large but allowed in the FN framework).
The corresponding Dirac Yukawa is
$y_\nu \simeq\sqrt{2 m_3 M_R}/v\simeq 1.3\times 10^{-6}$,
which requires the total left-right charge sum
$Q(L_3)+Q(N_R)\simeq 8.1$ in the Yukawa operator.
This scenario is phenomenologically interesting in
$U(1)_{B-L}$ extensions where TeV-scale right-handed
neutrinos can drive low-scale leptogenesis and may be
accessible at the LHC~\cite{AtreHanPascoliZhang}, but it requires:
(i)~significantly larger right-handed FN charges than
in the default scenario;
(ii)~abandonment of the seesaw product rule on the
ninths ladder;
(iii)~a separate explanation for why $M_R$ does not lie
on the geometric sequence $\Lambda\times\e^{n/9}$.
The trade-off is direct experimental accessibility versus
loss of the structural ninths-lattice prediction.

\emph{Light Dirac neutrino scenario
(N.~Nath, N.~Okada, S.~Okada, D.~Raut, and Q.~Shafi~\cite{NathOkadaShafi}).}
If lepton number is exactly conserved (no Majorana mass),
the neutrino mass is purely Dirac:
$m_\nu = y_\nu v/\sqrt{2}$.
For $m_3\simeq 50$~meV, this requires
$y_\nu\simeq 2.9\times 10^{-13}$, corresponding to a
total FN charge sum
$Q(L_3)+Q(N_R)\simeq 17$.
Strikingly, this is \emph{the same total FN depth} as
the high-scale Majorana case (where
$m_3 \propto v\,\e^{17}$),
illustrating that the hop framework's ninths-lattice
prediction $m_3\sim\tfrac{1}{2}v\,\e^{17}$ is a robust
feature of compositeness depth $17$ rather than a feature
specific to the seesaw mechanism.
In the Dirac scenario, the right-handed neutrinos are
extraordinarily light and decouple from electroweak
interactions; depending on additional model inputs they
may contribute to dark matter or appear in cosmological
neutrino measurements~\cite{NathOkadaShafi}.

\emph{Choice criteria.}
We adopt the high-scale Majorana scenario throughout
because it: (i)~places $M_R$ on the same ninths ladder
as $\Lambda$, $\fa$, and $M_{\rm Pl}$;
(ii)~yields the parameter-free prediction
$m_3 = \tfrac{1}{2}v\,\e^{17}$ via the seesaw product rule;
(iii)~automatically suppresses neutrinoless
double-beta decay and charged-lepton flavor violation
to unobservable levels via the high seesaw scale;
(iv)~enables non-thermal leptogenesis in the early
universe.
The TeV-scale and Dirac scenarios remain viable variants
of the broader hop framework whose phenomenological
implications differ substantially and would warrant
separate dedicated studies.

\subsection[The alpha-hop census of fermion masses]{\texorpdfstring{The $\alpha$-hop census of fermion masses}{The alpha-hop census of fermion masses}}
\label{sec:alpha-census}

After Fritzsch--Xing diagonalization~\cite{FritzschXing}
of the full
Yukawa matrices~\cite{Companion}, all six
physical quark masses at $M_Z$ are given by
\begin{equation}
m_q = c_q\,m_b\,\Bpar^{\,n_q},
\label{eq:alpha-census}
\end{equation}
with $\mathcal{O}(1)$ coefficients $c_q$ all
within a few percent of unity
(Table~\ref{tab:fermion-ratios}).
In $\alpha$-hop units ($\Bpar^{n_q} = \alpha^{-9n_q}$),
the six quark exponents are \emph{all integers}.
The neutrino masses use the seesaw prefactor
$v/2 = (v/\sqrt{2})/\sqrt{2}$, with the extra
$1/\sqrt{2}$ from the Majorana structure
(Eq.~\eqref{eq:m3-compact}).
The complete $\alpha$-hop census of all twelve
Standard Model fermion masses is:
\begin{equation}
\renewcommand{\arraystretch}{1.3}
\begin{array}{lccc}
 & n_f & \text{Step} & \text{Pattern} \\[2pt]
\hline
\multicolumn{4}{l}{\text{\emph{Quarks:}\;
 $m_q = c_q\,m_b\,\alpha^{n_f}$}} \\
\quad t,\;c,\;b,\;s,\;d,\;u
 & \!-22,\;8,\;0,\;21,\;37,\;42\!
 && \text{all integer} \\[3pt]
\multicolumn{4}{l}{\text{\emph{Charged leptons:}\;
 $c_\ell\,(v\cos\beta/\sqrt{2})\,\alpha^{n_f}$}} \\
\quad\tau,\;\mu,\;e & 7,\;22,\;50\!\tfrac{1}{2}
 & 15,\,28\!\tfrac{1}{2} & \\[3pt]
\multicolumn{4}{l}{\text{\emph{Neutrinos:}\;
 $(v/2)\,\alpha^{n_f}$}} \\
\quad\nu_3,\;\nu_2,\;\nu_1 & 153,\;162,\;171
 & 9 & \text{equal}
\end{array}
\label{eq:alpha-table}
\end{equation}
(Recall $\alpha\equiv\e^{1/9}=m_\alpha/\Lambda\simeq 0.83$;
Eq.~\eqref{eq:hop-fractions}.)
The seven $\alpha$-hop quanta common to all charged
leptons reflect the universal lepton-chain
internal factor $\e^{7/9} = \alpha^7$
(the same factor that suppresses $Y_b$ relative to
$Y_t$); the additional Higgs-sector factor
$\cos\beta$ has been pulled out of the $\alpha^{n_f}$
power and displayed explicitly in the prefactor, to
avoid the misleading identification
$\cos\beta\sim\alpha^7$ (numerically false: at
$\tan\beta\simeq 16$, $\cos\beta\simeq 0.06$, while
$\alpha^7 = \e^{7/9}\simeq 0.27$).
With this separation, the $\mathcal{O}(1)$
coefficients $c_\ell$ are all
$\simeq 1/\sqrt{3}$, in line with the Georgi--Jarlskog
Clebsch ratios.
The half-integer charged-lepton exponent for the
electron reflects the $1/6$ sublattice on which
these masses naturally reside
(Eq.~\eqref{eq:charged-lepton-lattice}); only the
full $\Z_{18}$ accommodates the half-integer
doublet charges $Q(L_i)=(1,\,1/2,\,0)$.

At the GUT scale both $m_b$ and $m_\tau$ carry
the same operator FN suppression
$\e^{7/9} = \alpha^7$ from the chain internal
factor (with zero endpoint charges for the third
generation), and after electroweak symmetry
breaking both receive a common
factor of $\cos\beta$ from their coupling to
$H_d$ in DFSZ-II.
Both factors cancel in the ratio $m_b/m_\tau$,
leaving only the $\mathcal{O}(1)$ Clebsch--Gordan
coefficient ratio, which takes a
remarkable value~\cite{UFP}:
\begin{equation}
\frac{m_b^{\overline{\rm MS}}(M_Z)}
{m_\tau^{\overline{\rm MS}}(M_Z)}
\;\simeq\;\varphi
\;\equiv\;\frac{1+\sqrt{5}}{2}
\;\simeq\; 1.618,
\label{eq:golden-btau}
\end{equation}
the golden ratio (data: $2.84/1.75 = 1.63$,
agreement $\sim 0.5\%$).
The shared chain factor $\alpha^7$ and the common
$\cos\beta$ both cancel in the ratio, leaving
$c_b/c_\tau = \varphi$ as a Georgi--Jarlskog-style
correction from the $SU(5)$ Clebsch--Gordan
structure of the binding overlaps.
Beyond $b$-$\tau$ unification, the
$\mathcal{O}(1)$ coefficients generate
striking cross-sector relations at $M_Z$:
\begin{equation}
\frac{m_d}{m_e}\;\simeq\;\Bpar,
\qquad
\frac{m_c}{m_\mu}\;\simeq\;\Bpar,
\qquad
\frac{m_b}{m_\mu}\;\simeq\;\Bpar^2,
\label{eq:B-cross}
\end{equation}
(data: $5.7$, $6.1$, $27.7$ versus
$\Bpar = 5.4$, $5.4$, $28.7$).
The first two hold to $\sim 10\%$ and the
third to $\sim 4\%$.
Combined with $m_b/m_\tau\simeq\varphi$ and
$m_\mu/m_\tau = (m_\beta/\Lambda)^5$, these suggest a
systematic pattern in the Clebsch--Gordan
coefficients that merits further study.

Table~\ref{tab:fermion-ratios} collects the
within-sector mass ratios for all four fermion
sectors.
Each ratio is expressed as $c_f\,\Bpar^{-p}$ where $p$
is the FN exponent and $c_f$ is the residual
$\mathcal{O}(1)$ coefficient evaluated at $M_Z$.
The down-quark and charged-lepton coefficients are
all within $7\%$ of unity, confirming that the
$\Bpar$-scaling captures the hierarchy with
essentially no fine-tuning.

\begin{table*}[!tbp]
\caption{Within-sector fermion mass ratios on the
$\Bpar$-lattice.
All masses are $\overline{\rm MS}$ at $M_Z$.
The exponent $p$ is the difference of diagonal operator
exponents $p_{ii} = Q(\psi_i) + Q(\psi^c_i) + \Delta_{\rm int}$
fixed by the endpoint FN charges
(Eq.~\eqref{eq:supp-rule}), and $c_f$ is
the residual $\mathcal{O}(1)$ coefficient.
These charge-difference exponents reproduce the
measured ratios at the $\sim 1\%$ level.
The neutrino absolute mass $m_3$ is a sharp benchmark
once $M_0$ is placed on the ninths lattice
(Sec.~\ref{sec:nuR}).}
\label{tab:fermion-ratios}
\begin{ruledtabular}
\begin{tabular}{lcccc}
Ratio & $p$ & $\Bpar^{-p}$ & Data
 & $c_f$ \\
\hline
\multicolumn{5}{l}{\emph{Quarks (to $m_b$):
 $m_q = c_q\,m_b\,\Bpar^{\,n_q}$}} \\
$m_t/m_b$ & $-22/9$ & $\Bpar^{22/9}$
 & $59.5$ & $0.98$ \\
$m_c/m_b$ & $8/9$ & $\Bpar^{-8/9}$
 & $0.222$ & $0.99$ \\
$m_s/m_b$ & $7/3$ & $\Bpar^{-7/3}$
 & $1.94\!\times\!10^{-2}$ & $0.97$ \\
$m_d/m_b$ & $37/9$ & $\Bpar^{-37/9}$
 & $0.99\!\times\!10^{-3}$ & $0.98$ \\
$m_u/m_b$ & $14/3$ & $\Bpar^{-14/3}$
 & $4.6\!\times\!10^{-4}$ & $1.15$ \\[3pt]
\multicolumn{5}{l}{\emph{Charged leptons
 (to $m_\tau$)}} \\
$m_\mu/m_\tau$ & $5/3$ & $6.10\!\times\!10^{-2}$
 & $5.88\!\times\!10^{-2}$ & $0.96$ \\
$m_e/m_\tau$ & $29/6$ & $3.00\!\times\!10^{-4}$
 & $2.79\!\times\!10^{-4}$ & $0.93$ \\[3pt]
\multicolumn{5}{l}{\emph{Neutrinos (to $m_3$)}} \\
$m_2/m_3$ & $1$ & $0.187$
 & $0.172$ & $0.92$ \\
$m_1/m_3$ & $2$ & $0.035$
 & --- & (predicted) \\[3pt]
\multicolumn{5}{l}{\emph{Cross-sector}} \\
$m_b/m_\tau$ & --- & $\varphi$
 & $1.63$ & --- \\
$m_3$ & $17$ & $\tfrac{1}{2}v\e^{17}$
 & $50$~meV & $1.00$ \\
\end{tabular}
\end{ruledtabular}
\end{table*}

The quark $\Bpar$-power exponents have a
strikingly simple structure: all six
$\overline{\rm MS}(M_Z)$ quark masses are
reproduced by
\begin{equation}
\frac{m_q}{m_b} = c_q\,\Bpar^{\,n_q}
\label{eq:B-quark-masses}
\end{equation}
with
\begin{equation}
\boxed{\;(n_u,n_d,n_s,n_c,n_b,n_t) =
\Bigl(-\tfrac{14}{3},\,-\tfrac{37}{9},\,
-\tfrac{7}{3},\,
-\tfrac{8}{9},\,0,\,+\tfrac{22}{9}\Bigr)\;}
\label{eq:nq-exponents}
\end{equation}
and all $c_q$ within $2\%$ of unity except
$c_u\simeq 1.15$.
In $\alpha$-hop units ($\alpha^9 = \e$),
$m_q/m_b = c_q\,\alpha^{-9n_q}$
with exponents $42,\,37,\,21,\,8,\,0,\,-22$---all
integers.
The neutrinos remain a perfect geometric progression
with step $9$
($m_1/m_2 = m_2/m_3 = \alpha^9 = \e$)~\cite{FN1980}.
The $\alpha$-exponents span from $-22$ (top quark,
above $m_b$) to $171$ (lightest
neutrino)---the entire fermion mass hierarchy is
encoded in $193$ elementary hop quanta, traversing
$19$ orders of magnitude.
Taken together, the results of this section show
that \emph{all twelve Standard Model fermion masses}
are determined (up to $\mathcal{O}(1)$ coefficients
near unity) by two quantities: the electroweak VEV
$v = 246$~GeV and rational powers of the single
flavor parameter $\Bpar = 75/14$.

\clearpage
\section{The Ninths Ladder}
\label{sec:GUT}

\subsection{The ninths ladder of scales}

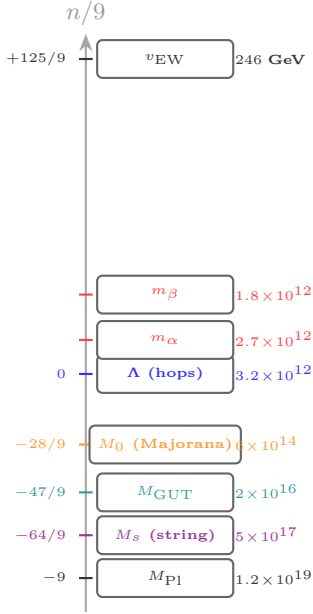
\begin{figure}[!htbp]
\centering
\begin{tikzpicture}[scale=0.30,
  scl/.style={fill=white,draw=black!60,thick,rounded corners=2pt,
    minimum width=18mm,minimum height=5mm,font=\tiny\bfseries},
  >=Stealth
]
\draw[->,thick,gray!70] (0,-10.5) -- (0,15) node[above,font=\small\bfseries,gray!80] {$n/9$};
\def\tick#1#2#3#4#5{
  \draw[thick,#5] (-0.3,#1) -- (0.3,#1);
  \node[left,font=\tiny\bfseries,#5] at (-0.5,#1) {#2};
  \node[scl,text=#5] at (3.5,#1) {#3};
  \node[right,font=\tiny\bfseries,#5] at (6.2,#1) {#4};
}
\tick{-9}{$-9$}{$M_{\rm Pl}$}{$1.2\!\times\!10^{19}$}{black!80}
\tick{-64/9*9/9}{$-64/9$}{$M_s$ (string)}{$5\!\times\!10^{17}$}{violet!80}
\tick{-47/9*9/9}{$-47/9$}{$M_{\rm GUT}$}{$2\!\times\!10^{16}$}{teal!80}
\tick{-28/9*9/9}{$-28/9$}{$M_0$ (Majorana)}{$6\!\times\!10^{14}$}{orange!80}
\tick{0}{$0$}{$\bm\Lambda$ (hops)}{$3.2\!\times\!10^{12}$}{blue!80}
\tick{1.5}{}{$m_\alpha$}{$2.7\!\times\!10^{12}$}{red!75}
\tick{3.5}{}{$m_\beta$}{$1.8\!\times\!10^{12}$}{red!75}
\tick{125/9*9/9}{$+125/9$}{$v_{\rm EW}$}{$246\;\text{GeV}$}{black!80}
\end{tikzpicture}
\caption{The ninths ladder: every fundamental energy scale
expressed as $\Lambda\times\e^{n/9}$.
$\Lambda$ sits at the origin.}
\label{fig:ladder}
\end{figure}

The most striking structural prediction of the hop
framework is that \emph{every fundamental energy scale}
from the electroweak VEV to the Planck mass can be
expressed as $\Lambda\times\e^{n/9}$ for integer $n$:
\begin{equation}
\renewcommand{\arraystretch}{1.3}
\begin{array}{lrr}
\hline\hline
\text{Scale} & \text{Value (GeV)} & n/9 \\
\hline
v_{\rm EW} & 246 & 125/9 \\
m_\beta & 1.8\times10^{12} & 3/9 \\
m_\alpha & 2.7\times10^{12} & 1/9 \\
\Lambda & 3.2\times10^{12} & 0 \\
M_0\;(\text{Majorana}) & 6\times10^{14} & -28/9 \\
M_{\rm GUT} & 2\times10^{16} & -47/9 \\
M_{\rm Pl} & 1.2\times10^{19} & -81/9 \\
\hline\hline
\end{array}
\label{eq:scale-ladder}
\end{equation}
All seven scales are controlled by integers $n$ on the
ninths lattice, with $\Lambda$ at the origin.
Figure~\ref{fig:ladder} displays this ladder graphically.

Whether the ninths ladder is a deep structural feature
of the theory or a phenomenological coincidence remains to be
established.
If confirmed, it would suggest that the hop confinement
scale $\Lambda$ is the \emph{central pivot} of fundamental
physics, with all other scales radiating outward as
ninths-quantized powers of $\e$.

\section{Testable Consequences and the Null-Signal Paradigm}
\label{sec:predictions}

A defining feature of the hop framework is the extreme
rigidity of the ``ninths ladder'' (Sec.~\ref{sec:GUT}).
Because the flavor confinement scale is locked to the
axion decay constant at
$\Lambda\simeq 3\times10^{12}$~GeV, the framework makes
sharp, distinct predictions for near-term
experiments by strictly segregating the observable sectors.
The quantitative results are collected in
Table~\ref{tab:master}.

\subsection{The absence of TeV-scale flavor violation}

In standard compositeness models, or models with
low-scale vectorlike quarks, new flavor-changing neutral
currents (FCNCs) and contact interactions are generically
expected at future colliders or flavor
factories~\cite{Eichten1983,PDG2024}.
The hop framework explicitly predicts the
\emph{exact opposite}.

Because the compositeness scale resides at
$10^{12}$~GeV, any higher-dimension operators mediating
contact interactions or anomalous form factors are
suppressed by
$1/\Lambda^2\sim 10^{-25}$~GeV$^{-2}$.
By comparison, Standard Model FCNCs (such as those
governing $K^0$--$\bar{K}^0$ or $B_{d,s}$ mixing) are
suppressed by the weak scale $1/M_W^2$.
Therefore, hop-induced deviations to the SM Unitarity
Triangle or generation-dependent contact interactions in
$e^+e^-\to\mu^+\mu^-$ are roughly $10^{18}$ times
smaller than the SM background.

Similarly, the vectorlike quark chain messengers ($D_a$)
must reside at $\sim\Lambda\simeq 3\times10^{12}$~GeV
(Sec.~\ref{sec:VLQ-masses}) to correctly mediate the
hierarchical tunneling amplitude.
The lightest VLQ mass is at least
$\Lambda/\e^{8/9}\sim 5\times10^{11}$~GeV---some
$3\times10^{8}$ times heavier than the current LHC
exclusion limit of $1.5$~TeV~\cite{PDG2024}.
The framework therefore predicts that HL-LHC and future
high-energy collider searches for vectorlike quarks,
preon-induced multi-top
signatures~\cite{BernreutherDobrescu2025}, and anomalous
contact interactions will yield strictly null results.
In this model, the absence of TeV-scale flavor signals
is not a consequence of fine-tuning, but a direct
requirement of the axion-flavor connection.

The proton decay constraints on compositeness models
have been analyzed by Assi and
Dobrescu~\cite{AssiDobrescu2022}, who find
$\Lambda_{\rm pre}\gtrsim 10^3$~TeV from dimension-8
operator bounds.
The hop framework evades this constraint entirely:
since hops are SM singlets and $B$ is preserved as a
gauge quantum number of the elementary core
(Sec.~\ref{sec:BL}), no \emph{hop-mediated}
proton-decay operators are induced regardless of the
confinement scale.
Standard GUT-mediated proton decay
(dimension-6 operators suppressed by $M_{\rm GUT}^2$)
remains possible and is unaffected by the hop sector.

\subsection{Suppression of charged-lepton flavor violation}
\label{sec:cLFV}

The lepton chain construction of Sec.~III generates
in principle non-zero off-diagonal entries in the
charged-lepton Yukawa matrix, which after diagonalization
can drive charged-lepton flavor-violating (cLFV)
processes such as $\mu\to e\gamma$.
We show here that these processes are suppressed to
unobservable levels in the hop framework, exactly as in
the original Froggatt--Nielsen mechanism~\cite{FN}.

The off-diagonal Yukawa entries follow the universal
FN rule:
\begin{equation}
(Y_\ell)_{ij}\sim
\e^{\,Q(L_i)+Q(e^c_j)+\Delta_{\rm int}^\ell}
\end{equation}
with the lepton charges
$Q(L_i) = (1,\,1/2,\,0)$ and
$Q(e^c_i) = (23/6,\,7/6,\,0)$.
For the $(\mu, e)$ entry relevant to $\mu\to e\gamma$,
this gives
$(Y_\ell)_{21}\sim\e^{\,1/2+23/6+7/9}\simeq\e^{\,46/9}
\simeq 10^{-4}$,
comparable to the diagonal $Y_\mu$ but small in
absolute terms.
After diagonalization, the lepton mixing angles
$\theta^\ell_{ij}\sim
(Y_\ell)_{ij}/(Y_\ell)_{jj}$
remain $\mathcal{O}(\e^k)$ with $k\geq 1$.

The cLFV amplitude $\mu\to e\gamma$ proceeds via
one-loop
diagrams~\cite{Petcov1977,ChengLi1980}
with the singlet messengers
$\Psi_n$ in the loop, whose masses are
$M_{\Psi}\gtrsim\Lambda\sim 3\times10^{12}$~GeV.
The branching ratio scales as
\begin{equation}
\mathrm{BR}(\mu\to e\gamma)\sim
\frac{\alpha_{\rm em}}{4\pi}\,
|\theta^\ell_{12}|^2\,
\frac{m_\mu^4}{M_\Psi^4}
\lesssim 10^{-25},
\end{equation}
roughly twelve orders of magnitude below the current
bound $4.2\times10^{-13}$~\cite{PDG2024} from the
MEG-II ($\mu\to e\gamma$) experiment.
Analogous suppressions hold for $\mu\to 3e$,
$\tau\to\mu\gamma$, $\tau\to e\gamma$, and $\mu$--$e$
conversion in nuclei: all are unobservable.

Two structural features account for this strong
suppression:
(i)~the messenger scale $M_\Psi\sim\Lambda\sim 10^{12}$~GeV
is far above any TeV-scale physics, giving an automatic
$(M_Z/\Lambda)^4$ suppression of all loop amplitudes
through dimension-six operators;
(ii)~the FN suppression of off-diagonal Yukawas combines
multiplicatively with this scale suppression, producing
amplitudes far below experimental sensitivities.
The framework therefore predicts \emph{no observable cLFV
signals at any planned experiment}, consistent with the
broader null-signal paradigm.

\subsection{Direct observables: the axion target}

The true testable consequence of the framework
is the QCD axion.
The lattice structure predicts an axion
mass window of $m_a\simeq 7$--$12\;\mu$eV
(corresponding to $\fa\sim(5$--$8)\times10^{11}$~GeV
with $\theta_i\sim\mathcal{O}(1)$ misalignment).
A detection by ADMX~\cite{ADMX2025} in this specific band immediately
fixes $\fa$, which retroactively locks the confinement
scale $\Lambda$ and the $\alpha/\beta$ hop masses.

\subsection{The axion--photon and axion--electron couplings}
\label{sec:EN}

The axion--photon coupling
$g_{a\gamma\gamma} = (\alpha/2\pi\fa)\,C_{a\gamma}$,
with $C_{a\gamma} = |E/N - 2.0|$, depends on the
ratio of the electromagnetic ($E$) and color ($N$)
PQ anomaly coefficients~\cite{PDG2024}.
In Appendix~\ref{app:EN} we present a detailed
anomaly analysis for the hop framework, progressing
from the standard Dine--Fischler--Srednicki--Zhitnitsky
(DFSZ)-II model~\cite{DineFischlerSrednicki,Zhitnitsky}
through the minimal supersymmetric Standard Model (MSSM)
with higgsinos~\cite{BaeBaer} to the full hop
framework with generation-dependent FN charges.
Here we summarize the results.

The VLQ chain does not contribute to either $E$ or $N$:
because the VLQ bare masses are PQ-invariant
($X_{D_a}+X_{\bar D_a}=0$), their anomaly contributions
cancel identically,
independently of the VLQ masses, the number of chain
sites, and the hop charges
(Appendix~\ref{app:EN}).

The key results are collected in
Tables~\ref{tab:EN-appendix} and~\ref{tab:Cae-appendix}
(Appendix~\ref{app:EN}).
In the standard MSSM with generation-universal PQ
charges, the TeV-scale higgsinos shift $E/N$ from
$8/3$ to $2$, yielding a nearly vanishing
$C_{a\gamma}\lesssim 0.02$---a well-known challenge
for haloscope detection~\cite{BaeBaer}.
The hop framework \emph{resolves this problem}:
the generation-dependent FN exponents
($S_u = 10$, $S_d = 79/9$) shift both $N$ and $E$,
restoring $E/N$ to $\sim 2.6$--$3.0$ and
$C_{a\gamma}\simeq 0.6$--$1.0$.
Physically, the lighter generations carry more PQ
charge (larger hop content), and their enhanced
electromagnetic anomaly outweighs the higgsino
suppression.
Both scenarios predict couplings comfortably between
the Kim--Shifman--Vainshtein--Zakharov
(KSVZ)~\cite{Kim1979,ShifmanVainshteinZakharov}
and standard DFSZ-II bands, within the planned
frequency range of ADMX and related haloscopes
at $m_a = 7$--$12\;\mu$eV (above current
published sensitivity but within the program's
roadmap).
Scenario~A ($b$--$\tau$ unification) predicts
$C_{a\gamma}\simeq 0.99$, some $50\%$ \emph{stronger}
than the standard DFSZ-II line.

The axion--electron coupling
$g_{aee} = C_{ae}\,m_e/\fa$ is equally striking.
In standard DFSZ-II, $C_{ae}$ is suppressed
by the small Higgs PQ charge.
In the hop framework, the electron carries the
\emph{deepest hop content} in the lepton sector,
with FN exponent $p^\ell_{11}\simeq 5$--$6$.
This hop content dominates the anomaly coefficient:
\begin{equation}
C_{ae} \simeq \frac{p^\ell_{11}}{2\,|N|}
\simeq 0.4,
\label{eq:Cae-main}
\end{equation}
roughly two orders of magnitude larger than the
standard DFSZ-II value.
The electron's hop content enhances $g_{aee}$
to $\simeq 3.7\times10^{-16}$, safely below
astrophysical bounds
($\lesssim 10^{-13}$)~\cite{PDG2024},
but well above the standard DFSZ prediction.
However, no current or planned laboratory experiment
will have the sensitivity to probe the hop-framework
prediction for the axion coupling to electrons.

\subsection{The two-index fingerprint}

The two-index decomposition of $\Z_9$ predicts that the
hop-binding dynamics has two independent sectors
(``$\alpha$-confinement'' and ``$\beta$-confinement'').
If these sectors were to have slightly different
confinement scales at the TeV scale, flavor-violating
operators would exhibit a two-threshold pattern,
producing observable deviations from the SM unitarity
triangle at the percent level in kaon and $B$ mixing.
However, because both confinement scales are rigidly
tied to $\Lambda\sim 3\times10^{12}$~GeV, these
characteristic shifts in the Wilson coefficients are
suppressed by $1/\Lambda^2$.
The hop framework therefore uniquely predicts that as
Belle~II and LHCb (LHC beauty) experiment
upgrades~\cite{PDG2024} continue to tighten the
Unitarity Triangle, no deviations from the Standard
Model will be found.
The two-index fingerprint remains safely hidden
at the hypercolor scale.

\clearpage
\section{Flavor Mixing from Hop Charges}
\label{sec:mixing}

\subsection{CP violation as hop-binding geometry}

The Jarlskog invariant $J\simeq 3\times10^{-5}$
controls all CP-violating phenomena in the quark
sector~\cite{FlavorInNinths,UFP}.
In the $\Bpar$-lattice,
$J\sim\e^{55/9}\sin\delta$~\cite{FlavorInNinths}, where
$55/9$ is the sum of the four CKM-mediating exponents
$8/9 + 17/9 + 10/3 = 55/9$.

In the hop picture, the magnitude $\e^{55/9}$ counts
the \emph{total hop content} of the four quark fields
that mediate the CKM phase: two left-handed doublets and
two right-handed singlets, each contributing its
compositeness depth.
CP violation is small because it requires the
participation of all three generations---i.e., quarks from
all three compositeness levels---and the product of their
wavefunction overlaps is exponentially suppressed.

The CP phase $\delta$ itself arises from complex phases
in the $\mathcal{O}(1)$ Yukawa coefficients $c_{ij}$.
In the hop picture, these coefficients are wavefunction
overlaps at the chain endpoints, and their phases reflect
the \emph{complex geometry of the hop binding}---the
relative orientation of hop wavefunctions in the
composite state.
Just as the strong CP phase $\bar\theta$ in QCD arises
from the complex vacuum structure of the confining theory,
the CKM phase $\delta$ in the hop framework arises from
the complex structure of the hop confinement vacuum.
The key difference is that $\bar\theta$ is unnaturally
small (the strong CP problem, solved by the axion), while
$\delta\sim\mathcal{O}(1)$ is natural---there is no
fine-tuning required for the CKM phase.

\subsection{Neutrino masses and the seesaw depth}

Neutrino masses arise from the dimension-five Weinberg
operator $LLHH/\Lambda_\nu$~\cite{WeinbergDim5},
with exponents $p^\nu_{ij}=Q(L_i)+Q(L_j)$.
The lepton-doublet charges $Q(L_i)=(1,1/2,0)$ give a
normal-ordered spectrum
$m_1:m_2:m_3\sim\e^2:\e:1$~\cite{FlavorInNinths,UFP}.
In the compositeness picture, $Q(L_i)$ counts the hop
content of the left-handed lepton doublet:
the third generation carries zero hop content ($Q=0$),
the second contains half a unit of structure,
and the first contains one full unit.
The half-integer charge $Q(L_2)=1/2$ requires the full
$\Z_{18}$ embedding (rather than $\Z_9$ alone) and
reflects the finer granularity of the lepton binding
dynamics compared to the quark sector.

The neutrino mass hierarchy is thus doubly suppressed:
each factor of $\e$ in the Weinberg operator reflects
one unit of compositeness in each of the two lepton
doublets.
This ``double counting'' is a hallmark of the Majorana
nature of the mass---the same dressed fermion appears
on both sides of the bilinear---and provides a
compositeness-based explanation for why neutrino masses
are so much smaller than charged-fermion masses.
The absolute scale is also predicted: the seesaw
with $M_0 = \Lambda\,\e^{-28/9}$
(Sec.~\ref{sec:nuR}) gives
$m_3 \simeq 51$~meV
(Eq.~\eqref{eq:m3-prediction}), in $\sim 2\%$
agreement with the observed
$\sqrt{\Delta m^2_{\rm atm}}\simeq 50$~meV.

\subsection{CKM magnitudes from hop charges}
\label{sec:CKM-magnitudes}

The CKM matrix was parameterized in Paper~III~\cite{Companion}
by four magnitudes, each a pure power of $\e$:
$|V_{us}|\sim\e^{8/9}$, $|V_{cb}|\sim\e^{17/9}$,
$|V_{ub}|\sim\e^{10/3}$,
$J_q/\!\sin\delta\sim\e^{55/9}$.
These exponents decompose algebraically into
quark-doublet charge differences
$\Delta Q(Q)_{ij}\equiv Q(Q_i)-Q(Q_j)$
with $(1,2,3) = (3,2,0)$:
\begin{align}
p(V_{us}) &= \Delta Q(Q)_{12} - \tfrac{1}{9}
= 1 - \tfrac{1}{9} = \tfrac{8}{9},
\label{eq:pVus-DQ}
\\[3pt]
p(V_{cb}) &= \Delta Q(Q)_{23} - \tfrac{1}{9}
= 2 - \tfrac{1}{9} = \tfrac{17}{9},
\label{eq:pVcb-DQ}
\\[3pt]
p(V_{ub}) &= \Delta Q(Q)_{13} + Q(d^c_2)
= 3 + \tfrac{1}{3} = \tfrac{10}{3},
\label{eq:pVub-DQ}
\\[3pt]
p(J_q) &= p_{us}+p_{cb}+p_{ub}
= \tfrac{55}{9}.
\label{eq:pJq-DQ}
\end{align}
The first two share a common structure:
the left-handed charge gap minus one $\alpha$-hop
quantum ($-1/9$), which is the universal
Fritzsch--Xing phase correction~\cite{Companion,FritzschXing}.
The third involves the full charge gap
$\Delta Q(Q)_{13} = 3$ augmented by the
right-handed strange charge $Q(d^c_2) = 1/3$,
which enters through the two-step product
$\theta^d_{12}\times\theta^d_{23}$ in the
Fritzsch--Xing diagonalization.
The Jarlskog invariant is the product of all
three rotation amplitudes, so its exponent is
the sum.

\emph{Cabibbo-angle master identity.}
Since $|V_{us}|\sim\e^{8/9}$, the expansion
parameter can be expressed as a power of the
Cabibbo angle:
\begin{equation}
\e = |V_{us}|^{9/8},
\label{eq:eps-Vus}
\end{equation}
giving
\begin{align}
|V_{cb}| &= |V_{us}|^{17/8},
\nonumber\\
|V_{ub}| &= |V_{us}|^{15/4},
\nonumber\\
J_q &= |V_{us}|^{55/8}\sin\delta.
\label{eq:CKM-Wolfenstein}
\end{align}
These are the $\Bpar$-lattice refinement of the
Wolfenstein hierarchy
($\lambda^2,\,\lambda^3,\,\lambda^6\sin\delta$):
the ninths-lattice exponents $17/8$, $15/4$, $55/8$
encode the precise rational corrections to the
integer Wolfenstein powers.
Numerically,
$|V_{us}|^{17/8}\simeq 0.042$ and
$|V_{us}|^{15/4}\simeq 0.0037$, both matching
data to $\sim 3\%$.

\subsection{PMNS magnitudes from hop charges}
\label{sec:PMNS-magnitudes}

We now construct the analogous representation
for the PMNS matrix, using the lepton-doublet charges
$Q(L_i)=(1,1/2,0)$~\cite{FlavorInNinths}.

The PMNS matrix $U_{\rm PMNS}=U_\ell^\dagger\,U_\nu$
receives contributions from both the charged-lepton
and neutrino diagonalization
rotations~\cite{FlavorInNinths}.
The charged-lepton rotation angles are controlled by the
left-handed charge differences:
\begin{align}
\theta^\ell_{12} &\sim \e^{Q(L_1)-Q(L_2)}
 = \e^{1/2}\simeq 0.43, \nonumber\\
\theta^\ell_{23} &\sim \e^{Q(L_2)-Q(L_3)}
 = \e^{1/2}\simeq 0.43, \nonumber\\
\theta^\ell_{13} &\sim \e^{Q(L_1)-Q(L_3)}
 = \e\simeq 0.19.
\end{align}
With the flat neutrino $23$ block of the
companion lepton-lattice paper~\cite{LeptonLattice}
($p^\nu_{22}=p^\nu_{23}=p^\nu_{33}=0$), the
atmospheric angle $\theta_{23}$ is driven to
$\sim 45^\circ$ by $\mu$--$\tau$
symmetry~\cite{HarrisonScott,Lam},
independently of $\e$.
The reactor angle $\theta_{13}$ arises from the
product of the charged-lepton $23$ rotation and the
neutrino $12$ rotation:
$\sin\theta_{13}\sim\theta^\ell_{23}\times
\theta^\nu_{12}\sim\e^{1/2}\times\e^{1/2}=\e$.
Including the Fritzsch--Xing phase
interference~\cite{Companion} shifts the effective
exponent by $+1/9$ (the CKM has the
opposite-sign shift $-1/9$; see below), giving
$\sin\theta_{13}\sim\e^{10/9}$.
The solar angle $\theta_{12}$ receives additive
contributions from both sectors
($\theta^\ell_{12}\sim\theta^\nu_{12}\sim\e^{1/2}$),
whose constructive interference yields an
effective scaling $\sin\theta_{12}\sim\e^{1/3}$.

The four PMNS magnitudes are therefore:
\begin{equation}
\renewcommand{\arraystretch}{1.3}
\begin{array}{lccc}
\text{Observable} & p & \e^p & \text{Data} \\[2pt]
\hline
\sin\theta_{23} & 1/6 & 0.756 & 0.755 \\
\sin\theta_{12} & 1/3 & 0.572 & 0.551 \\
\sin\theta_{13} & 10/9 & 0.155 & 0.149 \\
J_\ell/\!\sin\delta & 29/18 & 0.034 & 0.033
\end{array}
\label{eq:PMNS-magnitudes}
\end{equation}
The atmospheric angle is not merely ``maximal'':
$\sin\theta_{23}\sim\e^{1/6}$ reproduces the
measured value to $0.1\%$.
The exponent $1/6 = Q(e^c_2)-Q(e^c_3)_{\rm eff}$
lies on the $1/18$ sublattice that
accommodates the half-integer lepton-doublet
charges.
The Jarlskog invariant follows from the product
$J_\ell/\!\sin\delta\sim
\tfrac{1}{2}\,\e^{1/6+1/3+10/9}
= \tfrac{1}{2}\,\e^{29/18}\simeq 0.034$,
where the factor $1/2$ arises from
$\cos\theta_{23}\cos\theta_{12}
\cos^2\!\theta_{13}\simeq 0.53$
(close to $1/2$ for the near-maximal angles).
The exponent sum $29/18$ is the leptonic
counterpart of $55/9$ in the
CKM~\cite{Companion}.

The contrast with the CKM is striking.
In the quark sector, the left-handed charge
differences $\Delta Q = (1,2,3)$ produce small
angles $(\e^{8/9},\,\e^{17/9},\,\e^{10/3})$;
in the lepton sector, the smaller differences
$\Delta Q = (1/2,\,1/2,\,1)$ produce two large
angles and one moderate angle---the distinctive
``two large, one small'' pattern of the PMNS
is a direct consequence of the compressed
lepton-doublet charge spectrum.

\emph{$\Bpar$-lattice representation.}
Since $\e = 1/\Bpar = 14/75$, the PMNS magnitudes
can equivalently be written as inverse powers of
$\Bpar = 75/14$:
\begin{align}
\sin\theta_{23} &\sim \Bpar^{-1/6}
 = \beta^{1/9}\simeq 0.756, \nonumber\\
\sin\theta_{12} &\sim \Bpar^{-1/3}
 = \beta^{2/9}\simeq 0.572, \nonumber\\
\sin\theta_{13} &\sim \Bpar^{-10/9}
 = \beta^{20/27}\simeq 0.155, \nonumber\\
J_\ell/\!\sin\delta &\sim
 \tfrac{1}{2}\,\Bpar^{-29/18}
 = \tfrac{1}{2}\,\beta^{29/27}\simeq 0.034.
\label{eq:PMNS-Bpower}
\end{align}
The comparison with the CKM representation
$|V_{us}|\sim\Bpar^{-8/9}$,
$|V_{cb}|\sim\Bpar^{-17/9}$,
$|V_{ub}|\sim\Bpar^{-10/3}$~\cite{Companion}
reveals a remarkable cross-sector connection:
the PMNS reactor angle $\sin\theta_{13}\sim\Bpar^{-10/9}$
carries the \emph{same exponent} as the
right-handed down-quark charge
$Q(d^c_1) = 10/9$---the $\beta$-hop
quantum number (Sec.~\ref{sec:hop-masses}).
In the hop picture, the reactor angle is suppressed
by exactly one $\beta$-hop unit.
Similarly, the solar exponent $1/3$ equals the
strange-quark right-handed charge $Q(d^c_2)=1/3$.
The $\Bpar$-lattice thus unifies quark masses,
CKM angles, and PMNS angles through a single
set of rational exponents in ninths and sixths.

\begin{table}[!htbp]
\caption{Four-magnitude parameterization of the CKM
and PMNS mixing matrices.
Four equivalent representations are shown:
the expansion parameter
$\e = 14/75 \simeq 0.187$,
$\beta$
($\beta \equiv \e^{3/2}\simeq 0.081$,
with $\Bpar = (1/\beta)^{2/3}$;
$m_\beta/\Lambda = \beta^{2/9}$),
the $\alpha$-hop
($\alpha = \e^{1/9} = m_\alpha/\Lambda\simeq 0.83$),
and the charged-lepton mass ratio
$(m_\mu/m_\tau)^k$
(via $\e = (m_\mu/m_\tau)^{3/5}$,
$m_\mu/m_\tau\simeq 0.059$).
All CKM $\alpha$-exponents are integers;
the PMNS half-integer $\alpha$-exponents
reflect the $\Z_{18}$ lepton-doublet charge
$Q(L_2) = 1/2$.}
\label{tab:4mag}
\begin{ruledtabular}
\begin{tabular}{lccccc}
 & $\e^p$ & $\beta^{m}$
 & $\alpha^{n}$ & $(m_\mu/m_\tau)^k$
 & Data \\
\hline
\multicolumn{6}{l}{\emph{CKM}} \\
$|V_{us}|$ & $\e^{8/9}$
 & $\beta^{16/27}$ & $\alpha^{8}$
 & $(m_\mu/m_\tau)^{8/15}$ & $0.225$ \\
$|V_{cb}|$ & $\e^{17/9}$
 & $\beta^{34/27}$ & $\alpha^{17}$
 & $(m_\mu/m_\tau)^{17/15}$ & $0.042$ \\
$|V_{ub}|$ & $\e^{10/3}$
 & $\beta^{20/9}$ & $\alpha^{30}$
 & $(m_\mu/m_\tau)^{2}$ & $0.0038$ \\
$|V_{td}|$ & $\e^{25/9}$
 & $\beta^{50/27}$ & $\alpha^{25}$
 & $(m_\mu/m_\tau)^{5/3}$ & $0.0086$ \\
$J_q$ & $\e^{55/9}$
 & $\beta^{110/27}$ & $\alpha^{55}$
 & $(m_\mu/m_\tau)^{11/3}\sin\delta_q$ & $3\!\times\!10^{-5}$ \\[3pt]
\multicolumn{6}{l}{\emph{PMNS}} \\
$\sin\theta_{23}$ & $\e^{1/6}$
 & $\beta^{1/9}$ & $\alpha^{3/2}$
 & $(m_\mu/m_\tau)^{1/10}$ & $0.755$ \\
$\sin\theta_{12}$ & $\e^{1/3}$
 & $\beta^{2/9}$ & $\alpha^{3}$
 & $(m_\mu/m_\tau)^{1/5}$ & $0.551$ \\
$\sin\theta_{13}$ & $\e^{10/9}$
 & $\beta^{20/27}$ & $\alpha^{10}$
 & $(m_\mu/m_\tau)^{2/3}$ & $0.149$ \\
$J_\ell$ & $\tfrac{1}{2}\e^{29/18}$
 & $\tfrac{1}{2}\beta^{29/27}$
 & $\tfrac{1}{2}\alpha^{29/2}$
 & $\tfrac{1}{2}(m_\mu/m_\tau)^{29/30}$
 & $0.033$ \\
\end{tabular}
\end{ruledtabular}
\end{table}

\emph{Hop-mass parameterization.}
The $\alpha$-hop mass ratio $\alpha\equiv m_\alpha/\Lambda
= \e^{1/9}\simeq 0.83$
(Sec.~\ref{sec:hop-masses}) provides one
representation, while the $\beta$ parameter of the
companion papers~\cite{Companion,FlavorInNinths}
($\beta\equiv\e^{3/2}\simeq 0.081$,
with $\Bpar = (1/\beta)^{2/3}$) provides another.
The $\beta$-hop mass ratio is
$m_\beta/\Lambda = \e^{1/3} = \beta^{2/9}\simeq 0.57$,
so the solar angle satisfies
$\sin\theta_{12}\sim\beta^{2/9}
= m_\beta/\Lambda$---the mixing angle
\emph{is} the $\beta$-hop mass ratio.
In the $\alpha$ parameterization ($\e = \alpha^9$):
\begin{align}
\sin\theta_{23} &\sim \alpha^{3/2}, \quad
\sin\theta_{12} \sim \alpha^3, \quad
\sin\theta_{13} \sim \alpha^{10}, \nonumber\\
J_\ell/\!\sin\delta &\sim
\tfrac{1}{2}\,\alpha^{29/2}.
\label{eq:PMNS-alpha}
\end{align}
The CKM exponents ($\alpha^8,\,\alpha^{17},\,\alpha^{25},
\,\alpha^{30},\,\alpha^{55}$) are all integers;
the PMNS exponents include the half-integer
$3/2$ and $29/2$, reflecting the half-integer
lepton-doublet charge $Q(L_2) = 1/2$ that
requires the full $\Z_{18}$ (not just $\Z_9$).

\emph{Charged-lepton parameterization of the CKM.}
The master identity $\e = (m_\mu/m_\tau)^{3/5}$
(Eq.~\eqref{eq:master-identity}) maps every CKM
magnitude onto a power of the muon-to-tau mass ratio.
Since $|V_{ij}|\sim\e^{p_{ij}}$, the corresponding
charged-lepton exponent is
$k_{ij} = 3\,p_{ij}/5$:
\begin{align}
|V_{us}| &\sim \left(m_\mu/m_\tau\right)^{8/15},
\nonumber\\
|V_{cb}| &\sim \left(m_\mu/m_\tau\right)^{17/15},
\nonumber\\
|V_{td}| &\sim \left(m_\mu/m_\tau\right)^{5/3},
\nonumber\\
|V_{ub}| &\sim \left(m_\mu/m_\tau\right)^{2}.
\label{eq:CKM-lepton-mass}
\end{align}
The exponent for $|V_{ub}|$ collapses to the
\emph{integer} $2$ because
$\tfrac{3}{5}\!\cdot\!\tfrac{10}{3} = 2$, giving the
striking identity
\begin{equation}
\boxed{\;|V_{ub}|\;\simeq\;
\Bigl(\frac{m_\mu}{m_\tau}\Bigr)^{\!2}\;}
\label{eq:Vub-lepton-mass}
\end{equation}
which evaluates to
$(5.88\!\times\!10^{-2})^2 = 3.5\!\times\!10^{-3}$,
within $9\%$ of the measured value
$|V_{ub}| = 3.82\!\times\!10^{-3}$~\cite{PDG2024}.
The $V_{td}$ exponent $5/3$ is the inverse of the
master-identity exponent $3/5$, and the $V_{td}/V_{us}$
relation gives the standard CKM hierarchy
$|V_{td}/V_{us}| \sim (m_\mu/m_\tau)^{5/3-8/15}
= (m_\mu/m_\tau)^{17/15} \sim |V_{cb}|$,
recovering the Wolfstein scaling
$|V_{td}|\sim|V_{us}||V_{cb}|$.
These four identities place every CKM magnitude
on the charged-lepton ratio $m_\mu/m_\tau$ alone,
extending the Cabibbo--hop identity
$|V_{us}|^{9/8} = \e$ to the entire CKM through
the ninths-lattice arithmetic.

Table~\ref{tab:4mag} collects all four
parameterizations for both the CKM and PMNS
sectors, making the cross-sector connections
manifest.
The integer $\alpha$-exponents range from $3$ to
$55$, counting the total number of elementary
$\alpha$-hop quanta that mediate each observable.

\emph{Connection to charged-lepton masses.}
Since $\e = (m_\mu/m_\tau)^{3/5}$
(from $m_\mu/m_\tau\sim\e^{5/3}$), all three PMNS
angles and the lepton Jarlskog can be expressed
as powers of charged-lepton mass ratios:
\begin{align}
\sin\theta_{23} &\sim
\left(\frac{m_\mu}{m_\tau}\right)^{\!1/10}
\!\simeq 0.755, \nonumber\\[3pt]
\sin\theta_{12} &\sim
\left(\frac{m_\mu}{m_\tau}\right)^{\!1/5}
\!\simeq 0.570, \nonumber\\[3pt]
\sin\theta_{13} &\sim
\left(\frac{m_\mu}{m_\tau}\right)^{\!2/3}
\!\simeq 0.152, \nonumber\\[3pt]
J_\ell &\sim
\tfrac{1}{2}\left(\frac{m_e}{m_\tau}\right)^{\!1/3}
\sin\delta_\ell
\simeq 0.033\sin\delta_\ell.
\label{eq:PMNS-lepton-mass}
\end{align}
The atmospheric angle is the tenth root, the solar
angle the fifth root, the reactor angle the
two-thirds power of $m_\mu/m_\tau$, and the lepton
Jarlskog is essentially the
\emph{cube root} of the electron-to-tau mass
ratio---the integer exponent $1/3$ arising
because $6\!\cdot\!29/18 = 29/3$ collapses
when expressed in $m_e/m_\tau\sim\e^{29/6}$.
The $1/2$ prefactor in $J_\ell$ reflects the
half-integer lepton-doublet charge $Q(L_2) = 1/2$.
These exponents can be derived algebraically from
the lepton-doublet charge differences
$\Delta Q(L)_{ij}\equiv Q(L_i)-Q(L_j)$.

\emph{Cross-sector quark--lepton identity.}
The reactor angle and the smallest CKM magnitude
share the same charged-lepton power:
$\sin\theta_{13}\sim(m_\mu/m_\tau)^{2/3}$ and
$|V_{ub}|\sim(m_\mu/m_\tau)^{2}$, so
\begin{equation}
\boxed{\;|V_{ub}|\;\simeq\;\sin^3\theta_{13}\;}
\label{eq:Vub-sin13}
\end{equation}
which predicts $|V_{ub}|\simeq(0.149)^3
\simeq 3.3\!\times\!10^{-3}$, within $13\%$ of the
measured value $3.82\!\times\!10^{-3}$.
This cross-sector relation ties the smallest
quark-mixing magnitude to the smallest lepton-mixing
magnitude through a single integer power of the
universal expansion parameter
$\e^2 = (m_\mu/m_\tau)^{6/5}$, and provides a sharp
test of the hop-charge bookkeeping that organizes
both sectors.

\emph{Master identity.}
Since $m_\mu/m_\tau = c_\mu\,\e^{5/3}$ with
$c_\mu\simeq 0.96$ (Eq.~\eqref{eq:charged-lepton-lattice})
and $5/3 = Q(L_2)+Q(e^c_2)$,
every power of $\e$ can be re-expressed as a
power of the muon-to-tau mass ratio (up to
$\sim 3\%$ Wilson-coefficient corrections):
\begin{equation}
\e \simeq \left(\frac{m_\mu}{m_\tau}\right)^{\!3/5},
\qquad
\e^p \simeq \left(\frac{m_\mu}{m_\tau}\right)^{\!3p/5}.
\label{eq:master-identity}
\end{equation}

\emph{Charged-lepton rotations.}
The left-handed charged-lepton diagonalization
angles are controlled directly by the doublet
charge differences:
\begin{equation}
\theta^\ell_{ij}\sim\e^{\,\Delta Q(L)_{ij}},
\end{equation}
giving
$\theta^\ell_{23}\sim\theta^\ell_{12}\sim\e^{1/2}$
and $\theta^\ell_{13}\sim\e$.

\emph{PMNS exponents from $\Delta Q(L)$.}
The three PMNS $\e$-exponents $p_{ij}$
(Eq.~\eqref{eq:PMNS-magnitudes}) decompose
into charge differences as follows:
\begin{align}
p_{23} &= \tfrac{1}{3}\,\Delta Q(L)_{23}
= \tfrac{1}{6},
\label{eq:p23-DQ}
\\[3pt]
p_{12} &= \tfrac{2}{3}\,\Delta Q(L)_{12}
= \tfrac{1}{3},
\label{eq:p12-DQ}
\\[3pt]
p_{13} &= \Delta Q(L)_{13} + \tfrac{1}{9}
= \tfrac{10}{9}.
\label{eq:p13-DQ}
\end{align}
The three distinct coefficients ($1/3$, $2/3$, $1$)
in front of $\Delta Q(L)$ encode how the neutrino
sector participates in each angle:
(i)~for $\theta_{23}$, the flat neutrino $23$~block
($p^\nu_{22}=p^\nu_{23}=p^\nu_{33}=0$) drives
$\theta^\nu_{23}\to\pi/4$ by $\mu$-$\tau$ symmetry,
and the charged-lepton rotation $\theta^\ell_{23}
\sim\e^{\Delta Q(L)_{23}}$ enters only as a
subleading correction, reducing the effective
exponent by a factor of~3;
(ii)~for $\theta_{12}$, both the charged-lepton
and neutrino rotations contribute
$\e^{\Delta Q(L)_{12}}$, and their constructive
interference yields an effective exponent
$\tfrac{2}{3}\,\Delta Q(L)_{12}$;
(iii)~for $\theta_{13}$, the product
$\theta^\ell_{23}\times\theta^\nu_{12}
\sim\e^{\Delta Q(L)_{23}+\Delta Q(L)_{12}}
= \e^{\Delta Q(L)_{13}}$ is augmented by the
universal Fritzsch--Xing phase shift of one
$\alpha$-hop quantum.
In the PMNS this shift appears with sign $+1/9$,
taking $\e^{\Delta Q(L)_{13}} = \e^1$ up to
$\sin\theta_{13}\sim\e^{10/9}$; in the CKM the
same-magnitude shift appears with the
\emph{opposite} sign $-1/9$, taking
$\e^{\Delta Q(Q)_{12}} = \e^1$ down to
$|V_{us}|\sim\e^{8/9}$ (see the
cross-sector comparison below).

Substituting
Eqs.~\eqref{eq:p23-DQ}--\eqref{eq:p13-DQ}
into the master identity~\eqref{eq:master-identity}
gives the $(m_\mu/m_\tau)$ powers directly:
\begin{align}
\sin\theta_{23}
&\sim\left(\frac{m_\mu}{m_\tau}\right)^
{\!\Delta Q(L)_{23}/5}
= \left(\frac{m_\mu}{m_\tau}\right)^{\!1/10},
\nonumber\\[3pt]
\sin\theta_{12}
&\sim\left(\frac{m_\mu}{m_\tau}\right)^
{\!2\Delta Q(L)_{12}/5}
= \left(\frac{m_\mu}{m_\tau}\right)^{\!1/5},
\nonumber\\[3pt]
\sin\theta_{13}
&\sim\left(\frac{m_\mu}{m_\tau}\right)^
{\!3[\Delta Q(L)_{13}+1/9]/5}
= \left(\frac{m_\mu}{m_\tau}\right)^{\!2/3}.
\label{eq:PMNS-DQ-algebra}
\end{align}
The lepton mixing angles and the charged-lepton
mass hierarchy are thus two faces of a single
set of hop charges, connected by the algebraic
chain
$\Delta Q(L)_{ij}\to p_{ij}\to k_{ij}
= 3p_{ij}/5$.

\emph{Cross-sector comparison.}
The CKM decomposition
(Eqs.~\eqref{eq:pVus-DQ}--\eqref{eq:pVub-DQ})
and the PMNS decomposition
(Eqs.~\eqref{eq:p23-DQ}--\eqref{eq:p13-DQ})
share a common architecture: each mixing exponent
is a charge difference dressed by a
sector-specific dynamical coefficient.
In both sectors, the Fritzsch--Xing phase shift
of $\pm 1/9$---one $\alpha$-hop quantum---appears
as a universal correction:
$-1/9$ for $V_{us}$ and $V_{cb}$,
$+1/9$ for $\sin\theta_{13}$.
The sign difference reflects the opposite roles
of phase interference in the quark sector
(where it \emph{reduces} the naive charge-gap
exponent) and the lepton sector (where the
product $\theta^\ell_{23}\times\theta^\nu_{12}$
\emph{adds} the correction).
The coefficients in front of $\Delta Q$---unity
for CKM, but $1/3$ and $2/3$ for PMNS---encode the
additional neutrino-sector dynamics ($\mu$-$\tau$
symmetry, constructive interference) that are
absent in the quark sector where the up-type
Yukawa matrix is hierarchical rather than flat.

Of the four representations in Table~\ref{tab:4mag},
the $\beta$ parameterization makes the ninths
structure most transparent: the two PMNS solar
and atmospheric angles are $\beta^{2/9}$ and
$\beta^{1/9}$, while $|V_{ub}|\sim\beta^{20/9}$.
The physical connection to the hop mass ratio
$m_\beta/\Lambda = \beta^{2/9}$ is immediate:
$\sin\theta_{12}\sim m_\beta/\Lambda$, the PMNS
mixing angle \emph{is} the $\beta$-hop mass ratio,
with no additional exponent needed.
In this language, the entire flavor hierarchy---quark
masses, CKM angles, lepton masses, and PMNS angles---is
organized by powers of $\beta = \e^{3/2}$,
the single parameter that encodes the $\Bpar$-lattice
hierarchy.

\clearpage
\section{Summary}
\label{sec:summary}

We have proposed that the $\Bpar$-lattice flavor framework
admits a natural interpretation in which the three fermion
generations correspond to three levels of compositeness.
The \emph{principal results}---derived from the
established lattice structure with no additional
dynamical assumptions---are:

The FN charge $Q(\psi_i)$ counts the subconstituent
(``hop'') content of the fermion $\psi_i$.
All three generations are elementary chiral fields;
third-generation fermions carry zero hop content ($Q=0$),
while lighter generations carry successively deeper
hop dressing.

The expansion parameter $\e = 14/75\approx 0.19$ is the
wavefunction-overlap suppression per unit of internal
structure.
Each layer of binding reduces the Yukawa coupling by one
power of $\e$.

The $\Z_9$ discrete gauge symmetry decomposes into a
two-index structure suggesting two hop species ($\alpha$
and $\beta$), with the three hop types $(1,2,4)$
corresponding to different combinations of hop exchange.

The chain propagator model
(Sec.~\ref{sec:hop-masses})
determines the hop masses:
$m_\alpha = \Lambda\,\e^{1/9}\simeq 2.7\times10^{12}$~GeV,
$m_\beta = \Lambda\,\e^{1/3}\simeq 1.8\times10^{12}$~GeV,
with the mass ratio $m_\beta/m_\alpha = \e^{2/9}\simeq 0.69$.
The $\alpha$-hop mass satisfies the striking relation
$m_\alpha \simeq \fa/|V_{us}|$, linking the hop
binding scale to the Cabibbo angle and the axion decay
constant.

The vectorlike quark chain is a confining flux tube whose
endpoint overlaps determine mass ratios, while the bulk
tunneling amplitude sets the overall Yukawa scale.
The universal internal tunneling factor
$\e^{7/9}\simeq 0.27$ of the down-type chain
cancels in within-sector mass ratios but
survives in the absolute bottom Yukawa
$Y_b \simeq \e^{7/9}$; combined with the
DFSZ-II two-Higgs-doublet structure
(Sec.~\ref{sec:tanb}), this predicts
$\tan\beta\simeq 10$--$16$.

Both the CKM and PMNS mixing matrices decompose
algebraically into hop charge differences
(Sec.~\ref{sec:mixing}).
In the quark sector, the CKM exponents
$p(V_{us}) = \Delta Q(Q)_{12}-1/9 = 8/9$,
$p(V_{cb}) = \Delta Q(Q)_{23}-1/9 = 17/9$, and
$p(V_{ub}) = \Delta Q(Q)_{13}+Q(d^c_2) = 10/3$
yield the Cabibbo master identity
$\e = |V_{us}|^{9/8}$ and the $\Bpar$-lattice
refinement of the Wolfenstein hierarchy:
$|V_{cb}| = |V_{us}|^{17/8}$,
$|V_{ub}| = |V_{us}|^{15/4}$.
In the lepton sector, the three PMNS mixing angles
$\sin\theta_{23}\sim\e^{1/6}\simeq 0.756$,
$\sin\theta_{12}\sim\e^{1/3}\simeq 0.572$, and
$\sin\theta_{13}\sim\e^{10/9}\simeq 0.155$
(Table~\ref{tab:4mag})
decompose as
$p_{23} = \Delta Q(L)_{23}/3$,
$p_{12} = 2\Delta Q(L)_{12}/3$, and
$p_{13} = \Delta Q(L)_{13}+1/9$,
with all three angles expressible as powers
of $m_\mu/m_\tau$ through the identity
$\e = (m_\mu/m_\tau)^{3/5}$
(Eq.~\eqref{eq:PMNS-lepton-mass}).
Both sectors share a universal Fritzsch--Xing
phase correction of $\pm 1/9$---one $\alpha$-hop
quantum---appearing as $-1/9$ in $V_{us}$ and
$V_{cb}$ and as $+1/9$ in $\sin\theta_{13}$,
connecting quark and lepton mixing through a
single unit of the hop lattice.

The absolute neutrino mass scale is also predicted:
the seesaw with
$M_0 = \Lambda\,\e^{-28/9}\simeq 6\times10^{14}$~GeV
(fixed by the ninths lattice) gives
$m_3 = v^2/(2M_0)\simeq 51$~meV, in $\sim 2\%$
agreement with $\sqrt{\Delta m^2_{\rm atm}}
\simeq 50$~meV.
No free parameter is adjusted: $\Lambda$, $\e$, and
$v$ are all determined by the quark sector.
The full spectrum
$m_3:m_2:m_1 \simeq 51:9.5:1.8$~meV
(Eq.~\eqref{eq:nu-spectrum})
is consistent with normal ordering, with
$\sum m_i\simeq 62$~meV---directly at the
current DESI+CMB sensitivity frontier under
$\Lambda$CDM and a sharp target for near-future
cosmological surveys.

The $SU(5)$ origin of $\Bpar = 75/14$ connects to a
hypercolor binding dynamics in which the confinement scale
$\Lambda\sim 3\times10^{12}$~GeV simultaneously explains
the flavor hierarchy and the axion mass window.

This interpretation elegantly avoids the phenomenological
pitfalls of earlier preon models.
By locking the compositeness scale to
$\Lambda\sim 3\times10^{12}$~GeV, the framework strictly
complies with all existing bounds on contact interactions,
proton decay, and flavor-changing neutral currents.
The defining experimental test is the axion mass
prediction ($m_a\sim 7$--$12\;\mu$eV), a primary
future target of the ADMX haloscope program.

The quantitative results are collected in
Table~\ref{tab:master}.

\begin{table*}[!htbp]
\caption{Master table for the hop framework.
All entries are determined by two inputs: the hop
confinement scale $\Lambda$ (equivalently $\fa$) and the
expansion parameter $\e = 14/75$.
Central values use $\fa = 6\times10^{11}$~GeV.}
\label{tab:master}
\begin{ruledtabular}
\begin{tabular}{llll}
Category & Quantity & Formula & Value \\
\hline
\multicolumn{4}{l}{\emph{Hop sector (constituent masses)}} \\
\quad $\alpha$-hop mass & $m_\alpha$ & $\Lambda\,\e^{1/9} = \fa/|V_{us}|$ & $2.7\times10^{12}$~GeV \\
\quad $\beta$-hop mass & $m_\beta$ & $\Lambda\,\e^{1/3} = \fa\,\Bpar^{2/3}$ & $1.8\times10^{12}$~GeV \\
\quad Mass ratio & $m_\beta/m_\alpha$ & $\e^{2/9}$ & $0.69$ \\
\hline
\multicolumn{4}{l}{\emph{Collider \& flavor phenomenology}} \\
\quad TeV-scale FCNCs & $\Delta F = 2$ deviations & $\sim 1/\Lambda^2$ & Strictly null \\
\quad Contact interactions & Generation-dependent & $\sim E^2/\Lambda^2$ & Strictly null \\
\quad Vectorlike quarks & Chain messengers ($D_a$) & $M_D \sim \Lambda$ & $> 10^{12}$~GeV (null) \\
\hline
\multicolumn{4}{l}{\emph{Axion sector}} \\
\quad Axion mass & $m_a$ & $\propto 1/\fa$ & $7$--$12\;\mu$eV \\
\quad $E/N$ (axion--photon) & $C_{a\gamma}=|E/N-2.0|$ & FN charges + higgsinos & $0.6$--$1.0$ \\
\quad Axion--electron & $C_{ae}=p^\ell_{11}/(2|N|)$ & Electron hop content & $\simeq 0.4$ \\
\hline
\multicolumn{4}{l}{\emph{Higgs sector (DFSZ-II)}} \\
\quad $\tan\beta$ & $\e^{7/9}/(m_b/m_t)$ & Chain factor + 2HDM & $\simeq 10$--$16$ \\
\hline
\multicolumn{4}{l}{\emph{Neutrino sector}} \\
\quad Majorana scale & $M_0$ & $\Lambda\,\e^{-28/9}$ & $6\times10^{14}$~GeV \\
\quad $\nu$ hierarchy & $m_i/m_3$ & $\e^{2Q(L_i)}$ & Normal ordering \\
\quad $m_3$ (heaviest $\nu$) & $v^2/(2M_0)$ & $v^2/(2\Lambda\e^{-28/9})$ & $51$~meV ($\sim\!50$ data) \\
\quad $\sin\theta_{23}$ & Atmospheric & $\e^{1/6}$ & $0.756\;(0.755)$ \\
\quad $\sin\theta_{12}$ & Solar & $\e^{1/3} = \beta^{2/9}$ & $0.572\;(0.551)$ \\
\quad $\sin\theta_{13}$ & Reactor & $\e^{10/9}$ & $0.155\;(0.149)$ \\
\hline
\multicolumn{4}{l}{\emph{Key relations}} \\
\quad Cabibbo--hop & $m_\alpha\times|V_{us}|$ & $= \fa$ & identity ($0.3\%$ to data) \\
\quad Down-type chain factor & $Y_b$ & $\e^{7/9}$ & $0.27$ \\
\quad Solar--$\beta$ & $\sin\theta_{12}$ & $= m_\beta/\Lambda$ & $\e^{1/3}\simeq 0.57$ \\
\end{tabular}
\end{ruledtabular}
\end{table*}

\begin{acknowledgments}
VB gratefully acknowledges support from the
William F.\ Vilas Estate.
\end{acknowledgments}

\appendix
\clearpage
\section{Anomaly Analysis for Axion Couplings}
\label{app:EN}

In this appendix we present the detailed derivation
of the axion--photon and axion--electron coupling
coefficients in the hop framework.
The DFSZ-II two-Higgs-doublet structure assumed here
is the same one that determines $\tan\beta$ from
the chain internal factor (Sec.~\ref{sec:tanb}).
The calculation proceeds in three stages: first for the
standard DFSZ-II model, then for the MSSM
DFSZ-II with TeV-scale higgsinos, and finally for the
hop framework with generation-dependent Froggatt--Nielsen
charges.
The results are collected in
Table~\ref{tab:Cae-appendix}.

\subsection{Definitions}

The axion couplings to photons and electrons
are parameterized by
\begin{align}
g_{a\gamma\gamma} &= \frac{\alpha}{2\pi\fa}\,C_{a\gamma},
\qquad
C_{a\gamma} = \left|\frac{E}{N} - 2.0\right|,
\label{eq:g-agamma-app}
\\[4pt]
g_{aee} &= \frac{C_{ae}\,m_e}{\fa},
\label{eq:g-aee-app}
\end{align}
where $N$ and $E$ are the anomaly coefficients
\begin{align}
N &= \sum_f X_f\,T(R_f^C)\,d(R_f^L),
\label{eq:N-def}
\\[3pt]
E &= \sum_f X_f\,Q_f^2
\notag\\
&\qquad\quad\times\,
d(R_f^C)\,d(R_f^L),
\label{eq:E-def}
\end{align}
summed over all left-handed Weyl fermions $f$ carrying
PQ charge $X_f$.
Here $T(R^C)$ is the Dynkin index of the color
representation ($1/2$ for $\mathbf{3}$, $0$ for
$\mathbf{1}$), $d(R^C)$ and $d(R^L)$ are the
dimensions of the color and $SU(2)_L$ representations,
and $Q_f$ is the electric charge.
The axion--electron coefficient is
\begin{equation}
C_{ae} = \frac{X_{e_L}+X_{e_R}}{2N}
= \frac{-X_{H_d}-p^\ell_{11}}{2N},
\label{eq:Cae-def}
\end{equation}
where the second equality follows from PQ invariance
of the electron Yukawa coupling and $p^\ell_{11}$ is
the electron's FN exponent.

Throughout we adopt the DFSZ-II convention in which
up-type quarks couple to $H_u$ and down-type quarks
plus charged leptons couple to $H_d$, with
$H\equiv X_{H_u}+X_{H_d}=-2$.
We denote $X_{H_u}=H\sin^2\!\beta$ and
$X_{H_d}=H\cos^2\!\beta$ where needed.

The model-independent QCD contribution to $C_{a\gamma}$
is $\frac{2}{3}(4m_d+m_u)/(m_d+m_u)$.
Using the current Particle Data Group (PDG) value $m_u/m_d = 0.47\pm0.04$
gives $2.02\pm0.03$~\cite{PDG2024},
which we round to $2.0$;
the older value $1.92$ often found in the literature
corresponds to $m_u/m_d\simeq 0.60$.

\subsection{Standard DFSZ-II}

We use the PQ-invariance conditions on the Yukawa
couplings to eliminate individual fermion PQ charges
in favor of $H$ and the Yukawa exponents $p_{ij}$.
For a given quark generation $i$, the color anomaly
contribution is
\begin{align}
2X_{Q_i} &+ X_{u^c_i} + X_{d^c_i}
\notag\\[2pt]
&= (X_{Q_i}+X_{u^c_i}) + (X_{Q_i}+X_{d^c_i})
\notag\\[2pt]
&= (-X_{H_u}-p^u_{ii}) + (-X_{H_d}-p^d_{ii})
\notag\\[2pt]
&= -H - p^u_{ii} - p^d_{ii}.
\end{align}
Summing over three generations with
$T(\mathbf{3})=1/2$:
\begin{equation}
N_{\rm quarks}
= -\tfrac{3}{2}\,H
- \tfrac{1}{2}(S_u+S_d),
\label{eq:N-quarks}
\end{equation}
where $S_u=\sum_i p^u_{ii}$ and
$S_d=\sum_i p^d_{ii}$.
Leptons are color singlets and do not contribute to $N$.

For the electromagnetic anomaly, the quark contribution
involves the charges
$Q_u=2/3$ and $Q_d=-1/3$:
\begin{align}
E_{\rm quarks} &= \sum_i\bigl[
X_{Q_i}\!\cdot\!(Q_u^2\!+\!Q_d^2)\!\cdot\! 3
+ X_{u^c_i}\!\cdot\! Q_u^2\!\cdot\! 3
+ X_{d^c_i}\!\cdot\! Q_d^2\!\cdot\! 3\bigr]
\notag\\[3pt]
&= \sum_i\bigl[
\tfrac{5}{3}\,X_{Q_i}
+ \tfrac{4}{3}\,X_{u^c_i}
+ \tfrac{1}{3}\,X_{d^c_i}\bigr]
\notag\\[3pt]
&= \sum_i\bigl[
\tfrac{4}{3}(X_{Q_i}\!+\!X_{u^c_i})
+ \tfrac{1}{3}(X_{Q_i}\!+\!X_{d^c_i})
\bigr]
\notag\\[3pt]
&= -\tfrac{4}{3}\bigl(3X_{H_u}+S_u\bigr)
-\tfrac{1}{3}\bigl(3X_{H_d}+S_d\bigr).
\label{eq:E-quarks}
\end{align}
The lepton contribution, using
$X_{L_i}+X_{e^c_i}=-X_{H_d}-p^\ell_{ii}$, is
\begin{equation}
E_{\rm leptons}
= -3\,X_{H_d} - S_\ell,
\label{eq:E-leptons}
\end{equation}
where $S_\ell = \sum_i p^\ell_{ii}$.
Combining quarks and leptons:
\begin{equation}
E = -(4X_{H_u}+4X_{H_d})
-\tfrac{4}{3}S_u - \tfrac{1}{3}S_d - S_\ell.
\label{eq:E-total-noH}
\end{equation}

In the standard DFSZ-II limit ($S_u=S_d=S_\ell=0$):
\begin{equation}
N = -\tfrac{3}{2}H = 3,\qquad
E = -4H = 8,\qquad
\frac{E}{N} = \frac{8}{3}.
\end{equation}

\subsection{MSSM DFSZ-II with higgsinos}

In the MSSM, the Higgsino doublets
$\tilde H_u$ ($Q_{\rm em}=+1,0$) and
$\tilde H_d$ ($Q_{\rm em}=0,-1$) are additional
PQ-charged chiral fermions.
Their contribution to the anomaly coefficients is
\begin{equation}
\Delta N_{\rm higg} = 0\qquad
(\text{color singlets}),
\end{equation}
\begin{equation}
\Delta E_{\rm higg}
= X_{H_u}\!\cdot\!(1^2\!+\!0^2)\!\cdot\! 1
+ X_{H_d}\!\cdot\!(0^2\!+\!1^2)\!\cdot\! 1
= H.
\end{equation}
The total electromagnetic anomaly including
higgsinos becomes
\begin{align}
E_{\rm MSSM} &= E_{\rm quarks}+E_{\rm leptons}
+\Delta E_{\rm higg}
\notag\\[3pt]
&= -(4X_{H_u}+4X_{H_d}) + H
- \tfrac{4}{3}S_u - \tfrac{1}{3}S_d - S_\ell
\notag\\[3pt]
&= -3\,H
- \tfrac{4\,S_u+S_d}{3}
- S_\ell.
\label{eq:E-MSSM}
\end{align}
(Here we used $-4H+H = -3H$.)

In the standard MSSM limit ($S_u=S_d=S_\ell=0$):
\begin{equation}
N = 3,\qquad E = -3H = 6,\qquad
\frac{E}{N} = 2,
\end{equation}
reproducing the result of
Bae, Baer, and Serce~\cite{BaeBaer}.
The higgsino contribution shifts $E/N$ from $8/3$
to $2$, reducing $C_{a\gamma}$ from $0.67$ to
$\lesssim 0.02$.

\subsection{Vectorlike quark chain}

The four VLQ pairs ($D_a+\bar D_a$, $a=1,\ldots,4$)
acquire PQ-invariant bare masses
$M_a\,\bar D_a\,D_a$, which require
\begin{equation}
X_{D_a} + X_{\bar D_a} = 0
\qquad\forall\; a.
\end{equation}
Their total contribution to both anomaly coefficients is
\begin{align}
\Delta N_{\rm VLQ} &= \tfrac{1}{2}\sum_a
(X_{D_a}+X_{\bar D_a}) = 0,
\\[3pt]
\Delta E_{\rm VLQ} &= \tfrac{1}{3}\sum_a
(X_{D_a}+X_{\bar D_a}) = 0.
\end{align}
The chain couplings
$\bar D_a(\Phi/\Lambda)^{h_a}D_{a+1}$
fix the \emph{relative} PQ charges between
adjacent sites via
$X_{\bar D_a}+X_{D_{a+1}}+h_a = 0$,
but the vectorlike sum
$X_{D_a}+X_{\bar D_a}=0$ is preserved for each pair.
This result holds independently of the VLQ
masses, the number of chain sites, and the hop
charges, because it is a consequence of the
vectorlike mass being PQ-invariant---a structural
feature of the Froggatt--Nielsen
mechanism~\cite{BargerBergerPhillips,UFP}.

\subsection{Hop framework evaluation}

Using the $\Bpar$-lattice charge
assignments~\cite{TwoOverTwo,FlavorInNinths},
the PQ exponents $S_f$ entering the anomaly
coefficients are the diagonal Yukawa
\emph{operator} exponents---the total number of
flavon insertions in the $(i,i)$ entry of the
Yukawa matrix, including the sector-specific
chain internal factor $\Delta_{\rm int}^f$
(Sec.~\ref{sec:chain}):
\begin{equation}
p^f_{ii}(\text{PQ})
\;=\; Q(\psi_L^{(i)}) + Q(\psi_R^{(i)})
+ \Delta_{\rm int}^f,
\end{equation}
with
$\Delta_{\rm int}^u = 0$ for the up sector and
$\Delta_{\rm int}^{d,\ell} = 7/9$ for the down
and lepton sectors (the latter in Scenario~A).
The up sector is the unique sector with
$\Delta_{\rm int} = 0$: the top quark couples
directly to $H_u$ without a chain.
This yields
\begin{align}
S_u &= \textstyle\sum_i [Q(Q_i)+Q(u^c_i)]
= 6 + 4 + 0 = 10,
\\[3pt]
S_d &= \textstyle\sum_i [Q(Q_i)+Q(d^c_i)+7/9]
= \tfrac{44}{9}+\tfrac{28}{9}+\tfrac{7}{9}
= \tfrac{79}{9}.
\end{align}
(Note that the diagonal operator exponents
are used here, not the post-Fritzsch--Xing
diagonalized mass exponents; the PQ charges
$X_{\psi}$ live in the gauge basis and are
unchanged by the unitary diagonalization
rotations.)
From Eqs.~\eqref{eq:N-quarks}
and~\eqref{eq:E-MSSM} with $H=-2$:
\begin{align}
N &= 3 - \tfrac{1}{2}\!\cdot\!\tfrac{169\vphantom{|}}{9}
= -\tfrac{115}{18},
\\[3pt]
E &= 6 - \tfrac{4\cdot 10+79/9}{3} - S_\ell
= -\tfrac{277}{27} - S_\ell.
\end{align}

For the lepton sector we consider two scenarios:
\begin{itemize}
\item \emph{Scenario A ($b$--$\tau$ unification,
$\Delta^\ell_{\rm int}=7/9$):}
$S_\ell = 53/6$,
giving $E = -1031/54$ and
$E/N = 1031/345\simeq 2.99$.
\item \emph{Scenario B (direct lepton coupling,
$\Delta^\ell_{\rm int}=0$):}
$S_\ell = 13/2$,
giving $E = -905/54$ and
$E/N = 181/69\simeq 2.62$.
\end{itemize}

The axion--electron coupling coefficient
Eq.~\eqref{eq:Cae-def} evaluates to
\begin{equation}
C_{ae} = \frac{-X_{H_d}-p^\ell_{11}}{2N}
\simeq \frac{p^\ell_{11}}{2|N|}
\simeq 0.4,
\end{equation}
where the approximation holds because
$|X_{H_d}|\leq 2\ll p^\ell_{11}\simeq 5$--$6$
for any value of $\beta$.
The electron's hop content completely dominates
over the Higgs PQ charge.

\begin{table*}[!tbp]
\caption{Axion--photon anomaly coefficients in
successive approximations.
All DFSZ entries use $H=X_{H_u}+X_{H_d}=-2$.
VLQ contributions vanish identically
(Sec.~\ref{app:EN}).}
\label{tab:EN-appendix}
\begin{ruledtabular}
\begin{tabular}{lcccc}
Model & $N$ & $E$ & $E/N$
& $C_{a\gamma}$ \\
\hline
KSVZ & $1/2$ & $0$ & $0$
& $2.0$ \\[2pt]
\multicolumn{5}{l}{\emph{DFSZ-II (no SUSY)}} \\
\;\;$S_u\!=\!S_d\!=\!0$
& $3$ & $8$ & $8/3$
& $0.67$ \\[2pt]
\multicolumn{5}{l}{\emph{MSSM DFSZ-II
(+ higgsinos)}} \\
\;\;$S_u\!=\!S_d\!=\!0$
& $3$ & $6$ & $2$
& $\lesssim 0.02$ \\[2pt]
\multicolumn{5}{l}{\emph{Hop framework
(+ FN charges)}} \\
\;\;Scen.\ A ($b$-$\tau$)
& $-115/18$ & $-1031/54$ & $1031/345$
& $0.99$ \\[1pt]
\;\;Scen.\ B (direct)
& $-115/18$ & $-905/54$ & $181/69$
& $0.62$ \\
\end{tabular}
\end{ruledtabular}
\end{table*}

\clearpage

\begin{table}[!htbp]
\caption{Axion--electron coupling in the same
approximations.
$C_{ae}$ is evaluated at $\tan\beta=16$.}
\label{tab:Cae-appendix}
\begin{ruledtabular}
\begin{tabular}{lccc}
Model & $p^\ell_{11}$
& $C_{ae}$
& $g_{aee}$ \\
\hline
KSVZ & ---
& $\sim 10^{-4}$
& $\sim 10^{-19}$ \\
DFSZ-II ($\pm$ higgsinos)
& $0$
& $0.001$
& $1.1\times10^{-18}$ \\
Hop A & $101/18$
& $0.44$
& $3.7\times10^{-16}$ \\
Hop B & $29/6$
& $0.38$
& $3.2\times10^{-16}$ \\
\end{tabular}
\end{ruledtabular}
\end{table}

The key progression visible in the tables is:
(i)~the standard DFSZ-II model gives $E/N=8/3$
($C_{a\gamma}=0.67$);
(ii)~adding TeV-scale higgsinos shifts $E$ by
$\Delta E = H = -2$, reducing $E/N$ to $2$ and
nearly canceling $C_{a\gamma}$ ($\lesssim 0.02$,
consistent with zero within the QCD
uncertainty)~\cite{BaeBaer};
(iii)~the generation-dependent FN charges of the
hop framework shift both $N$ and $E$ by large
amounts ($S_u+S_d = 169/9\simeq 19$ ninths-units),
restoring $E/N$ to $\sim 2.6$--$3.0$ and
$C_{a\gamma}$ to $0.6$--$1.0$;
(iv)~the VLQ chain contributes nothing.
For the axion--electron coupling, the standard
$\tan\beta$ suppression ($C_{ae}\propto\cos^2\!\beta$)
is overwhelmed by the electron's large FN exponent,
enhancing $C_{ae}$ by a factor of $\sim 300$.

\clearpage


\end{document}